\documentclass[pre,aps,eqsecnum,groupedaddress,showkeys,showpacs,amsmath,nofootinbib]{revtex4}
\usepackage{amssymb}
\usepackage{amsmath}
\usepackage{graphicx}
\usepackage{amsbsy}
\usepackage{color}
\usepackage{verbatim}
\usepackage{bm}

\begin{document}

\title{Flory theory for Polymers }

\author{Somendra M. Bhattacharjee}
\affiliation{Institute of Physics, Bhubaneswar, 751 005 India}
\email{somen@iopb.res.in}

\author{Achille Giacometti}
\affiliation{Dipartimento di Scienze Molecolari e Nanosistemi, Universit\`a Ca' Foscari Venezia,
Calle Larga S. Marta DD2137, I-30123 Venezia, Italy}
\email{achille@unive.it}

\author{Amos Maritan}
\affiliation{Dipartimento di Fisica e Astronomia, Universit\`{a} di Padova, via
Marzolo 8 I-35131 Padova}
\affiliation{CNISM, Unit\`a di Padova, Via Marzolo 8, I-35131 Padova, Italy}
\affiliation{Sezione INFN, Universit\`a di Padova, I-35131 Padova, Italy}
\email{maritan@pd.infn.it}

%%%%%%%%%%%%%%%%%%%%%%%%%%%%%%%%%%%%%%%%%%%%%%%%%%%%%%%%%%%%%%%%%%%%%%%%%%%%%%%
\date{\today}
\begin{abstract}
  We  review various simple analytical theories for
  homopolymers within a unified framework.  The common
  guideline of our approach is the Flory theory, and  its
  various avatars, with the attempt of being reasonably
  self-contained.  We expect this review to be useful as an
  introduction to the topic at the graduate students level.
\end{abstract}
%%%%%%%%%%%%%%%%%%%%%%%%%%%%%%%%%%%%%%%%%%%%%%%%%%%%%%%%%%%%%%%%%%%%%%%%%%%%%%%
%\pacs{...}
%\keywords{ ...}

%%%%%%%%%%%%%%%% FIG1%
\newcommand{\figone}{%
\begin{figure}[htbp]
\centering
\includegraphics[width=0.5\textwidth]{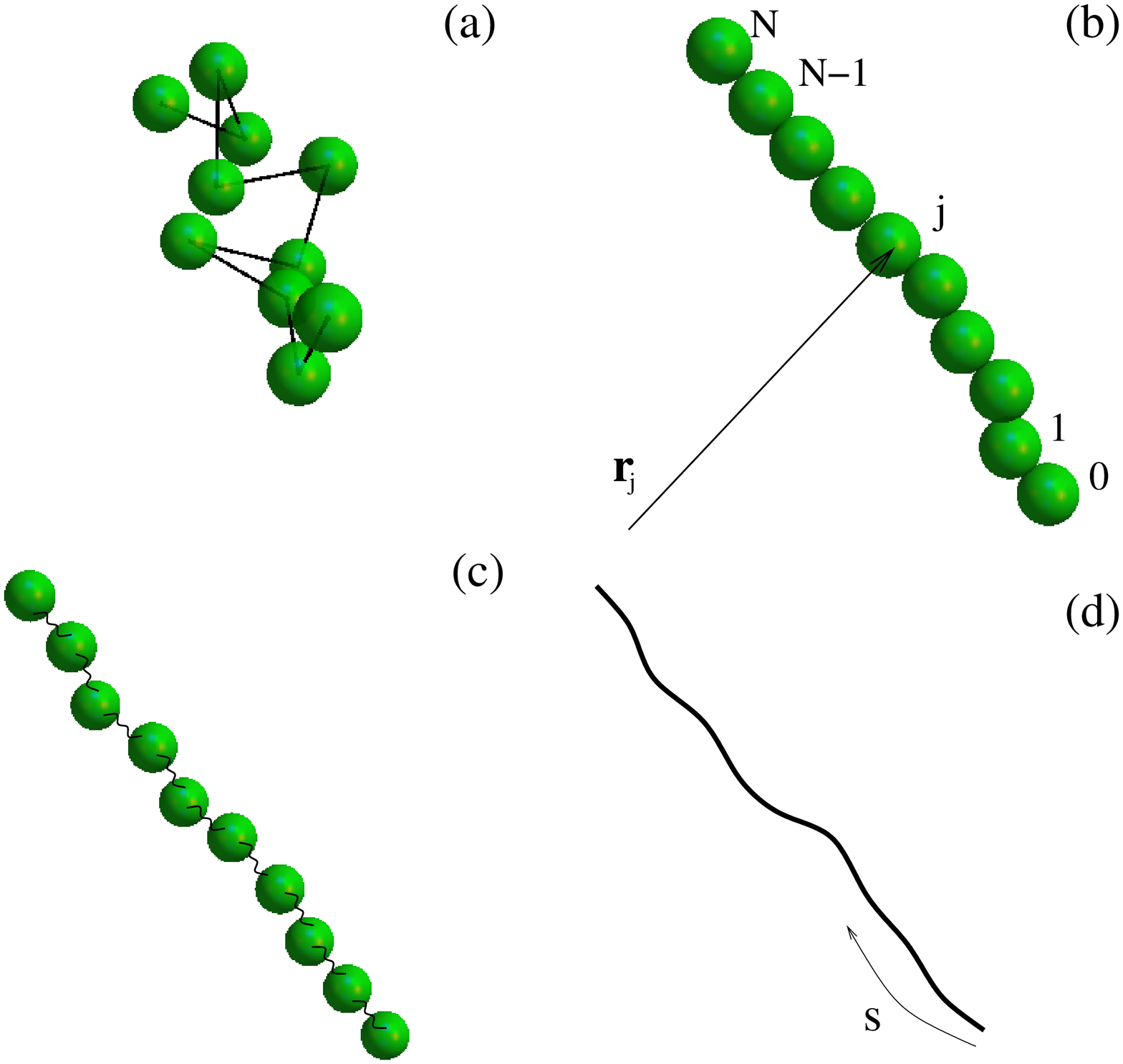}
\caption{Various representations of a polymer. (a) Freely jointed
  chain: Rigid bonds with full free rotations.  The beads and bonds
  may cross without any penalty. (b) A collection of $N$ tethered
  spheres (monomers) at positions $\mathbf{r}_j$, with
  $j=0,1,\ldots,N$. The size of the monomers could be indicative of the
  excluded volume interaction of the monomers. (c) A bead spring model
  where the harmonic springs take care of the polymer connectivity.
  (d) Continuum model - no details of the polymeric structure is
  important.  The location of a monomer is given by a length $s$.}
\label{fig:fig1}
\end{figure}
}%
%%%%%%%%%%%%%%

%%%%%%%%%%%%%% FIG2%
\newcommand{\figtwo}{%
\begin{figure}[htbp]
\begin{center}
\includegraphics[width=0.50\textwidth]{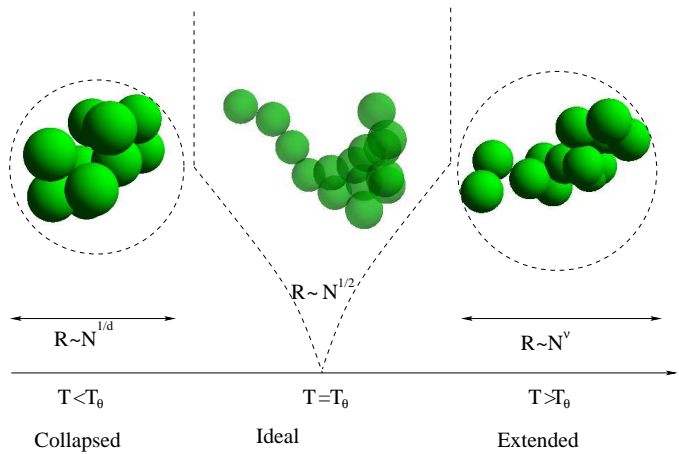}
\end{center}
\caption{Schematic phase diagram of an isolated homopolymer. At high
  temperature $T > T_{\theta}$, the polymer is in a swollen  phase
  (right), whereas one expects a compact globule at sufficiently low
  temperatures $T < T_{\theta}$ (left). These two regimes are
  separated by a transition regime at $T=T_{\theta}$ (center) where
  the polymer behaves more or less  as a Gaussian  chain, 
  at least in $d>3$.
  }
\label{fig:fig2}
\end{figure}
}%
%%%%%%%%%%%%%
%%%%%%%%%%%%%% FIG3%
\newcommand{\figthree}{%
\begin{figure}[htbp]
\begin{center}
\includegraphics[width=0.35\textwidth]{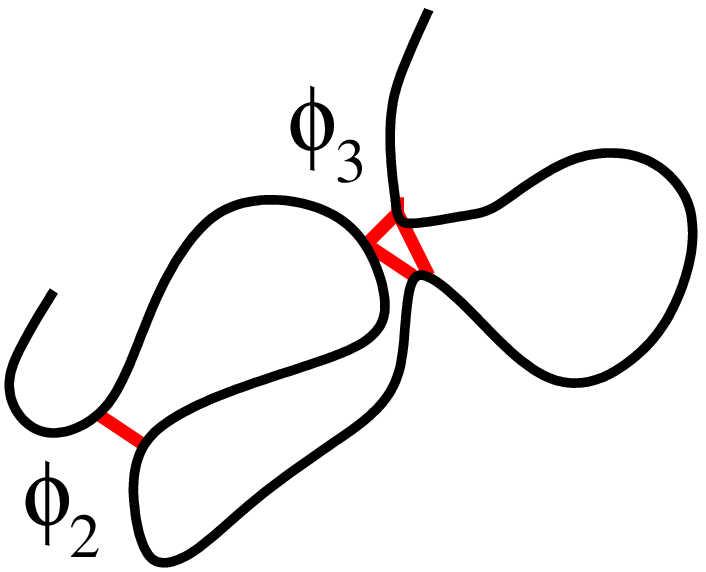}
\end{center}
\caption{Schematic representation of the two-body $\Phi_2$ and the three-body $\Phi_3$ 
interactions. For contact interactions 
$\Phi_2({\bf{r}},{\bf{r}^{\prime}} ) = u \delta({\bf{r}},{\bf{r}^{\prime}} )$,
$\Phi_3({\bf{r}},{\bf{r}^{\prime}},{\bf{r}^{\prime\prime}} ) =v \delta^d\left(\mathbf{r}-
  \mathbf{r}^{\prime}\right) \delta^d\left(\mathbf{r}^{\prime}-
  \mathbf{r}^{\prime\prime}\right)$. 
}
\label{fig:fig3}
\end{figure}
}%
%%%%%%%%%%%%%

%%%%%%%%%%%% FIG4 %
%
\newcommand{\figfour}{%
\begin{figure}[htbp]
\begin{center}
\includegraphics[width=0.5\textwidth,clip]{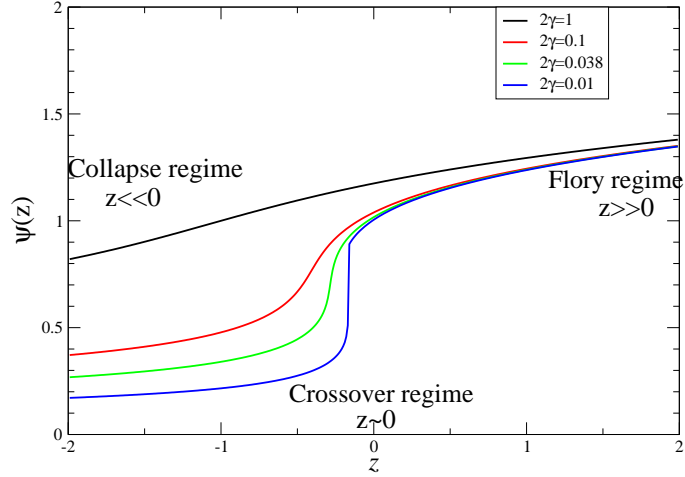}
\end{center}
\caption{Numerical solution of Eq.(\ref{elastic:eq8}) as in Fig1 of Ref.\cite{deGennes75} with the
same values of the parameters $2 \gamma=0.01,0.038,0.1,1$.}
\label{fig:fig4}
\end{figure}
}%
%%%%%%%%%%%%%%%%

%%%%%%% FIG5 %%%%%%%%%%%%%%%%%%%%%%%
\newcommand{\figfive}{%
\begin{figure}[htbp]
\begin{center}
\includegraphics[width=6cm,clip]{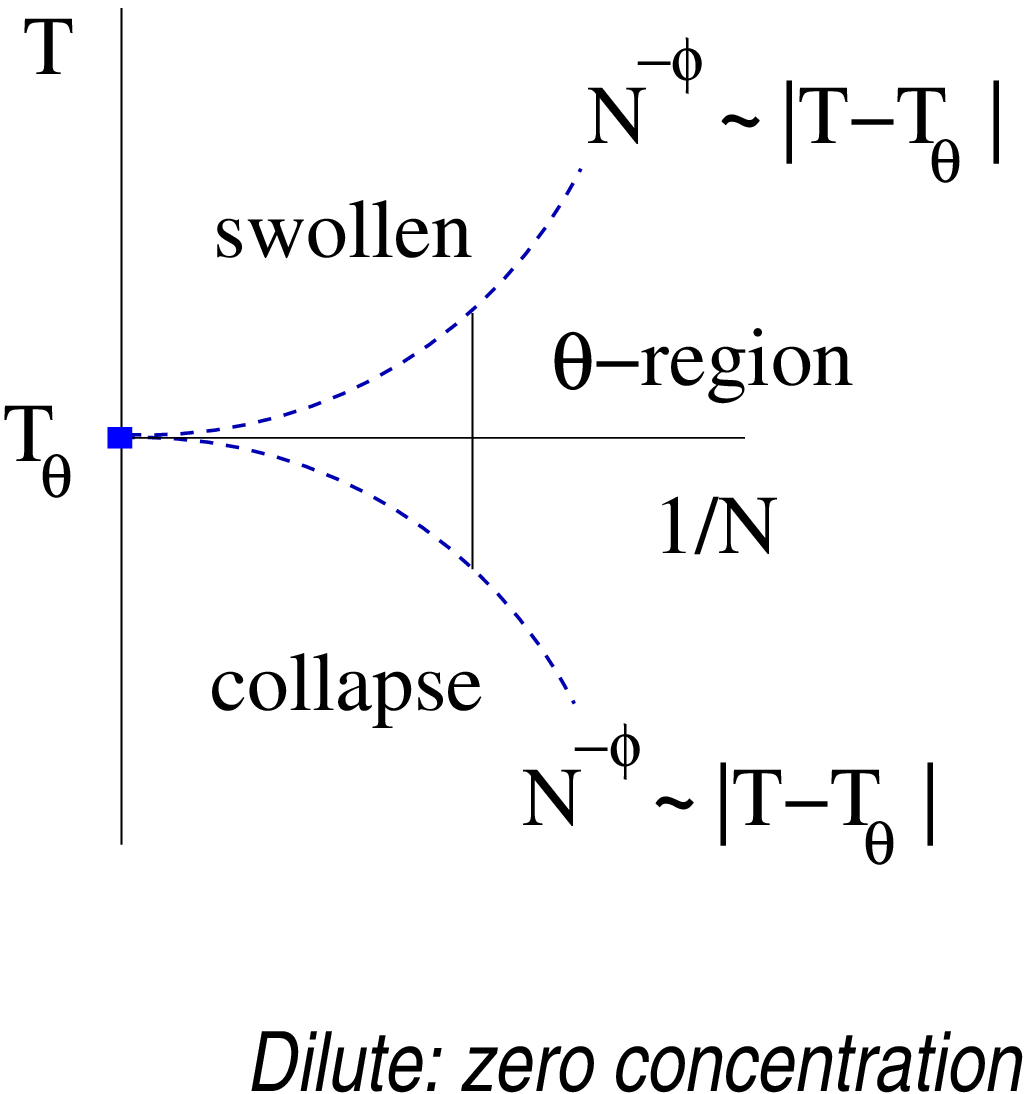}
\end{center}
\caption{Theta point region in a $T$ vs $1/N$ plot (single chain).  
The theta point is at $T= T_{\theta}, 1/N=0$. There is a cross-over 
region determined by the crossover 
exponent $\phi$, emanating from the theta point  (marked theta region) within 
which the theta point behaviour could be seen for shorter chains.
Beyond the dashed line for $T> T_{\theta}$, one sees the swollen behaviour for 
long chains while below a similar line for $T<T_{\theta}$ one sees a collpase 
phase.   The vertical solid line gives the width of the theta region for a 
finite chain.  This is used in Fig. \ref{fig:fig8}.
}
\label{fig:fig5}
\end{figure}
}
%%%%%%%%%%%%%%%%%%%

%%%%%%%%%%%%%% FIG6 %%%%%%%%%%%%
%
\newcommand{\figsix}{%%
\begin{figure}[htbp]
\begin{center}
\includegraphics[width=12cm,clip]{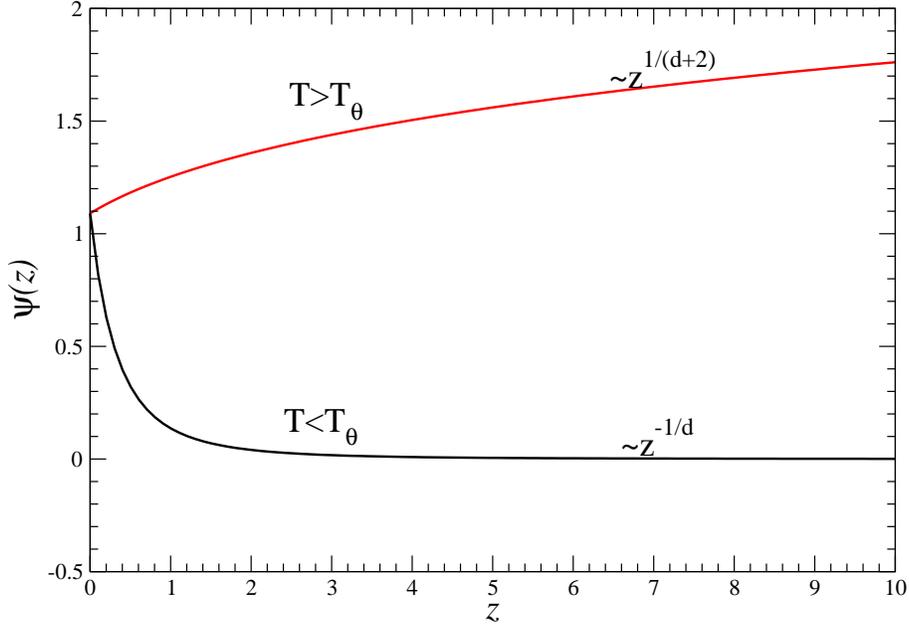}
\end{center}
\caption{Plot of $R/(b N^{\nu_{\theta}})=\Psi_{\pm}(z)$ as a function of $z$
as given by Eq.(\ref{case1:eq7}) in $d=3$ when $T> T_{\theta}$ and .$T< T_{\theta}$}
\label{fig:fig6}
\end{figure}
}%
%%%%%%%%%%%%%%%%%

%%%%%%%%%%%% FIG7 %%%%%%%%%%%%
%
\newcommand{\figsept}{%%
\begin{figure}[htbp]
\begin{center}
\includegraphics[width=6cm]{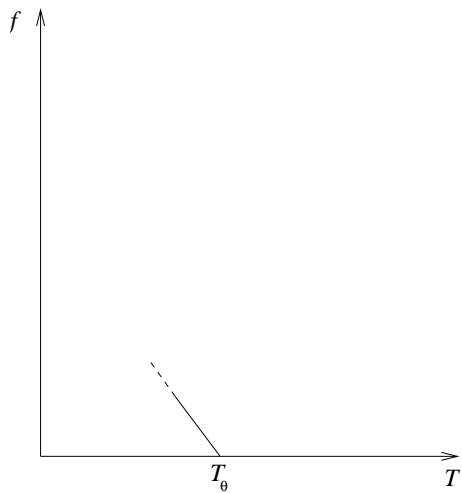}
\end{center}
\caption{Schematic phase diagram, in the force-temperature plane, as predicted by the Flory theory}
\label{fig:fig7}
\end{figure}
}
%%%%%%%%%%%%%%%%%

%%%%%% FIG 8 %%%%%%%%%%%%%%%
\newcommand{\figeight}{%
\begin{figure}[htbp]
\begin{center}
\includegraphics[width=12cm,clip]{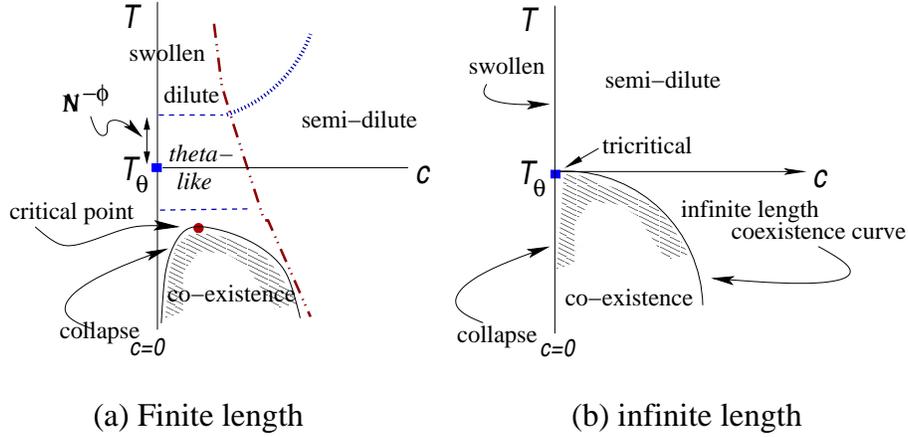}
\end{center}
\caption{$T$ vs $c$ (concentration) diagram for (a)  a finite chain and 
(b) infinite chain length.  In (a) there is no true  theta point 
but instead a  critical points for phase separation of polymer 
solutions. 
With concentration, 
there is a crossover line for theta like polymer solution to repulsive polymer 
solution (blue horizontal dashed lines). The dilute to semidilute crossover 
takes place at the overlap 
concentration where the polymer spheres in the dilute solution start to touch
each other.  This is indicated by a red dash-dot  line.  The blue hashed line 
indicates the  variation of the overlap concentration with $N$ (the locus of 
a typical point on the red dashed line). 
Near the critical point the solution behaviour is controlled by the 
concentratin fluctuations.  
In (b) there is a theta point (for $c=0$) which is the end 
point of the 
line of critical points of phase separation. This is a tricritical point.
There is a  phase coexistence line from the theta point for infinitely 
long polymers.  On this line the osmotic pressure is zero.  The critical 
point is coincident with the $c=0$ theta point. For any nonzero concentration 
it is a semi-dilute solution. The swollen and the collpased states exit only 
on the $c=0$ line.
}
\label{fig:fig8}
\end{figure}
}
%%%%%%%%%%%%%%%%%%%

%%%%%%%%%%%%%%%
%% Table I %%
%%%%%%%%%%%%%%%
\newcommand{\tableone}{
\begin{table}[htbp]
\centering
\begin{tabular}{lcccccccc}
\hline
\hline
Flory Regime &&$T > T_{\theta}$&& $\alpha > 0$&& $2 \le d \le 4$ && $\nu_{F}=\frac{3}{d+2}$ \\
Flory Regime &&$T > T_{\theta}$&& $\alpha > 0$&& $d > 4 $ && $\nu_{G}=\frac{1}{2}$ \\
Theta Regime &&$T = T_{\theta}$&& $\alpha = 0$&& $2 \le d \le 3 $ && $\nu_{\theta}=\frac{2}{d+1}$ \\
Theta Regime &&$T = T_{\theta}$&& $\alpha = 0$&& $d>3 $ && $\nu_{G}=\frac{1}{2}$ \\
Compact Regime &&$T < T_{\theta}$&& $\alpha < 0$&& $\forall$ $d$ && $\nu_{c}=\frac{1}{d}$ \\
\hline
\hline
\end{tabular}
\caption[]{Summary of the exponents predicted by Flory theory in various regimes as reported in Sec. \ref{subsec:steepest}.}
\label{tab:tab1}
\end{table}
}
%%%%%%%%%%%%%%%

%%%%%%%%%%%%%%%%%%%%%%%%%%%%
\newcommand{\tableLN}{%
\begin{table}[htbp]
\centering
\begin{tabular}{|c|c|c|}
\hline
                  & minimal model       & with small scale\\
\hline
microscopic length &   None              &   $b$              \\
\hline
polymer length     &   $L$                 &   $N b^2$          \\
\hline
size           &                         &                  \\
(gaussian polymer)&  $R= d\sqrt{L}$        &  $R=d b \sqrt{N}$      \\ 
\hline
2-body parameter &  $z=c \; u\;  L^{(4-d)/2}$   &  $\alpha=c \; u\; b^{4-d}$\\
\hline
3-body parameter &  $w=c \;   v L^{(3-d)}$  &  $\gamma=\; c \; v b^{6-2d}$\\
\hline
Size             &  $R\sim L^{1/2} \; z^p$ & $R\sim b^{1-2\nu} L^{\nu}$\\
(interacting)     &                        & $\sim b N^{\nu}$\\
\hline 
\end{tabular}
\caption{The dictionary for the variables involving $\{L,u, v\}$ and $\{b, N, u, v\}$. Here $c$
represents some appropriate constant, not necessarily same everyehere. }
\label{tab:tab2}
\end{table}
}%

\maketitle

\tableofcontents

%%%%%%%%%%%%%%%%%%%%%%%%%%%%%%%%%%%%%%%%%%%%%%%%%%%%%%%%%%%%%%%%%%%%%%%%%%%%
\section{Introduction}
\label{sec:introduction}
%%%%%%%%%%%%%%%%%%%%%%%%%%%%%%%%%%%%%%%%%%%%%%%%%%%%%%%%%%%%%%%%%%%%%%%%%%%%%
Polymer physics, with an old and venerable history, spanning more than
60 years, now occupies an important position in basic physics,
providing conceptual support to wide varieties of problems 
\cite{Flory53,Flory69,yamakawa,deGennes79,Doi86,Freed87,desCloizeaux90,lifshitz78,Grosberg94,Rubinstein03,Hughes,Vanderzande,Giacomin,Raphael92,Kamien93,Orland94,bhattacharjee05}.

A polymer, from a physicist's point of view, is a set of units, called
monomers, connected linearly as a chain.  Such polymers are the
natural or synthetic long chain molecules formed by bonding monomers
chemically as in real polymers or bio-polymers like DNA, proteins etc,
but they need not be restricted to those only.  Polymers could also be
the line defects in superconductors and other ordered media, the
domain walls in two dimensional systems and so on. Even if
non-interacting, a polymer by virtue of its connectivity brings in
correlations between monomers situated faraway along the chain. This
makes a polymer different from a collection of independent monomers.
The basic problem of polymer physics is then to tackle the inherent
correlations due to the long length of the string like object.

A gas of $N$ isolated monomers at any nonzero temperature $T$ would
like to occupy the whole available volume to maximize the entropy but
that would not be the case when they are connected linearly as a
polymer.  This brings in a quantity very special to polymers, namely
the equilibrium size of a polymer, in addition to the usual
thermodynamic quantities.  Traditionally thermodynamic quantities, at
least for large $N$, are expected to show extensivity, i.e.,
proportionality to the number of constituent units, but the size of a
polymer in thermal equilibrium need not respect that.  In other words,
if the length of a polymer is doubled, the size need not change by the
same factor.  Consequently even the usual thermodynamic quantities
would have an extra polymer length dependence which will not
necessarily be extensive but would encode the special polymeric
correlations mentioned above.  How the equilibrium size of a polymer
changes or scales, as its length is increased, whether this dependence
shows any signature of phase transitions with any external parameter
like temperature, and the consequent effects on other properties are
some of the questions one confronts in the studies of polymers.

The success of exact methods, scaling arguments and the
renormalization group crafted the statistical physics approach to
polymer physics into a well defined and recognized field.  One of the
first, and most successful, theoretical approaches to thermophysical
properties of polymers, is the celebrated Flory theory, that will be
the central topic of this review.  This simple argument was a key step
in the history of critical phenomena, especially, in seeing the
emergence of power laws and the role of dimensionality.  For the
special effects of long range correlations that develop near a
critical point, one needs a fine tuning of parameters like
temperature, pressure, fields etc, to be close to that special point.
In contrast, the simple Flory theory showed that a polymer exhibits
critical features, power laws in particular, and a dimensionality
dependence beyond the purview of perturbation theories, all without
any requirement of fine-tuning.  Here is an example of self-organized
criticality - a phenomenon where a system shows critical-like features
on its own without any external tuning parameter - though the name was
coined decades after the Flory theory.

Various monographs
\cite{Flory53,Flory69,yamakawa,deGennes79,Doi86,Freed87,desCloizeaux90,Grosberg94,Rubinstein03,Hughes,Vanderzande,Giacomin,Raphael92,Kamien93,Orland94,bhattacharjee05,Edwards65,SMB91},
covered different aspects of methodologies and techniques.  This
notwithstanding, our aim is to bring out the nuances present in the
Flory theory and to place it in the current context, to appreciate why
this theory stands the test of time as compared to other mean-field
theories.

This review is organized as follows.  After a recapitulation of the
basic facts of a noninteracting polymer and the simple Flory theory in
Sec. \ref{sec:elementary}, we introduce the Edwards continuum
model \cite{Edwards65,SMB91} (Section \ref{sec:edwards}) and the mean
field approximation to its free energy (Section \ref{sec:flory}).
This forms the basis for discussing the Flory approximation through a
saddle-point method (Section \ref{subsec:steepest}).  The results for
the three regimes of a polymer (swollen, theta and compact), and the
transition behaviour can also be found in the same Section.  How the Flory theory
fares when compared with the current view of scale invariance,
universality and scaling is discussed in Sec.
\ref{sec:flory-theory-modern} and the role of a microscopic length
scale discussed there.  A few
modifications\cite{Ptitsyn68,deGennes75}, and a simple extension to
include external forces applied to one extreme of the polymer, are
discussed in Sections \ref{subsec:elastic} and \ref{subsec:inclusion},
respectively.

While the original Flory theory describes the size at a fixed
temperature, as the number of monomers increases, it is possible to go
beyond power laws in the current framework. The analysis allows one to
discuss the temperature dependence of the size at a fixed number of
monomers (assumed to be sufficiently large). This cross-over effect is
discussed in Section\ref{sec:crossover}.  A particularly interesting
case appears to be the two-dimensional case, discussed in Section
\ref{sec:explicit}, where the scaling function can be computed
exactly. Section \ref{subsec:uniform} also includes the uniform expansion method
\cite{Doi86} along with its relationship with a perturbative approach 
\cite{Edwards79a,Muthukumar84,Muthukumar86}.

Besides the three states mentioned above, there is an obvious state of
a polymer, namely a stretched or a rod like state.  This state can be
achieved by a force at one end, keeping the other end fixed, or by
assigning a penality for bending.  In absence of any interaction,
there is no transition from this rod-like state to any of the other
states.  But still, for completeness, the universal features of the
crossover behaviour needs to be discussed.  This is done in the last
part of the paper.  It is devoted to the semiflexible chain, where
bending rigidity competes with entropy.  The response of the polymer
when a pulling force is applied to an extremum is discussed in Section
\ref{sec:semiflexible}, with an eye on the interpolating formula
between flexible and semiflexible
regimes\cite{Ha95,Ha97,Rosa03,Marko95}.  Ancillary results for the
structure factor and the end-to-end distance will also be presented in
Section \ref{subsec:structure}.

Several technical issues are relegated to the Appendixes.  A few
Gaussian transformations that are frequently employed are listed in Appendix \ref{app:hubbard}.  A
discussion on the central limit theorem as applied to polymers and a
possible deviation can be found in Appendix \ref{app:distribution}.  In
Appendix \ref{sec:perturbation}, the theoretical framework of perturbation
theory\cite{Edwards79a,Muthukumar84}, is introduced at the simplest
possible level, and the lowest order calculation is explicitly
performed to show how the method works. Finally, for completness, Appendices \ref{app:structure_gaussian},
\ref{app:exact}, \ref{app:green}, \ref{app:integral} include the explicit derivation
of some results that are used in the main text.

We end this introduction with a few definitions.  If all the monomers,
and therefore the bonds, can be taken as similar, then the polymer is
called a homopolymer.  If there is any heterogeneity either in
monomers or in bonds, it will be a heteropolymer.  In case of two
types of monomers arranged in a regular pattern, the polymer is called
a co-polymer. Two different types of polymers connected together is an
example of a block-copolymer.  This review focuses on the homopolymer
case only.

We use the symbol $\sim$ to denote the dependence on certain
quantities, ignoring prefactors and dimensional analysis, while the
symbol $\approx$ is to be used for approximate equality.

%%%%%%%%%%%%%%%%%%%%%%%%%%%%%%%%%%%%%%%%%%%%%%%%%%%%%%%%%%%%%%%%%%%%%%%%%%%%%

%%%%%%%%%%%%%%%%%%%%%%%%%%%%%%%%%%%%%%%%%%%%%%%%%%%%%%%%%%%%%%%%%%%%%%%%%%%%%
\section{Elementary version of The Flory theory}
\label{sec:elementary}
%%%%%%%%%%%%%%%%%%%%%%%%%%%%%%%%%%%%%%%%%%%%%%%%%%%%%%%%%%%%%%%%%%%%%%%%%%%%%
\subsection{Gaussian Behaviour}\label{sec:gaussian-behaviour}
\subsubsection{Freely Jointed chain}\label{sec:freely-jointed-chain}
%%%%%%%%%%%%%%%%%%%%%%%%%%%%%%%%%%%%%%%%%%%%%%%%%%%%%%%%%%%%%%%%%%%%%%%%%%%%
Consider an isolated homopolymer formed by $N+1$ monomers at positions
$\{\mathbf{r}_0,\mathbf{r}_1,\ldots,\mathbf{r}_N \}$ in space, and let
$b$ be the monomer-monomer distance (sometimes also referred to as the
Kuhn length). This is depicted in Fig.\ref{fig:fig1}.

\figone

We further introduce the bond variable $ {\bm \tau}_j ( \left \vert
  {\bm \tau}_{j} \right \vert = b)$  and the end-to-end distance $\mathbf{R}$,
\begin{equation}
\label{elementary:eq1}
  {\bm \tau}_j=\mathbf{r}_j-\mathbf{r}_{j-1},\quad {\rm and\ }
\mathbf{R}= \mathbf{r}_{N}-\mathbf{r}_{0}=\sum_{j=1}^N {\bm \tau}_{j}.
\end{equation}
A flexible polymer is defined as one for which the bond vectors are
completely independent so that each bond can orient in any direction
in space irrespective of the orientations of the others.  This freedom
is expressed as an absence of any correlation between \textit{any} two
different bonds, that is
\begin{eqnarray}
\label{elementary:eq2}
\left \langle {\bm \tau}_i \cdot {\bm \tau}_j \right \rangle 
     &=& b^2 \delta_{ij}.
\end{eqnarray}
This is the basis of the freely-jointed chain(FJC).  As the
monomer-monomer distance is fixed, the average in
Eq.(\ref{elementary:eq2}) is an average over all possible
orientations.  This 
ensemble averaging is denoted by the angular brackets $\langle
\ldots\rangle$.  A more realistic model, where there is an
orientational correlation between successive bonds, called worm-like
chain model (WLC) (or Kratky-Porod model), is the paradigm of the
stiff polymer, and will be discussed later on.

\subsubsection{Size of a polymer}
\label{sec:size-polymer}
A use of Eqs.(\ref{elementary:eq1}) and (\ref{elementary:eq2}) leads to
\begin{eqnarray}
\label{elementary:eq3}
\left \langle R^2 \right \rangle 
    &=& \sum_{i,j=1}^N \left \langle {\bm \tau}_i \cdot {\bm \tau}_j
                         \right \rangle  
     = N b^2,
\end{eqnarray}
so that the size $R$, measured by the root mean square (rms)
end-to-end distance of a polymer, depends on its length $N$ as
\begin{equation}
  \label{eq:2}
R \sim b N^{\nu},    
\end{equation}
with $\nu=1/2$ for the FJC.  The exponent $\nu$ is called the size
exponent.  We are using the rms value as the size of the polymer
because by symmetry (i.e. isotropy) $\langle {\bf R}\rangle=0$.  A
judicious choice of origin can always remove a non-zero average of any
probability distribution, whereas it would be impossible to make the
variance zero. Hence the importance of the rms value as a measure of
the size.

The behavior described by Eq.(\ref{eq:2}) can be also read as follows.
If a sphere of radius $R$ is drawn with its center in a random
position along the chain, the total length of the polymer contained in
the sphere is about $R^{d_{F}}$, with $d_{F}=1/\nu$ being what is
known as the {\it fractal dimension}. So, the fractal dimension of our
non-interacting polymer is $d_{F}=2$.

The  probability distribution $P(\mathbf{R},N)$
of the end-to-end distance  is a Gaussian (see Appendix
\ref{app:distribution} for details) and  in $d=3$ it is (see
Eq.\ref{distribution:eq12}) 
\begin{eqnarray}
\label{elementary:eq4}
P\left(\mathbf{R},N\right)  &\approx& \left(\frac{3}{2 \pi N b^2}
\right)^{3/2} \exp\left[-\frac{3}{2} \frac{R^2}{N b^2} \right]. 
\end{eqnarray}
The standard deviation, that determines the width of this distribution, 
gives the rms size $R$ of Eq. \eqref{eq:2}.

A chain characterized by the Gaussian behavior (\ref{elementary:eq4})
is also called an \textit{ideal or phantom} chain.  It also goes by
the names of a Gaussian polymer, a non-self-interacting polymer.
These names are used interchangeably.

The size of a polymer discussed above is an example of a critical-like
power law whose origin can be traced to correlations.  Even-though the
bonds are uncorrelated, the monomers are not.  This can be seen from
Eqs. ~\eqref{elementary:eq1} and ~\eqref{elementary:eq2} as the
positions of monomers $i$ and $j$ satisfy
\begin{equation}
  \label{eq:3}
  \left \langle [\mathbf{r}_j -\mathbf{r}_i]^2 \right \rangle =
  \sum_{l,m=i+1}^{j} \left \langle {\bm \tau}_l \cdot  
{\bm \tau}_m  \right \rangle = (j-i) b^2.  
\end{equation}
Generalizing Eq. \eqref{elementary:eq4}, the conditional probability
density of monomer $j$ to be at ${\bf r}'$ if the $i$-th monomer is at
${\bf r}$ is given
\begin{equation}
  \label{eq:9}
  P({\bf r}',j|{\bf r},i)\propto \exp\left[-\frac{3}{2} \frac{({\bf
        r'-r})^2}{|j-i| b^2} \right]. 
\end{equation}
The distribution becomes wide as $j-i$ increases and it is not
factorizable.  This is to be contrasted with the case of
noninteracting monomers without polymeric connections. There this
joint probability distribution is the product of the individual
probability densities and hence devoid of any correlations 
\footnote{Two random variables $x, y$ are correlated, i.e. 
            $\langle xy \rangle \ne \langle x \rangle \langle y
            \rangle$ if and only if $P(x,y) \ne p(x) p(y)$.}.
The behaviour of an ideal chain as formulated here is purely
entropic in origin because all the configurations are taken
to have the same energy.

If one generalizes Eq.(\ref{elementary:eq2}) by substituting
$\delta_{ij}$ by a general correlation $g_{ij}$ which (a) depends only
on $\vert i-j \vert$, and (b) is such that $\sum_{j}g_{ij} < \infty$,
then the results, like $R^2\sim N$, remain essentially the same, since
Eq.(\ref{elementary:eq3}) is modified by a multiplicative constant.
In this case, the decay length of the correlation $g_{ij}$ gives the
Kuhn length.

\subsection{Non-Gaussian Behaviour}
\label{sec:non-gauss-behav}
To go beyond the Gaussian behaviour, let us introduce the repulsive
interaction of the monomers, {\it e.g.}, the athermal excluded volume
interaction.  The question is how this repulsion of the monomers
affects the size of the polymer.  Does it just change the amplitude in
Eq. ~\eqref{eq:2} or it changes the exponent?  A change in the
exponent needs to be taken more seriously than in the amplitude
because the latter is equivalent to a change in the unit of
measurement while the former changes the fractal dimension of the
polymer.

\subsubsection{Simple Flory theory}
\label{sec:simple-flory-theory}
A simple way to accounting for the fact that non-consecutive spheres
(i.e.  monomers) cannot interpenetrate, is provided by a hard-sphere
repulsion, that is proportional to the excluded volume $v_{exc}$ of
each pair of monomers, times the number of monomer pairs ($N^2$) per
unit of available volume ($R^3$), that is
\begin{equation}
  \label{eq:5}
{\rm repulsive\  energy} \sim v_{exc}\frac{N^2}{R^3}.  
\end{equation}
The total free energy $F_N(R)$ of the system can then be quickly
estimated as follows \cite{deGennes79,Fisher67}.

From Eq.(\ref{elementary:eq4})
\begin{equation}
  \label{eq:4}
S_N(R)=k_B \log P(R,N) \sim -\frac{ R^2}{N b^2},  
\end{equation}
is the entropy of the chain \footnote{The entropy should be $k_B
  \log[$ number of chains of $N$ monomers and end-to-end distance $=R$
  $]$. However, the number in the argument of the logarithm is
  proportional to $P(R,N)$, and so -- apart from an additive,
  $N-$dependent, constant -- we get Eq.(\ref{eq:4}) }, where $k_B$ is
the Boltzmann constant, so that at temperature $T$ one has
\begin{eqnarray}
\label{elementary:eq11}
F_N\left(R\right) &=&  F_0 +e_0 \frac{R^2}{N b^2}+e_1
v_{exc} \frac{N^2}{R^3}, 
\end{eqnarray}
$e_0$ and $e_1$ are $T$-dependent constants and $F_0$ is the remaining
part of the free energy.  Eq. \eqref{elementary:eq11} is to be
interpreted as the free energy of a polymer chain of $N$ monomers with
excluded volume interaction \textit{if} it had a size of radius $R$.
The size of an unconstrained polymer would come from a minimization of
$F_N(R)$ with respect to $R$ which amounts to
equating the two $R$-dependent terms in Eq.(\ref{elementary:eq11}).
The size still has the form of  Eq. \eqref{eq:2}, but with
\begin{equation}
\label{elementary:eq12}
\nu=3/5.
\end{equation} 
This $\nu$ is called the Flory exponent. This is the most elementary
version of the Flory theory that experienced a remarkable success in
explaining the experimental evidence in swelling of real polymers.
This success is thought to be accidental, but we shall see later on
that more systematic arguments  do lead to Eqs.  \eqref{elementary:eq11}
and \eqref{elementary:eq12}.

The above argument can be generalized to arbitrary dimensions $d$.
The entropy term as given by Eq. \eqref{eq:4} is independent of $d$,
but the excluded volume term in Eq. \eqref{eq:5} would be replaced by
$N^2/R^d$, $R^d$ being the volume occupied by the polymer.  A
minimization of the free energy then gives the Flory exponent as 
$\nu=\frac{3}{d+2}$.  However, for $d>4$, it gives a size exponent less
than $1/2$, which is not possible, because a repulsion cannot make a
polymer more compact than a free chain.   One therefore expects 
the free chain value $\nu=1/2$,  so that the general Flory exponent would be
\begin{equation}
  \label{eq:6}
  \nu=\left\{ \begin{array}{lrr}
              \frac{3}{d+2},& {\rm{for}}\  d\leq 4,& \qquad\qquad
              {\rm (swollen\  phase)}\\[3pt] 
              \frac{1}{2},  & {\rm{for}}\  d> 4, & 
                \end{array}\right.
\end{equation}
which agrees with the known exact results like, $\nu=1$ for $d=1$,
$\nu=3/4$ for $d=2$, $\nu=1/2$ for $d>4$ and is very close to the best
estimate $\nu \approx 0.588$ known for $d=3$\cite{Nienhuis,Zinn-Justin90}.

\subsubsection{Collapse}
\label{sec:collapse}
The case of attractive interaction may also be mentioned here.  With
attraction, and hard-core repulsion, the monomers would like to stay as
close as possible.  This gives a more or less compact packing of
spheres so that the monomer density inside a sphere enclosing the
polymer is O$(1)$ in $N$.  Note that the density for the repulsive case
$N/R^d\sim N^{1-d\nu}\to 0,$ for large $N$. A compact phase, also
called a globule, would then have 
\begin{equation}
  \label{eq:7}
  R\sim N^{1/d},\ i.e.,\ \nu=\frac{1}{d}.\qquad\qquad{\rm (compact)}
\end{equation}
The collapsed state is not a unique state  and the polymeric nature is
important in determining its overall property.
 
One expects a generic phase diagram, as
schematically depicted in Fig.\ref{fig:fig2}, with a theta point at
$T=T_{\theta}$, a high temperature ($T > T_{\theta}$) swollen or
coiled phase
and a low temperature ($T < T_{\theta}$) compact phase.
This will be discussed in detail in Section \ref{sec:flory}.

\figtwo

%%%%%%%%%%%%%%%%%%%%%%%%%%%%%%%%%%%%%%%%%%%%%%%%%%%%%%%%%%%%%%%%%%%%%%%%%%%%%

%%%%%%%%%%%%%%%%%%%%%%%%%%%%%%%%%%%%%%%%%%%%%%%%%%%%%%%%%%%%%%%%%%%%%%%%%%%%%il
\section{The Edwards continuum model }
\label{sec:edwards}
%%%%%%%%%%%%%%%%%%%%%%%%%%%%%%%%%%%%%%%%%%%%%%%%%%%%%%%%%%%%%%%%%%%%%%%%%%%%

\subsection{Discrete Gaussian model}
\label{sec:discr-gauss-model}
The central limit theorem, as explained in Appendix A,
allows us to describe a polymer by  the distribution
$W(\mathbf{r}_0,\ldots,\mathbf{r}_N)$ of $N$ bonds, ${\bm
  \tau}_1=\mathbf{r}_1-\mathbf{r}_0$,$\ldots$, ${\bm
  \tau}_N=\mathbf{r}_N-\mathbf{r}_{N-1}$, each having a Gaussian
distribution, as 
\begin{subequations}
\begin{eqnarray}
\label{elementary:eq13}
W\left(\mathbf{r}_0,\ldots,\mathbf{r}_N\right) 
   &=& \prod_{j=1}^N p\left({\bm \tau}_j\right)
     = \prod_{j=1}^N \left\{\left(\frac{1}{2 \pi b^2}\right)^{d/2}
        \exp \left[-\frac{1}{2} \frac{\tau_j^2}{b^2} \right]\right\}, \\ 
   &=& Z_G^{-1}  {\exp\left[-\beta H_G\right]}, \label{eq:10}
\end{eqnarray}
where we have introduced the Gaussian Hamiltonian
\begin{eqnarray}
\label{elementary:eq14}
\beta H_G &=& 
\frac{1}{2b^2} \sum_{j=1}^N
\tau_j ^2 =
\frac{1}{2b^2} \sum_{j=1}^N
\left(\mathbf{r}_j-\mathbf{r}_{j-1} \right)^2, 
\end{eqnarray}
with the  partition function $Z_G=(2 \pi b^2 )^{Nd/2}$.
\end{subequations}

The Gaussian Hamiltonian is another representation of a polymer where
the monomers are connected by harmonic springs (Fig.
\ref{fig:fig1}c).  At any nonzero temperature, the equipartition
theorem gives $\langle \tau_j^2\rangle/b^2=d$, which allows the bonds
to have a nonzero rms length.  The size of the polymer is given by
$\langle R^2\rangle=d b^2 N$.

The Gaussian Hamiltonian, being quadratic, makes analytical
calculations simpler compared to the FJC case with the rigid bond
constraints.  In contrast, the extensibility of the springs allows the
polymer to have a size $R >Nb$ with a nonzero probability as seen from
Eq. \eqref{elementary:eq4}. However, the probabilities being in the
tail of the Gaussian distribution, are too small to contribute to the
average.  Consequently most of the physical behaviour will be
controlled by the configurations around the peak of the distribution
and not by rare extreme configurations.  With this caveat in mind, the
Gaussian Hamiltonian can be used in most cases, unless certain
stretched states become important.

There is a subtle difference between this Gaussian Hamiltonian
approach and FJC of the previous section. Unlike FJC,
here we are associating energies to conformations and
the behaviour is not strictly entropic in origin.  However the ``springs''
help us in maintaining the polymeric connectivity and
the total elasticity of the Gaussian polymer would be the same as the
entropic elasticity of the ideal chain.  In that respect, the
elasticity of the Gaussian chain, Eq. \eqref{eq:4}, could be termed as
entropic in origin.  

\subsection{Continuum model}
\label{sec:continuum-model}
A simple-minded way of taking the continuum limit $N \to\infty, b \to
0$ with the length $Nb$ a constant, would lead to a vanishing $\langle
R^2 \rangle$ as defined by Eq.(\ref{elementary:eq3}). This is
avoided by introducing a curvilinear coordinate $s =jb^2, 0\le s \le
L=Nb^2$ for the monomer and a vector position $\mathbf{r}(s)$
associated with it.  The Gaussian Hamiltonian of
Eq.(\ref{elementary:eq14}) takes the limiting form
\begin{eqnarray}
\label{edwards:eq0a}
\beta H_G 
  = \frac{1}{2}\sum_{j=1}^N  b^2 \frac{1}{b^4}
      \left(\mathbf{r}_j-\mathbf{r}_{j-1} \right)^2 
&\rightarrow& 
 \beta H_{L}^{(0)}
    =\frac{1}{2} \int_{0}^{L} ds 
          \left(\frac{\partial \mathbf{r}}{\partial s} \right)^2.
\end{eqnarray}
In the above form, one end point of the polymer can be
anywhere in the whole volume available and it would contribute a
volume factor to the partition function, of no concern to us.  We may
get rid of this perfect-gas like redundant factor by fixing one
monomer preferably the end-point at $s=0$ at origin. 
The continuum limit of the corresponding distribution
Eq.(\ref{elementary:eq13}) is given by
\begin{eqnarray}
\label{edwards:eq0b}
W\left(\mathbf{r}_0,\ldots,\mathbf{r}_N\right) &\rightarrow&
W\left[\mathbf{r} \left(s\right) \right]= \frac{1}{Z_{0}}   
\exp\left[-\beta H_L^{(0)} \left[\mathbf{r}\left(s\right)\right]\right],
\end{eqnarray}
with the  ``configurational partition function'' written formally as \cite{note1}
\begin{eqnarray}
  Z_0 &=& \int {\cal D} \mathbf{r} \  \exp\left[-\beta
    H_L^{(0)}\left[\mathbf{r}\left(s\right)\right]\right]  
  \delta^{d}\left(\mathbf{r}\left(0\right)\right).  
\label{edwards:eq2}
\end{eqnarray}
The notation $\int{\cal D} \mathbf{r}$ represents a formal sum over
all possible paths or polymer configurations, but it is ill-defined if
taken literally as a $b\to 0$ limit of the measure expected from Eq.
(\ref{elementary:eq13}).   This continuum language, patterned after the
path integral representation in Quantum Mechanics \cite{Kleinert90},
was introduced by Edwards \cite{Edwards65,Doi86,SMB91}.  The path
integral, also known as the Wiener measure in the context of
diffusion, is to be interpreted as a limit of the discrete sum.  With
appropriate care, the limit process may be traded with standard
integrals, as will be done in this review.

Some more caution is needed here in interpreting the continuum
Hamiltonian.  Although $s$ is introduced as a curvilinear coordinate
measuring the arc-length or contour length along the polymer, the
string in the continuum (Fig \ref{fig:fig1} (d)) is not to be taken as
a space curve.  For a space curve $|\partial {\bf r}/\partial s| =1$,
which is not enforced in Eq. \eqref{edwards:eq0a}.  In this
interpretation, $s$ remains a measure of the contour-length obtained
from the bead numbers, but the string remains Gaussian at the smallest
scale. One may bypass this problem by assigning a new axis for $s$ so
that the polymer is viewed as a $d+1$ dimensional string.  To avoid
the pitfalls of the Gaussian behaviour at all length scales, it may be
necessary to put a lower cut-off in Eq.~\eqref{edwards:eq0a}.  Unless
necessary this is not to be specified explicitly.

\subsection{Interactions:  The Edwards model}
\label{sec:inter-edwards-model}
Next we consider a more general description, where the polymer can
also interact.  Since a polymer is generally in a solvent, the
interactions need not be the actual microscopic interactions of the
monomers.  If a polymer dissolves in a solvent, a monomer would be
surrounded mostly by the solvent molecules.  If we integrate out the
solvent part from the problem, it would look like the monomers staying
away from each other.  This situation of a polymer in a good solvent
can be described by an effective repulsion among the monomers.  On the
other hand if a polymer precipitates out from a solution, then there
is a preference for the monomers to avoid the solvent molecules.  This
is the case of a polymer in a bad solvent whose effective description
requires an attraction between the monomers.  In this spirit of
effective interactions, it suffices to consider the polymer as the
sole object with interactions among the monomers, which could depend
on temperature, solvent quality and other parameters of the original
problem.  As pseudo-interactions, these need not be restricted to
pairwise interactions only. A schematic representation of two body 
$\Phi_2({\bf{r}},{\bf{r}}^{\prime})$ and
three body interactions
$\Phi_3({\bf{r}},{\bf{r}}^{\prime},{\bf{r}}^{\prime\prime})$ is shown
in Fig. \ref{fig:fig3}.

\figthree 

A polymer in a good solvent can be described by a simple choice of a
pairwise contact repulsive interaction, represented by a delta
function, $\Phi_2=u \delta^d\left(\mathbf{r}\right)$, with a coupling
parameter $u>0$ so that ignoring all higher order terms,
\begin{equation}
  \label{eq:8}
 \beta H_{L}\left[\mathbf{r}\left(s\right) \right]= \frac{1}{2}
 \int_{0}^{L} ds \left( \frac{\partial \mathbf{r}}{\partial
     s}\right)^2  
+ \frac{1}{2}  u\: \int_{0}^{L} ds_1  \int_{0}^{L} ds_2 \;
\delta^{d}\left(\mathbf{r}\left(s_2\right)-\mathbf{r}\left(s_1\right)\right).
\end{equation}
The first term on the right side of Eq.(\ref{eq:8}) is the usual
Gaussian term, representing polymer connectivity, whereas the term
penalizes any two-monomer contact.  This particular form is known as
the Edwards Hamiltonian\cite{SMB91} and is a representation of a self
avoiding walk or a polymer with excluded volume interaction. This is
also called the minimal model for a polymer.

To describe the collapse, i.e., the poor solvent case, we need $u<0$
for attraction and for stability a repulsive three-body interaction.  With the choice of the usual three-body contact
pseudo-potential $\Phi_3=v \delta^d\left(\mathbf{r}-
  \mathbf{r}^{\prime}\right) \delta^d\left(\mathbf{r}^{\prime}-
  \mathbf{r}^{\prime\prime}\right)$, penalizing any three monomer
contact, the Edwards Hamiltonian becomes
\begin{eqnarray}
  \beta H_{L}\left[\mathbf{r}\left(s\right) \right]&=& \frac{1}{2}
  \int_{0}^{L} ds \left( \frac{\partial \mathbf{r}}{\partial s}\right)^2
  + \frac{1}{2}  u\:  \int_{0}^{L} ds_1  \int_{0}^{L} ds_2 \;
  \delta^{d}\left(\mathbf{r}\left(s_2\right)-\mathbf{r}\left(s_1\right)\right)\nonumber\\
  &&\ +\frac{1}{3!} v \int_{0}^{L} ds_1  \int_{0}^{L} ds_2  \int_{0}^{L} ds_3 \;
  \delta^d\left(\mathbf{r}(s_1)- \mathbf{r}(s_2)\right)  
  \delta^d\left(\mathbf{r}(s_2)-\mathbf{r}(s_3)\right).   \label{eq:20} \\
  &\equiv& \beta H_{L}^{(0)}+ \beta V_L\label{edwards:eq6} 
\end{eqnarray}
where
$ \beta H_{L}^{(0)}$ is given by Eq. (\ref{edwards:eq0a}) 
and $\beta V_L$ represents the
interaction part of the dimensionless Hamiltonian. These minimal
models involve only  three parameters $L,u$ and $v$.

Since a polymer can be precipitated out of a solution by cooling (Sec.
\ref{sec:elementary}), it is assumed that a temperature $T_{\theta}$
exists such that
$$u \propto (T -T_{\theta})/T_{\theta},$$
so that $T>T_{\theta}$ ($u>0$) corresponds to the repulsive case
(a polymer in a good solvent) while $T<T_{\theta}$ ($u<0$) for
the attractive case (the bad solvent case). The transition point $T=T_{\theta}$ is 
the theta point, as discussed in Fig. \ref{fig:fig2}.

\subsection{Green Functions}
\label{sec:green-function}
The problem associated with $b\to 0, N\to\infty$ for partition
functions ($b^{Nd}$ in Eq. \eqref{eq:10}) is avoided by normalization
by $Z_0$. The probability that the free polymer has an end-to-end
distance vector ${\mathbf{R}}$ is written as
\begin{eqnarray}
\label{edwards:eq9}
G_{L}^{(0)}\left(\mathbf{R}\right) 
&=&Z_0^{-1} \int_{\mathbf{R}}  {\cal D} \mathbf{r}  
\ \exp\left[-\beta H_L^{(0)} \left[\mathbf{r}\left(s\right)\right]\right]
= \left({2 \pi L }
\right)^{-d/2} e^{-R^2/(2L)},
\end{eqnarray}
where a short hand notation 
\begin{equation}
  \label{eq:12}
  \int_{\mathbf{R}} {\cal D} \mathbf{r} \equiv \int{\cal D} \mathbf{r}  
  \    \delta^d \left(\mathbf{R}- \left[\mathbf{r}\left(L\right) -
      \mathbf{r}\left(0\right) \right] 
  \right)\  \delta^{d}\left(\mathbf{r}\left(0\right)\right),   
\end{equation}
is used to indicate the sum over all paths with fixed end-to-end
distance ${\mathbf{R}}$ with the $s=0$ end fixed at the origin. 
The result, Eq. (\ref{edwards:eq9}), is the $d$-dimensional analogue
of the probability distribution in Eq.  \eqref{elementary:eq4}.
For the interacting case, the normalized
partition (Green) function is
\begin{eqnarray}
\label{edwards:eq7}
G_{L}\left(\mathbf{R}\right) &=& Z_0^{-1}
  \int_{\mathbf{R}} {\cal D} \mathbf{r} 
\ \exp\left[-\beta H_{L} \left[\mathbf{r}\left(s\right)\right] \right],
\end{eqnarray}
which is related to the probability of the end-to-end vector being
${\mathbf{R}}$, but for the normalization.  
$G$ is called the {\em Green  function} or a {\em propagator} while
$G_{L}^{(0)}\left(\mathbf{R}\right)$ is the ``free'' propagator.

The partition function of Eq. \eqref{edwards:eq2} corresponds to the
case where the polymer end at length $L$ is free, while the sum
expressed by Eq. \eqref{eq:12} corresponds to a constrained ensemble,
the ensemble of all configurations with the same end-to-end distance
${\mathbf{R}}$.  This is a fixed-${\mathbf{R}}$ ensemble.  Its
conjugate ensemble is the fixed force ensemble where a force is
applied at the free end.  In a fixed-${\mathbf{R}}$ ensemble, the
force required at the open end point is a fluctuating quantity whose
average gives the force required to maintain that distance. In the
fixed force ensemble, the end-to-end distance fluctuates and the
variance of this fluctuation is related to the elastic constant or
response function of the polymer.  This case will be taken up in Sec.
\ref{subsec:inclusion}.  As per standard arguments of statistical
mechanics, the results are supposed to be independent of ensemble
used.  However polymers provide many examples of non-equivalence of
these two ensembles \cite{Tintah99,kapri}.

%%%%%%%%%%%%%%%%%%%%%%%%%%%%%%%%%%%%%%%%%%%%%%%%%%%%%%%%%%%%%%%%%%%%%%%%%%%%%il
%%%%%%%%%%%%%%%%%%%%%%%%%%%%%%%%%%%%%%%%%%%%%%%%%%%%%%%%%%%%
\section{Flory theory in  a Modern Perspective}
\label{sec:flory-theory-modern}
%%%%%%%%%%%%%%%%%%%%%%%%%%%%%%%%%%%%%%%%%%%%%%%%%%%%%%%%%%%%%5
The significance of the Flory theory can be brought out by looking at
it from a modern point of view.  The failure of the mean field theory
in phase transitions and critical phenomena led to the ideas of
universality and scaling and the idea of studying problems at
different length scales, like renormalization group \cite{Freed87,Zinn-Justin90}.  The Flory
theory, even though believed to be a mean-field type theory, showed
all the aspects of the modern theory, in fact much more than a mean
field theory is expected to do.  In this section, we discuss the link
between the Flory theory and the idea of scale invariance and
universality, and the crossover behaviour.

\subsection{Scaling analysis}
\label{sec:scaling-analysis}
The appearance of power laws as in the $N$ dependence of the size of a
polymer is associated with scale invariance or the absence of any
typical scale.  To see this, compare the two functions, $f_l(x)=A
x^{-\alpha}$ and $f_s(x)=B e^{-x/\xi}/x^{\alpha}$ for $x\to\infty$.
For small $\alpha$, it is not possible to define any scale for
$f_l(x)$, apart from the size of its domain over which it is
normalized, while $f_s(x)$ is characterized by a scale $\xi$ of $x$.
If $x$ is measured on a different scale, i.e.  $x^{\prime}=\lambda x$,
then we see that by changing the prefactor $A^{\prime}=A
\lambda^{-\alpha}$, the functional form of $f_l$ remains invariant.
On this new scale one would still see the same power law behaviour, no
matter what the scale factor $\lambda$ is.  Compare this with $f_s$.
If $\lambda x\gg \xi$, then $f_s$  becomes too small to be
rejuvenated by increasing the coefficient $B$.  This is generally true
for any non-power law function.  On the other hand, if under a scale
change $x\to \lambda x$, a function $f(x)$ behaves as $f(\lambda
x)={\lambda}^{-p} f(x)$, then by choosing $\lambda=1/x$, we get a
power law form $f(x)\sim x^{-p}$. Therefore, a continuous scale
invariance (i.e.  any value of $\lambda$) implies power laws and vice
versa.

As an example, consider the probability distribution of the end-to-end
distance ${\bf R}$ of a polymer of length $L$.  This probability
$P({\mathbf R},L)$ depends, in principle, on all the parameters of the
problem, especially the starting microscopic length scales, in a way
consistent with dimensional analysis.  However for the large distance
behaviour, if the ratio of the bond length and the total length goes
to zero, one expects a dependence on $L$ only.

If we change the scale of measuring length by a factor $\lambda$, then
as per dimensional analysis all distances, small and large, need to be
scaled.  However if we choose to scale only the large lengths keeping
the microscopic scales unchanged, what we get is a dependence of
$\lambda$ solely coming from the $L$ part.  Such a transformation is
called a scale transformation.  A scale transformation for polymers
shows that if $L\to\lambda L$, then $R$ is scaled by $\lambda^{\nu}$,
even though in real life both $\langle R^2 \rangle^{1/2}$ and
$L^{1/2}$ are to be measured as lengths.  There is no violation of
dimensional analysis because a dimensionally correct form is $\langle
R^2 \rangle^{1/2}=b^{1-2\nu} L^{\nu}$.   If both $b$ and $L$ are
scaled as  $b\to\lambda^{1/2} b$ and $L\to\lambda L$, $\langle R^2
\rangle^{1/2}$ will also be scaled by $\lambda^{1/2}$ like $b$.  This
distinction between a scale transformation to get the scaling keeping
microscopic scales fixed and a transformation that scales all the
lengths as in dimensional analysis, is exploited in the
renormalization group approach.  In contrast, the Flory theory type
approaches take advantage of this difference by assuming that only the
large scales matter. We may amplify this by considering a particular
example.

Let us take the example of $P({\mathbf R},L)$ obtained from a
microscopic Hamiltonian defined earlier.  We change only $R$ and $L$,
as $L\to \lambda L$ and $R\to \lambda^{\nu} R$.  While doing this,
keep all other scales untouched, and therefore these will not be displayed.
Then $P$ can be claimed to show {\it scaling } if
\begin{equation}
  \label{eq:31}
  P({\mathbf R},L)={\lambda}^{X} P({\lambda}^{\nu}{\mathbf R},\lambda L),
\end{equation}
for any $\lambda$.  We are then free to choose $\lambda=1/L$ to write
\begin{equation}
  \label{eq:32}
  P({\mathbf R},L)=\frac{1}{L^X} \ {\cal P}\left(\frac{\mathbf
      R}{L^{\nu}}\right)  
    \equiv \frac{1}{L^X} \ P\left(\frac{\mathbf R}{L^{\nu}},1\right),
\end{equation}
where ${\cal P}(x) $ is called a scaling function. As advertised, this
form emphasizes the large scales only, by suppressing the dependence
on the small scales of the problem.  As a probability, the
normalization $\int d^d r P=1$ can be used to deduce that $X=d\nu$.
The scaling analysis therefore predicts the form of the probability distribution
as
\begin{equation}
  \label{eq:33}
   P({\mathbf R},L)=\frac{1}{L^{d\nu}} \ {\cal P}\left(\frac{\mathbf
      R}{L^{\nu}}\right),
\end{equation}
which agrees with the Gaussian distribution for $\nu=1/2$ (see
Eq.(\ref{elementary:eq4})).  We now use the result that the
equilibrium size is given by $R_0\sim N^{\nu}$, to argue in a
different way. If only the large scale like $R_0$ matters, then
dimensional analysis suggests that, being a density,
$P({\mathbf{R},N})\propto R_0^{-d}$, and the $R$ dependence has to be
in a dimensionless form.  With $R_0$ as the only scale, the argument
has to be ${\mathbf{R}}/R_0$.  This single scale assumption then tells
us $P({\mathbf R},L)=R_0^{-d} \ {\cal P}\left({\mathbf
    R}/{R_0}\right)$, in agreement with Eq. (\ref{eq:33}).

A different way to analyze the scale invariant behaviour is to do a
scale transformation of the underlying variables.
Let us start with the Edwards Hamiltonian Eq. \eqref{eq:8} and
scale the length of the polymer by a factor $\lambda$, i.e.,
$s=\lambda s^{\prime}$ so that ${\bf{r}}=\lambda^{\nu} {\bf{r}}^{\prime}$, where
$\nu$ is the polymer size exponent to be determined.  The Hamiltonian
now takes the form
\begin{equation}
  \label{eq:28}
  H\left[\mathbf{r}\left(s\right) \right]= 
     \frac{1}{2} \lambda^{2\nu-1} \int_{0}^{L/\lambda } ds 
          \left( \frac{\partial \mathbf{r}}{\partial s}\right)^2  
            +  \lambda^{2-d\nu} u \int_{0}^{L/\lambda } ds_1  
                   \int_{0}^{L/\lambda} ds_2 \;
  \delta^{d}\left(\mathbf{r}\left(s_2\right)-\mathbf{r}\left(s_1\right)\right),
\end{equation}
suppressing the primes on the variables.  We tacitly assumed that $u$
does not scale.  For $L\to\infty$, the first term is scale invariant
for the Gaussian value $\nu=1/2$.  This is ensured by the construction
of the Hamiltonian in terms of ${\bf{r}}$ and $s$.  However for this $\nu$ we
find that the scaled interaction $u^{\prime}=\lambda^{2-d/2} u$
increases with increasing $\lambda$ for $d<4$.  This suggests that a
Gaussian chain is unstable in presence of the interaction in the
continuum limit $\lambda \to \infty$ with $L\to\infty$.  Such a term
that grows on rescaling is called a {\it relevant term}.  The question is if
the Gaussian behaviour is unstable, whether there is a different scale
invariant stable behaviour.  For the interaction to be important, we
then demand that $\nu$ be such that both the terms scale in the same
way so that $H$ gets an overall scale factor.  This requires
$2\nu-1=2-d\nu$ or $\nu=\nu_F=3/(d+2)$, as given in Eq.(\ref{eq:6}).  In short, the scale invariance of the
Hamiltonian of a noninteracting polymer gives the Gaussian value
$\nu=1/2$ while the scale invariance for a repulsive polymer gives the
Flory exponent.  By taking $L\to\infty$, the polymer is made
scale-invariant at all large scales.  In this situation, the exponent
is visible in the scaling of space and length.  On the other hand, if
$L$ is finite but large, we expect the two terms to contribute equally
even at the largest possible scale, viz., $\lambda=L$.  One then
recovers the Flory exponent because the two integrals, assumed
convergent, are $O(1)$ in $L$ as the integral limits are from $0$ to
$1$.
 
What is actually required is the scale invariance of the free energy,
to include the effects of entropy, not the Hamiltonian per se.  This
introduces corrections that require an additional scale factor for
$u$.  As it so happens, this correction (vertex correction) is small
for polymers and the scale invariance of the Hamiltonian gives such a
close estimate.

%%%%%%%%%%%%%%%%%%%%%%%%%%%%%%%%%%%%%%%%%%%%%%%%%%%%%%%%%%%%%%%%%%%%%%%%%%%%%%%%%%%%%%%%%%%

\subsection{Scaling functions and interpolation}
\label{sec:interp-flow}
%The following section is meant for more advanced study and can be skipped at first reading.

Based on the Flory theory presented in the previous section, we see
the importance of the interactions $u$ and $v$ in determining the size
of a polymer but most importantly, the dependence on $N$ seems to be
universal in the sense that the exponent does not depend on $u,v$, but
any nonzero $u$ or $v$ change the Gaussian behaviour.

The dependence of the size on the interactions can be written in a
form consistent with dimensional analysis.  Taking the dimension of
${\bf r}$, the position vector, as length $\sf{L}$, and the
Hamiltonian as dimensionless, we have $[L]= {\sf{L}}^2$ (see Sec.\ref{sec:continuum-model}).  Although
$L$ is a measure of the polymer length it is dimensionally like a
surface, because of the fractal dimension ($d_F=2$) mentioned in Sec
II.  Since $[\delta^d({\bf{r}})]={\sf{L}}^{-d}$, the dimensions of
$u$ and $v$ in Eqs. \eqref{eq:8}, and \eqref{eq:20}  are respectively
${\sf{L}}^{d-4}$ and ${\sf{L}}^{2(d-3)}$.  With the help of
$L$, one may construct the dimensionless interaction parameters
\begin{equation}
  \label{eq:16}
  z= c_1 u L^{(4-d)/2},\quad {\rm and}\quad w= c_2 v L^{3-d},
\end{equation}
with dimensionless constants $c_1,c_2$ chosen as per convenience.
The size can be written as
\begin{equation}
  \label{eq:21}
  R^2 =  d\; L\ \psi^2(z,w).
\end{equation}
where the function $\psi$ gives the interpolation behaviour from the
Gaussian to the swollen chain ($u>0$). 
We suppress the $w$ dependence in this
discussion.  Eq.  \eqref{eq:21} of course subsumes that the short
distance scale $b$, or the range of interaction of $\Phi_2,\Phi_3$ do
not appear in the expression for the large $N$ limit.  This is a
highly nontrivial assertion which we shall assume to hold good for the
time being.  The renormalization group approach tackles this issue but
we do not get into that here. $\psi$ is often called a scaling function or 
a crossover function.

One question may arise here.  Just as Eq. \eqref{eq:21} is meant for a 
crossover from the Gaussian to the swollen chain, could we have done 
it the other way round?
We have seen that a polymer shows three different sizes, a theta chain
separating the swollen phase ($u>0$) and the collapsed phase ($u<0$).
There are situations (e.g., $d<3$) for which none of these is
Gaussian.  For a temperature or solvent induced phase transition of a 
polymer, what should be the reference point to define the scaling   function?
In such cases, it is the unstable state that is to be taken as
the primary size from which the interpolation formula or $\psi(z)$ has
to be constructed.  The Gaussian behaviour is unstable in presence of 
any $u$, but   the swollen state is the stable one for $u>0$, justifying 
the form written in Eq. \eqref{eq:21}.  
It then has to be modified by substituting $L^{1/2}$ by $L^{\nu_{\theta}}$ 
for the collapse transition.  This will be taken up
later in Sec. \ref{sec:crossover}.

One way to express the interpolation behaviour is to study the
behaviour of the size exponent as the parameters are varied.  Let us
consider the repulsive regime with $w=0$.  In this regime the power
law behaviour is observable in realistic systems only for large
lengths.  The approach to the asymptotic value can be determined by
studying the slope of $\psi$ with $L$ in a log-log plot.  The
effective exponent can be defined by a log-derivative
\begin{equation}
  \label{eq:18}
  \sigma(z)  = L \frac{\partial\ }{\partial L} \ln
  \psi=\frac{\epsilon}{2}\; z \frac{\partial\ }{\partial z} \ln
  \psi,\qquad (\epsilon=4-d), 
\end{equation}
with the $L$ derivative taken at a fixed $u$.  For large $z$ or $L$,
$\sigma(z)$ should approach a constant that from Eq. \eqref{eq:21}
would give $R\sim L^{(1/2) + \sigma}$, i.e., $\nu=(1/2)+\sigma$.

The function $\psi$ is analytic in the range $0<z<\infty$, because for
a finite chain the partition function, being a finite sum, cannot show
any singularity.  Also $\psi(0,0)=1$, by definition.  It is therefore
fair to expect a leading behaviour
\begin{equation}
  \label{eq:22}
  \psi(z)= (1+ a z)^p,
\end{equation}
so that 
\begin{equation}
  \label{eq:23}
   \psi(z)\stackrel{z\to\infty}{\sim} z^p, \quad {\rm{while}}\quad
   \psi(z)\stackrel{z\ll 1}{\sim} 1 + p az+\frac{p(p-1)}{2}
   (az)^2+..., 
\end{equation}
and
\begin{equation}
  \label{eq:24}
  \sigma(z)=\frac{\epsilon}{2}\ p \;
  \frac{az}{1+az}\stackrel{z\to\infty}{\longrightarrow}
  \frac{\epsilon}{2}\ p. 
\end{equation}
It seems that the large $z$ behaviour, of our immediate interest, can
be obtained from the small $z$ expansion of $\psi(z,0)$.  There are
various approaches to get $\psi$.  The perturbative renormalization
group approach, not discussed here, tries to get a well-behaved series
for $\sigma$, at least for small $\epsilon$, by starting with a power
series expansion in $z$.  A perturbative approach for $\psi$ will be
taken up in App \ref{sec:perturbation}.  In contrast to these, the
Flory theory is a {\it nonperturbative approach} to get $\psi$ or
$\sigma$ in the large $z$ limit directly.

As an example, we may quote the series for the scaling function obtained in a
double expansion in $z$ and $\epsilon$ \cite{Zinn-Justin90}  as
\begin{equation}
  \label{eq:25flow}
  \psi^{2}(z)= 1 + \frac{2}{\epsilon} z - \frac{6}{\epsilon^2} z^2
  +...  \approx (1+ 8z/\epsilon)^{1/4}, 
\end{equation}
keeping only highest order of $1/\epsilon$ in the coefficient of
powers of $z$.  This series then gives  $\sigma=\epsilon/16$, and  
$$\nu=\frac{1}{2} +\frac{\epsilon}{16} +{\rm O}(\epsilon^2).$$
For comparison, the Flory value  has the $\epsilon$-expansion,
\begin{equation}
  \label{eq:1flow}
  \nu_F=\frac{3}{d+2}= \frac{1}{2} +\frac{\epsilon}{12}+... . 
\end{equation}

\tableLN

\subsubsection{Why exponent?}
\label{sec:why-exponent}
The reason for focusing on the exponent $\nu$ can now be explained.
The occurrence of an exponent different from $\nu=1/2$ is noteworthy
because $R$ and $L^{\nu}$ are dimensionally different.  While the
difference owes its origin to the interactions, but, still, the
exponent obtained above does not depend on the parameters of the
interaction.  Even for a general short range interaction, instead 
of a contact potential 
in Eq.  (\ref{edwards:eq6}), the Flory argument in Sec.
\ref{sec:elementary} would produce the same $\nu$ as for the minimal
model.  This is universality.

There are two ways to motivate this universality.  One way is to use
$z$ as in Eq. (\ref{eq:16}), where the interaction parameter is made
dimensionless by $L$.  For any other short range interaction $\Phi_2$,
one may define a $z$-like appropriate dimensionless parameter e.g., by
taking $u=\int d^dr \Phi_2({\bf r})$.  In this case, for $L\to\infty$,
$z\to\infty$ for any value of $u>0$ if $d<4$, and the same asymptotic
limit is reached for all interactions.  The second way would be to use
a microscopic parameter like the bond-length, the range of interaction
or the size of a monomer, let's call it $b$, to define a dimensionless
interaction parameter $\alpha=u b^{4-d}$.  If on successive rescaling
of $b$ (``coarse-graining'') $\alpha$ approaches a fixed value
$\alpha^*$, then all short-range repulsions are ultimately described
by $\alpha^*$.  The emergence of $b$, as an extra length scale, then
allows a form $\langle R^2\rangle \sim L (L/b^2)^{2\nu-1}$, with $\nu$
determined by $\alpha^*$.  Here, $b$ appears as the saviour of an
apparent violation of dimensional analysis.

Although the Flory theory does not require the microscopic length scales
like $b$, we shall use both the versions, often by using $z$ to write 
$R\sim L^{1/2} z^{p},$  for some appropriate $p$ and often by introducing 
$b$ to make the power of $L$ explicit, as $R\sim b^{1-2\nu} L^{\nu}$.
The dictionary between the two sets with and without $b$ is summarized
in Table \ref{tab:tab2}.

%%%%%%%%%%%%%%%%%%%%%%%%%%%%%%%%%%%%%%%%%%%%%%%%%%%%%%%%%%%%%%%%%%%%%%%%%%%%%%%%%%

%%%%%%%%%%%%%%%%%%%%%%%%%%%%%%%%%%%%%%%%%%%%%%%%%%%%%%%%%%%%%%%%%%%%%%%%%%%%%%%%%%
\section{Flory Mean Field Theory}
\label{sec:flory}
%%%%%%%%%%%%%%%%%%%%%%%%%%%%%%%%%%%%%%%%%%%%%%%%%%%%%%%%%%%%%%%%%%%%%%%%%%%%%%%%%
For an interacting polymer, the partition function, from Eq.
\eqref{edwards:eq7}, can be written as
\begin{eqnarray}
\label{edwards:eq8}
\frac{Z}{Z_0}&=& \int d^d \mathbf{R} \,G_{L}\left(\mathbf{R}\right)= e^{-\beta \Delta F},
\end{eqnarray}
with $\Delta F=-k_BT(\ln Z -\ln Z_0)$ as the excess free energy due to
interaction.  For the excluded volume case of Eq. \eqref{eq:8}, this
excess free energy is called the free energy of swelling.

For the fixed-${\mathbf{R}}$ ensemble,  the Helmholtz free energy  is
\begin{eqnarray}
\label{edwards:eq14a}
F_L\left(\mathbf{R}\right)=E_L\left(\mathbf{R}\right)-TS_L\left(\mathbf{R}\right).
\end{eqnarray}
where the energy $E_L(\mathbf{R})$ is
defined from a ratio similar to Eq. \eqref{edwards:eq8}
but in terms of $G_{L}$,and $G_{L}^{(0)}$ as
\begin{eqnarray}
\label{edwards:eq12}
\exp\left[-\beta E_{L}\left(\mathbf{R}\right)\right]\equiv\frac{G_{L}\left(\mathbf{R}\right)}{G_{L}^{(0)}\left(\mathbf{R}\right)}&=&
\left \langle \exp \left[  -\beta V_L\left[\mathbf{r}\right] \right]
\right \rangle_{\mathbf{R}}^{(0)},
\end{eqnarray}
(the superscript $^{(0)}$ indicating an averaging with respect to
$H_L^{(0)}$), and the corresponding Boltzmann
entropy 
$S_L(\mathbf{R}) = k_B \ln [ G_L^{(0)}(\mathbf{R})\; Z_0 ]$.

The partition function  $Z$ of Eq. \eqref{edwards:eq8} is then given by
\begin{eqnarray}
\label{edwards:eq14b}
Z&=&Z_0 \int d^d \mathbf{R}\ G_{L}^{(0)}\left(\mathbf{R}\right) 
\ \langle e^{-\beta V_L\left[\mathbf{r}\right]}\rangle_{\mathbf{R}}^{(0)}=\int d^d \mathbf{R} \exp\left[-\beta F_L\left(\mathbf{R}\right) \right].
\end{eqnarray}

The Flory approximation discussed below attempts to get an approximation form for
 $F_L({\bf R})$.

%%%%%%%%%%%%%%%%%%%%%%%%%%%%%%%%%%%%%%%%%%%%%%%%%%%%%%%%

%%%%%%%%%%%%%%%%%%%%%%%%%%%%%%%%%%%%%%%%%%%%%%%%%%%%
%%%%%%%%%%%%%%%%%%%%%%%%%%%%%%%%%%%%%%%%%%%%%%%%%%%%%%%%%%%%%%%%%%%%%%%%%%%%%%%%
\subsection{Mean Field Approximation}
\label{subsec:mfa}
%%%%%%%%%%%%%%%%%%%%%%%%%%%%%%%%%%%%%%%%%%%%%%%%%%%%%%%%%%%%%%%%%%%%%%%%%%%%%%%%
Let us  introduce the monomer density
\begin{equation}
\label{edwards:eq4}
\rho\left(\mathbf{r}\right) =  \int_{0}^{L}ds \; \delta^{d}
\left(\mathbf{r}-\mathbf{r}\left(s\right) \right),\quad {\rm
  with}\quad \int d^d \mathbf{r} \; \rho(\mathbf{r}) = L, 
\end{equation}
so that $\rho$ is related to the number
concentration.  The polymer Hamiltonian in Eq.(\ref{edwards:eq6}) can
be recast in terms of the concentration in the form
\begin{eqnarray}
\beta V_L\left[\mathbf{r}\right]
&=&     \frac{u}{2!} \int d^d \mathbf{r}\  \rho^2\left(\mathbf{r}\right) 
     +  \frac{v}{3!} \int d^d \mathbf{r} \  \rho^3\left(\mathbf{r}\right),\label{eq:17} 
\end{eqnarray}
which can be verified by direct substitution of Eqs. \eqref{edwards:eq4}.

The mean field assumption is based on the (Gibbs-Bogoliubov) inequality
\begin{eqnarray}
\label{mfa:eq1}
\left \langle e^{-\beta V_{L}\left(\mathbf{r}\right)} \right \rangle_{\mathbf{R}} &\ge&  
e^{-\beta \left \langle V_{L}\left(\mathbf{r}\right) \right \rangle_{\mathbf{R}}}, 
\end{eqnarray}
and to a maximization of the right-hand-side with respect to a
parameter, as elaborated below, with the additional 
approximation 
\begin{equation}
  \label{mfa:eq1n}
\langle \rho \rho \ldots \rho
\rangle_{\mathbf{R}}=\langle \rho \rangle_{\mathbf{R}} \langle \rho
\rangle_{\mathbf{R}} \ldots \langle \rho \rangle_{\mathbf{R}}
\ldots,  
\end{equation}
so that the two and the three-body potential terms are reduced to a
product of $\langle\rho\left(\mathbf{r}\right)\rangle_{\mathbf{R}}$'s.
This factorization ignores all effects of density-density
correlations.  Since $\left \langle \rho\left(\mathbf{r}\right) \right
\rangle_{\mathbf{R}}$ gives the spatial variation of the density of
monomers of a single polymer, with the average density $L/R^d$, the
$r$-dependence can be taken in a scaling form
\begin{eqnarray}
\label{mfa:eq2}
\left \langle \rho\left(\mathbf{r}\right) \right
\rangle_{\mathbf{R}}&=& \frac{L}{R^{d}} 
\ \Theta\left(\frac{r}{R}\right)
\end{eqnarray}
where $\Theta(x)$ is a well-behaved function.  The assumption that has
gone in writing this form is that the behaviour of the density for
large $L$ is determined solely by the large distance scale $R$ and not
on the polymer-specific microscopic scales.  The prefactor takes care
of the dimensionality of the density so that the $r$-dependence has to
be in a dimensionless form. Under the assumption that only the large
scale $R$, the size of the polymer, matters, the dimensionless
argument of the function has to be $r/R$.  A uniform density sphere of
radius $R$ would have $\Theta(x)=$constant for $0<x\leq 1$ and zero
otherwise but there is no need to assume a uniform distribution of
monomers.

The mean-field expression for the Helmholtz free energy, from
Eq.(\ref{edwards:eq14a}) using Eq. (\ref{mfa:eq1}), \eqref{mfa:eq1n}
and \eqref{mfa:eq2}, gives the standard form of the Flory free energy
\begin{subequations}
\begin{eqnarray}
\label{mfa:eq7}
\beta F_{L}\left(R\right)&=&\frac{1}{2} \frac{R^2}{L} -
\frac{d}{2}
 \log \left(\frac{1}{2\pi L} \right) +  
R^d \left[\widetilde{u}   \left(\frac{L}{R^d}\right)^2 + 
\widetilde{v}
\left(\frac{L}{R^d}\right)^3 + \ldots \right]  \\
&\approx&\frac{1}{2} \frac{R^2}{ L} +  \widetilde{u}  \frac{L^2}{R^d}+
\widetilde{v}  \frac{L^3}{R^{2d}}  \label{mfa:eq8} 
\end{eqnarray}
\end{subequations}
where \begin{eqnarray}
\label{mfa:eq7b}
\widetilde{u}= \frac{1}{2!} u S_d \theta^{2}_{d-1}, &\quad {\rm and}\ & \widetilde{v}=
\frac{1}{3!}  v S_d \theta^{3}_{d-1}, 
\end{eqnarray}
are dependent on the density via the moments,
\begin{eqnarray}
\label{mfa:eq3}
\theta^{l}_k&=& \int_{0}^{\infty} d x \ x^{k} \; \Theta^{l}\left(x\right),
\end{eqnarray}
with $S_d = 2 \pi^{d/2}/\Gamma\left(d/2\right)$ coming from the
$(d-1)$-dimensional angular integrals.  Some explicit values in $d=3$ 
(corresponding to $k=2$) will be derived later on.  The resemblance of
this free energy, Eq. \eqref{mfa:eq8} with the simple Flory argument
of Eq. \eqref{elementary:eq11} should not go unnoticed.  

The logarithmic term appearing in Eq.(\ref{mfa:eq7}) yields
sub-dominant contributions under all circumstances, and will be
neglected in the most of the paper.  However this sub-dominant
log-term is an extremely important property of a polymer as it
signifies the probability of a polymer forming a loop with
${\mathbf{R}}=0$.  The coefficient ($d/2$ in this case) is called the
reunion exponent.  This exponent has certain universality
\cite{Mukherji} and is one of the characteristic exponents for
polymers.  Such a subleading term actually controls many polymeric
thermodynamic phase transitions and the order of the transition in the
$L\to\infty$ limit.  The most well-known example in this class is the
DNA melting\cite{poland}.
%%%%%%%%%%%%%%%%%%%%%%%%%%%%%%%%%%%%%%%%%%%%%%%%%%%%%%%%%%%%%%%%%%%%%%%%%%%%%%%%%%%%%%%%%%%%%%%%%%%%%%%%%%%%%%%%%%%%%%%%%%
\subsection{Solution through steepest descent method}
\label{subsec:steepest}
%%%%%%%%%%%%%%%%%%%%%%%%%%%%%%%%%%%%%%%%%%%%%%%%%%%%%%%%%%%%%%%%%%%%%%%%%%%%%%%%%%%%%%%%%%%%%%%%%%%%%%%%%%%%%%%%%%%%%%%%%
Our goal is to extract the dominant contribution to the integral
(\ref{edwards:eq14b}) using a steepest descent (saddle-point) method.
This method has the advantage of being systematically improvable.  In
addition we are also interested in the result for the end-to-end
distance
\begin{eqnarray}
\left \langle R^2 \right \rangle &=& \frac{1}{Z} \int d^d \mathbf{R}
\, R^2 \exp\left[- \beta F_{L} \left(R\right) \right]. 
\label{steepest:eq2}
\end{eqnarray}
This clearly amounts to considering the expansion around the minimum
$\mathbf{R}^{*}$
\begin{eqnarray}
\label{steepest:eq3}
\beta F_{L} \left(R\right) &=& \beta F_{L} \left(R^{*}\right)  + \frac{1}{2!} 
\left.\frac{\partial^2}{\partial R^2} \left[\beta F_{L} \left(R\right)
  \right]\right|_{R=R^{*}}\left(R-R^{*}\right)^2+ \ldots 
\end{eqnarray}
where the steepest descent condition $\partial \left[\beta F_{L}
  \left(R\right) \right]/\partial R|_{R=R^{*}}=0$ yields to lowest
order
\begin{eqnarray}
\label{steepest:eq4}
\frac{R^{2}}{d L} - 2  \widetilde{v}  \frac{L^{3}}{R^{2d}}+ \ldots
&=& \widetilde{u}  \frac{L}{R^{d}}. 
\end{eqnarray}
 
Anticipating the emergence of non-Gaussian value of $\nu$, a
short-distance scale $b$ (e.g. the bond length used earlier) can be
introduced to define a dimensionless variable $\mathbf{x}$
\begin{eqnarray}
\mathbf{R} &=& b N^{\nu} \mathbf{x},\quad {\rm where} \ N=L/b^2.
\label{flory:eq1}
\end{eqnarray}
With this $x$, 
the partition function and 
the size from
Eqs.(\ref{edwards:eq14b}) and (\ref{steepest:eq2}) can be expressed as 
 \begin{eqnarray}
 \label{flory:eq2}
 Z &=& b ^{d} N^{\nu d} \int d^d \mathbf{x}\, e^{-f(x,N)}\equiv
    b^d N^{\nu d} \widehat{Z},\quad {\rm and}\quad \left \langle R^{2} \right \rangle = b ^{2} N^{2 \nu }
\frac{1}{\widehat{Z}} \int d^d \mathbf{x}\, x^{2} \ e^{- f(x,N)},
\end{eqnarray}
where 
\begin{eqnarray}
\label{flory:eq3}
f(x,N)= \frac{1}{2} 
x^{2}  N^{2 \nu -1} + \alpha \frac{N^{2-\nu d}}{x^{d}}
+\gamma \frac{N^{3-2\nu d}}{x^{2d}},
\end{eqnarray}
and
\begin{equation}
  \label{flory:neq1}
\quad 
  \alpha=\widetilde{u} \ b^{4-d}, \quad \gamma=\widetilde{v}\ b^{6-2d}.
\end{equation}
are dimensionless.  The similarity of the powers of $N$ in $f(x,N)$
Eq.  (\ref{flory:neq1}) with the powers of $\lambda$ in Eq.
\eqref{eq:28} should be noted.

One may compare $\alpha, \gamma$ of Eq. (\ref{flory:neq1}) with the
dimensionless form $z,w$ introduced earlier in Eq. \eqref{eq:16}. The
latter are made dimensionless with $L$, the length, while here the
small scale $b$ is used for that purpose (see also Table \ref{tab:tab1}).  Although $Z$ and $\langle
R^2\rangle$ have been written above with an explicit $b$, it is
possible to avoid this arbitrary scale $b$ altogether.  By defining
$\mathbf{R}=L^{1/2}\; z^{(2\nu-1)/\epsilon}\; \mathbf{x},$ both $Z$
and $\langle R^2\rangle$ can be written in terms of $L,z,w$ without
any $b$.
 
The integrals involved in the above  expressions behave differently in
different temperature regimes.  These are discussed below. 

%%%%%%%%%%%%%%%%%%%%%%%%%%%%%%%%%%%%%%%%%%%%%%%%%%%%%%%%%%%%%%%%%%%%%%%%%%%%%%%%%%%%%%%%%%%%
\subsubsection{Flory regime($ \alpha > 0$, $ T > T_{\theta}$)}
\label{subsubsec:flory}
%%%%%%%%%%%%%%%%%%%%%%%%%%%%%%%%%%%%%%%%%%%%%%%%%%%%%%%%%%%%%%%%%%%%%%%%%%%%%%%%%%%%%%%%%%%
Let us first consider the good solvent case with $u>0$.  We still have
to worry about the dimensionality dependence. There are two
possibilities discussed one by one.
\begin{description}
\item{a)} Case $2<d<4$.  Matching the first and second terms in the
  exponential of (\ref{flory:eq2}) we find the Flory exponent seen
  earlier, $\nu=\nu_{F}={3}/({d+2})$.  In this case, the third term
  becomes sub-dominant in the $N \gg 1$ limit as $N^{3-2 \nu_{F} d}
  =N^{-3(d-2)/(d+2)}\ll 1$.  The size, 
is then given by Eqs.(\ref{flory:eq2}) and
  (\ref{flory:eq3}), 
with $\nu=\nu_F$ and 
\begin{eqnarray}
\label{flory:eq5}
f(x,N)=N^{\frac{4-d}{d+2}}\  \left(\frac{1}{2} x^2 + \frac{\alpha}{x^{d}}\right).
\end{eqnarray}

The function $f(x,N)$ can then be expanded around the minimum
$x^{*}=(\alpha d)^{1/(d+2)}$ resulting to lowest order
\begin{eqnarray}
\label{flory:eq9}
\left \langle R^{2} \right \rangle &\approx& b^{2} N^{ 2\nu_F}
\alpha^{2/(d+2),}
\end{eqnarray}
absorbing some unimportant constants in the definition of $b$. This is
the Flory regime.

\item{b)} Case $d>4$.  In this case the above analysis is no longer
  valid as $N^{\frac{4-d}{d+2}}\ll 1$ in the large $N$ limit and the
  steepest descent method cannot be applied.%, to Eqs. (\ref{flory:eq7})
  %and (\ref{flory:eq5}).  
We then go back to Eq.(\ref{flory:eq2}) and
  assume that there is no $N$ dependence in the entropic term. This
  amounts to setting $N^{2\nu-1}=1$, or $\nu\equiv\nu_G=\frac{1}{2}.$
  Then all the other terms in (\ref{flory:eq2}) vanish in the $N \gg
  1$ limit and Eqs.(\ref{flory:eq2}) and (\ref{flory:eq3}) give
\begin{eqnarray}
\left \langle R^2 \right \rangle \approx b^{2}N  
\label{flory:eq12}
\end{eqnarray}
\end{description}
Eqs.(\ref{flory:eq9}) and (\ref{flory:eq12}) can then be written as 
\begin{eqnarray}
\label{flory:eq13}
\left \langle R^2 \right \rangle \approx C b^2
N^{2\max\left(\nu_F,1/2\right)}, \ {\rm or}\ R\approx C^{1/2}b^{1-\nu} L^{\nu},
\end{eqnarray}
where $C=\alpha^{2/(d+2)}$ when $d<4$ and $C=1$ when $d \ge 4$, and
$\nu= \max\left(\nu_F,1/2\right)$.

That $d=4$ is special is seen from the power of $N$ in Eq.
(\ref{flory:eq5}) and from the fact that $u$ is dimensionless in $d=4$
so that $z$ becomes large as $L\to\infty$ for $d<4$.
%%%%%%%%%%%%%%%%%%%%%%%%%%%%%%%%%%%%%%%%%%%%%%%%%%%%%%%%%%%%%%%%%%%%%%%%%%%%%%%%%%%%%
\subsubsection{Theta regime ($\alpha=0$, $T=T_{\theta}$)}
\label{subsubsec:theta}
%%%%%%%%%%%%%%%%%%%%%%%%%%%%%%%%%%%%%%%%%%%%%%%%%%%%%%%%%%%%%%%%%%%%%%%%%%%%%%%%%%%%
In this case, the term proportional to $\alpha$ is absent in both
Eqs.(\ref{flory:eq2}) and (\ref{flory:eq3}), and again there exist two
different regimes depending on the dimensionality of the system
\begin{description}
\item{a)} Case $1<d<3$. Matching the first and the third terms we find
\begin{eqnarray}
\nu_{\theta} &=& \frac{2}{d+1}
\label{theta:eq1}
\end{eqnarray}
which coincides with the Gaussian value $\nu_G=1/2$ in $d=3$ but is
different from it in $d=2$.  However, the value $\nu=2/3$ for $d=2$
differs from the known exact value $\nu=4/7$ \cite{flavio91,flavio88}.
The Flory theory has also been extended to theta points on
fractals\cite{flavio91b} and dilute lattices\cite{bkc95}.

Substituting in Eq.(\ref{flory:eq2}) we find
\begin{eqnarray}
f(x,N)= N^{\frac{3-d}{d+1}}\  \left(\frac{1}{2} x^2 + \frac{\gamma}{x^{2d}}\right)
\label{theta:eq3}
\end{eqnarray}
It is worth noticing that the matching choice ($1^{\text{st}}$ and
$2^{\text{nd}}$ terms for Flory regime $\alpha>0$, $1^{\text{st}}$ and
$3^{\text{rd}}$ terms for the $\theta$-regime $\alpha=0$) is {\it
  unique}, as any alternative choice would lead to inconsistent
results.

Because $N^{\frac{3-d}{d+1}} \gg 1$ in this regime, we can apply the
steepest descent method along the lines previously shown thus yielding
the $\theta$-regime.

Following the same reasoning as in the previous case, to leading order
in the steepest descent expansion, we find
\begin{eqnarray}
\label{theta:eq6}
  \left \langle R^2 \right \rangle 
         \approx  b^{2} N^{2 \nu_{\theta}} \left(2
           \gamma\right)^{1/\left(d+1\right)}, 
         & {\rm or}& 
   \left \langle R^2 \right \rangle \approx  L w^{1/\left(d+1\right)}.
\end{eqnarray}
\item{b)} Case $d>3$. Again, the only possibility is to choose $\nu$
  such that the entropic part has no $N$ dependence, and this again
  leads to the Gaussian result, (\ref{flory:eq12}).
\end{description}
The final result for this case is then
\begin{eqnarray}
\label{theta:eq7}
\left \langle R^2 \right \rangle 
\approx C_{\theta} b^2 N^{2\max\left(\nu_{\theta},1/2\right)},\quad
\end{eqnarray}
where $C_{\theta}=(2\gamma)^{1/(d+1)}$ when $d<3$ and $C_{\theta}=1$
when $d \ge 3$.

Here, $d=3$ turns out to be special, unlike the good solvent case
which has $d=4$ as the special dimensionality.  This is apparent from
the power of $N$ in Eq. (\ref{theta:eq3}) and the fact that $v$ is
dimensionless for $d=3$.  For $d<3$, $w$ becomes large with length of
the polymer.

%%%%%%%%%%%%%%%%%%%%%%%%%%%%%%%%%%%%%%%%%%%%%%%%%%%%%%%%%%%%%%%%%%%%%%%%%%%%%%%%%%%%%%%%
\subsubsection{Compact regime ($\alpha<0$, $T<T_{\theta}$)}
\label{subsubsec:compact}
%%%%%%%%%%%%%%%%%%%%%%%%%%%%%%%%%%%%%%%%%%%%%%%%%%%%%%%%%%%%%%%%%%%%%%%%%%%%%%%%%%%%%%%%
We now go back to Eqs(\ref{flory:eq2}) and (\ref{flory:eq3}) where we
set $\alpha=-\vert \alpha\vert$.  In this case, the term proportional
to $\gamma$ becomes very important to guarantee the convergence of the
integral. As the term proportional to $\alpha$ cannot be dropped, the
only remaining possibility is to match these two terms. This leads to
the result $\nu_c = \frac{1}{d}$, as noted in Eq. \eqref{eq:7}.

In this case $N^{2 \nu_{c} -1}=N^{(2-d)/d} \ll 1$ and the Gaussian
term is sub dominant in the $N \gg 1$ limit.  Note, however, that
unlike previous cases, it {\it cannot} be dropped as it ensures the
convergence of the integral at large $x$ \footnote{Note that this resembles the role of the
{\it irrelevant dangerous} variables in renormalization group
theory.}. This should
also be taken into account on the integration domain of $R$ since it
should not extend beyond $N$.  It is nevertheless irrelevant for the
computation of average quantities such as $\langle R^2 \rangle$ given
by Eq.(\ref{flory:eq3}).

To leading order, one finds
\begin{eqnarray}
\label{compact:eq3}
\left \langle R^{2} \right \rangle \approx b^{2} N^{2 \nu_{d}} \left(
  \frac{2 \gamma}{\left \vert \alpha \right \vert} \right)^{2/d},
\end{eqnarray}
for any $d$.  This value $\nu_d=1/d$ is consistent with the idea of a
compact sphere with density $\sim O(1)$.  To prevent a complete
collapse, one needs a repulsive interaction and the three body term
$v>0$ helps in stabilization of the phase.
%%%%%%%%%%%%%%%%%%%%%%%%%%%%%%%%%%%%%%%%%%%%%%%%%%%%%%%%%%%%%%%%%%%%%%%%%%%%%%%%%%%%%%%%%%%%%%%%%
%%%%%%%%%%%%%%%%%%%%%%%%%%%%%%%%%%%%%%%%%%%%%%%%%%%%%%%%%%%%%%%%%%%%%%%%%%%%%%%%%%%%%%%%%%%%%%%%%%%%%%%%%%%%%%%%
\section{Additional remarks on Flory theory}
\label{sec:additional}
%%%%%%%%%%%%%%%%%%%%%%%%%%%%%%%%%%%%%%%%%%%%%%%%%%%%%%%%%%%%%%%%%%%%%%%%%%%%%%%%%%%%%%%%%%%%%%%%%%%%%%%%%%%%%
%%%%%%%%%%%%%%%%%%%%%%%%%%%%%%%%%%%%%%%%%%%%%%%%%%%%%%%%%%%%%%%%%%%%%%%%%%%%%%%%%%%%%%%%%%%%%%%%%%%%%%%%%%%%%
\subsection{Summary of Flory predictions}
\label{subsec:summary}
%%%%%%%%%%%%%%%%%%%%%%%%%%%%%%%%%%%%%%%%%%%%%%%%%%%%%%%%%%%%%%%%%%%%%%%%%%%%%%%%%%%%%%%%%%%%%%%%%%%%%%%%%%%%
For each of the swollen and the theta cases, there is a critical
dimensionality $d_c$ above which the interactions are not significant
enough to cause a change in the Gaussian behaviour.  This critical
dimensionality is $d_c=4$ for the excluded volume interaction and
$d_c=3$ for the theta point.  These are the dimensions at which the
interaction constants $u$ and $v$ are dimensionless.  For $d<d_c$, the
interactions cannot be ignored, no matter how small, but its magnitude
does not play any role.  The size exponents are not dependent on the
strength of the interaction, so long it is nonzero and positive.  The
values of the exponents predicted by Flory theory in the above three
regimes and different dimensionalities are  recalled in Table
\ref{tab:tab1}.

\tableone
%%%%%%%%%%%%%%%%%%%%%%%%%%%%%%%%%%%%%%%%%%%%%%%%%%%%%%%%%%%%%%%%%%%%%%%%%%%%%%%%%%%%%%%%%%%%%%%%%%%%%%%%%%%
\subsection{Modification of the entropic term and Flory interpolation formula}
\label{subsec:elastic}
%%%%%%%%%%%%%%%%%%%%%%%%%%%%%%%%%%%%%%%%%%%%%%%%%%%%%%%%%%%%%%%%%%%%%%%%%%%%%%%%%%%%%%%%%%%%%%%%%%%%%%%%%%
As remarked earlier, the equilibrium size of a polymer coil is
determined by the balance between polymer interactions and polymer
elasticity, which is entropic in nature.  To derive an interpolation

formula that would be applicable away from the asymptotic large $N$
regime, an intuitively appealing argument can be made.  The free
energy in the fixed $R$ ensemble, based on the Gaussian distribution
and the interactions, is given by Eq. \eqref{mfa:eq7}.  To this we may
add an extra entropy coming from the possibility of placing one end
point of the polymer anywhere in the volume $R^d$.  This entropy is of
the form $\sim \ln R^d$. 
Therefore the modified Flory free energy, obtained by adding this
extra entropic contribution to the form given in Eq.(\ref{mfa:eq7}), is 
$\beta F_{L}\left(R\right)|_{\rm modified} = \beta
F_{L}\left(R\right) - d \ln R$,  
where $\beta F_L(R)$ is given by  Eq.(\ref{mfa:eq7}).

In terms of  the swelling factor,
$\psi^2 $
that compares the size of a polymer with the corresponding Gaussian
size as defined in Eq. (\ref{eq:21}), the modified  free energy can be
expressed as
\begin{eqnarray}
  \beta \overline{F}_{L}\left(\psi\right) 
        &\approx& \frac{d}{2} \psi^2 -d \log\left(\psi\right) 
  +\frac{d+1}{d} \frac{z}{\psi^{d}} + \frac{w}{\psi^{2d}}.
\label{elastic:eq6}
\end{eqnarray}
Here  $z,w$ defined in Eq. \eqref{eq:16}, with  $c_1,c_2$ involving the
$\theta$'s of Eq. \eqref{mfa:eq7b}, are
\begin{eqnarray}
\label{elastic:eq9}
z=\frac{d^{(2-d)/2}}{2(d+1)}S_d \theta^{2}_{d-1} u L^{\epsilon/2},
\quad w=\frac{1}{3!}d^{-d} S_d \theta^{3}_{d-1} v L^{3-d}.
\end{eqnarray}

The corresponding steepest descent equation yields
\begin{eqnarray}
  \psi^{d+2}- \psi^d 
  - 2 w \frac{1}{\psi^d} &=& \frac{d+1}{d}\  z.
\label{elastic:eq7}
\end{eqnarray} 
Note that, this is basically Eq.(\ref{steepest:eq4}) with an extra
term (the second term on the left-hand-side of Eq.(\ref{elastic:eq7}))
that stems from the modification.  In particular, for $d=3$, the form
is \cite{Ptitsyn68,deGennes75}, is
\begin{eqnarray}
\psi^{5}\left(z\right)- \psi^{3}\left(z\right) -\frac{2
  \widetilde{v}}{\psi^{3}\left(z\right)} &=& \frac{4}{3} z.
\label{elastic:eq8}
\end{eqnarray}
with $z$ as in Eq. \eqref{elastic:eq9} for $d=3$.   A comparison with
the value of $z$ used  in Eq. \eqref{mfa:eq7b} (see also Eq. \eqref{flory:neq1}) yields a well
defined value for $\theta^{2}_2$ whose general expression appeared in
Eq.(\ref{mfa:eq3}), that is
\begin{eqnarray}
\label{elastic:eq10}
\theta^{2}_2 &=& \frac{9}{\pi^2} \sqrt{\frac{1}{2 \pi}}
\end{eqnarray}
so that inserting Eq.(\ref{mfa:eq7b}) into Eq.(\ref{elastic:eq9}) we
get
\begin{eqnarray}
\label{elastic:eq11}
z&=& \left(\frac{3}{2 \pi}\right)^{3/2}  u L^{1/2}
\end{eqnarray}
The behavior of $\psi(z)$ is given in Fig. \ref{fig:fig4} which
displays the 'loop' for sufficiently low values of  parameter $2
\gamma$ (see also Eq. \eqref{flory:neq1}).

For $\gamma=0 (w=0),$ a power series solution for $\psi(z)$ can be
constructed as
\begin{eqnarray}
  \label{eq:26}
  \psi\left(z\right)&=&1 +\frac{2}{3} z - \frac{14}{9} z^2 + \frac{160}{27} z^3 +\ldots,
\end{eqnarray}
which could be verified, order by order, by direct substitution in
Eq. (\ref{elastic:eq8}).    Furthermore, this  gives
  \begin{equation}
    \label{eq:26sq}
    \psi^2(z)=1 +\frac{4}{3} z -2.66667  z^2 + 9.77778  z^3 +\ldots.
\end{equation}
This solution may be compared with a brute-force computation of the
perturbative series 
in Ref. \cite{Muthukumar84},
  (see Appendix \ref{sec:perturbation}) 
\begin{eqnarray}
\label{first:eq5}
\psi^2\left(z\right)
    &=& 1+ \frac{4}{3} z - 2.0754... z^2 + 6.2968... z^3
         -25.057... z^4
           + \ldots,
\end{eqnarray}
where $z$ is as given in Eq.(\ref{elastic:eq11}). 

The interpolation formula given by Eq. \eqref{elastic:eq8} was
verified for polystyrene in cyclohexane \cite{Swislow80}.  A slightly
different form wiht a constant term on the right hand side was however
found for polymethyl methacrylate \cite{nakata97}.

\figfour

\subsection{Modification of the Flory estimate}
\label{sec:modif-flory-estim}
The idea of the fractal dimension introduced in \ref{sec:size-polymer}
suggests that the number of monomers in a sphere of radius $R$ is
$R^{d_F}$ and therefore the energy term should be $R^{2d_F-d}$, $d_F$
to be determined.  This is not consistent with the estimate of the
repulsive energy $N^2/R^d$ which is crucial for the Flory exponent.  A
more refined argument would however favour the Flory estimate
\cite{Maritan87}.

The polymer chain can be thought of consisting of smaller blobs of $n$
monomers within which the effect of repulsion is not significant and
the spatial size can be taken as $\xi^2\sim n$.  The chain then
consists of $N/n$ such blobs. This coarse-grained polymer will have
$(N/n)^2 (R/\xi)^{-d}$ contacts of blobs.  There is a need to know the
number of overlap of monomers of the two fractal blobs.  The
dimensionality of the intersections of two fractals follow a rule of
addition of co-dimensions.  For a $D$--dimensional fractal embedded in
a $d$--dimensional space, the co-dimension is the dimensionality of
complementary space and is $d-D$.  The additivity rule says the
co-dimension of the intersections of two fractals is the sum of the
co-dimensions of the two.  In other words the fractal dimension $D$ of
the points of contact of two fractals of dimensions $D_1$ and $D_2$
would obey, $d-D= (d-D_1)+(d-D_2)$, i.e., $D=D_1+D_2-d$.  As per this
rule the number of contacts of the two blobs will be $\xi^{4-d}$.
Therefore, the repulsive energy will be
$$\frac{N^2}{n^2}\ \frac{\xi^d}{R^d}\ \xi^{4-d}=\frac{N^2}{R^d},$$
recovering the Flory estimate.

%%%%%%%%%%%%%%%%%%%%%%%%%%%%%%%%%%%%%%%%%%%%%%%%%%%%%%%%%%%%
%%%%%%%%%%%%%%%%%%%%%%%%%%%%%%%%%%%%%%%%%%%%%%%%%%%%%%%%%%%%%
\subsection{Explicit computation of the scaling function in $d=2$}
\label{sec:explicit}
%%%%%%%%%%%%%%%%%%%%%%%%%%%%%%%%%%%%%%%%%%%%%%%%%%%%%%%%%%%%%%%%%%%%%%%%%%%%%%%
A rather interesting case occurs in $d=2$ where the scaling function
$\psi(z)$ can be computed explicitly.  We go back to Eqs.
\eqref{flory:eq2} for $d=2$ and $\gamma=0$.  With a change of variable
$\mathbf{R} = b N^{\nu_F} \mathbf{x}$ similar to Eq.(\ref{flory:eq1})
we get from Eq.(\ref{flory:eq2}) (in polar coordinates)
\begin{eqnarray}
\label{explicit:eq2}
\left \langle R^{n} \right \rangle &=& b^{n} N^{n \nu_F} \frac{\int_0^{\infty} dx\, x^{n+1} 
  \exp\left[-f_2(x,N)\right]}{\int_0^{\infty} dx\, x 
  \exp\left[-f_2(x,N)\right]}
\end{eqnarray} 
where
\begin{eqnarray}
  f_2(x,N)&=& \frac{1}{2} N^{1/2} x^2 +\alpha N^{1/2} \frac{1}{x^2},
\label{explicit:eq4}
\end{eqnarray}
in the Flory regime $\nu_F=3/4$ ( for which $\gamma$ and higher terms  are
irrelevant in the $N\gg 1$ limit).  Here, $f_2$  is Eq. \eqref{flory:eq5} in
$d=2$.

Let us introduce the ratio
\begin{eqnarray}
  \chi_2\left(N,\alpha\right)&=& \frac{\left \langle R^{3} \right \rangle}{\left \langle R \right \rangle},
\label{explicit:eq5}
\end{eqnarray}
which is also a definition of ``polymer extension'' analogous to
$\langle R^2\rangle$.  The scaling function is then
$\psi^2=\chi_2/(b^2N)$ which can be written as
\begin{eqnarray}
  \psi^2\left(N,\alpha\right) &=&  N^{1/2} \frac{\int_{0}^{\infty} dx\, x^{4} 
    \exp\left[-N^{1/2} \left(\frac{x^2}{2}+\frac{\alpha}{x^2} \right)\right]}{\int_{0}^{\infty} dx\, x^{2} 
    \exp\left[-N^{1/2} \left(\frac{x^2}{2}+\frac{\alpha}{x^2} \right)\right]}.
\label{explicit:eq6}
\end{eqnarray}
Both integrals are easily evaluated within a steepest descent
procedure with saddle point $x_0^2=(2\alpha)^{1/2}$. The asymptotic
form in the large $N$ limit is
\begin{eqnarray}
 \psi^2\left(N,\alpha\right)&=&  N^{1/2} \alpha^{1/2}\sim\sqrt{z},
\label{explicit:eq7}
\end{eqnarray}
consistent with the large $z$ behaviour of Eq. \eqref{elastic:eq7} for
$d=2$.

On the other hand an exact calculation can also be carried out by
using the following identities
\begin{eqnarray}
\label{explicit:eq8}
\int_{0}^{\infty} dx \, \exp\left[-b x^2 -\frac{a}{x^2} \right]&=&
\frac{1}{2} \sqrt{\frac{\pi}{b}} \exp\left[-2 \sqrt{ab}\right] \\
\label{explicit:eq9}
\int_{0}^{\infty} dx \, x^2 \exp\left[-b x^2 -\frac{a}{x^2} \right]&=&
\left(\frac{1}{2b}+\sqrt{\frac{a}{b}}\right) \frac{1}{2} \sqrt{\frac{\pi}{b}} \exp\left[-2 \sqrt{ab}\right]\\
\label{explicit:eq10}
\int_{0}^{\infty} dx \, x^4 \exp\left[-b x^2 -\frac{a}{x^2} \right]&=&
\left(\frac{3}{4b^2}+\frac{a}{b}+\frac{3}{2b}\sqrt{\frac{a}{b}}\right) 
\frac{1}{2} \sqrt{\frac{\pi}{b}} \exp\left[-2 \sqrt{ab}\right]\nonumber \\
\end{eqnarray}
The integral (\ref{explicit:eq8}) is proven in Ref.\cite{Schulman},
remaining two can be evaluated immediately by taking derivatives with
respect to $b$. Then we find the exact result
\begin{eqnarray}
  \psi^2\left(N,\alpha \right) &=& \sqrt{N\alpha}\  \frac{1+3 X+ 3X^2}{1+X},\quad{\rm with}\quad X=\frac{1}{ \sqrt{2\alpha N}},
\label{explicit:eq11}
\end{eqnarray}
so that for $N \gg 1$
\begin{eqnarray}
 \psi^2\left(N,\alpha\right)&=& = \sqrt{N\alpha}
\left[1+\frac{1}{\sqrt{\alpha N}}+\mathcal{O}\left(\frac{1}{\alpha N}\right) \right]
\label{explicit:eq12}
\end{eqnarray}
in agreement with Eq.(\ref{explicit:eq7}), while for $\alpha\to 0$, 
$\psi^2\to 3/\sqrt{2}$. 

In view of the exactness of the Flory exponent in $d=2$, it would be
extremely important to see how this exact result on the crossover
function fares with real experiments.  This could be accessible in
studies on polymers adsorbed on a surface or planar interface.

%%%%%%%%%%%%%%%%%%%%%%%%%%%%%%%%%%%%%%%%%%%%%%%%%%%%%%%%%%%%%%%%%%%%%%%%%%%%%%%%%%%%%%%%%%%%%%%%%

%%%%%%%%%%%%%%%%%%%%%%%%%%%%%%%%%%%%%%%%%%%%%%%%%%%%%%%%%%%%%%%%%%%%%%%%%%%%%%%%%%%%%%%%%%%%%%%%%%%%%%%%%%%%%%%%%%%%%%%%%%%%%%%%%%%
%\subsection{Explicit solution in the $d=3$ case}
%\label{sec:solution}
%%%%%%%%%%%%%%%%%%%%%%%%%%%%%%%%%%%%%%%%%%%%%%%%%%%%%%%%%%%%%%%%%%%%%%%%%%%%%%%%%%%%%%%%%%%%%%%%%%%%%%%%%%%%%%%%%%%%%%%%%%%%%%%%%%%
%%%%%%%%%%%%%%%%%%%%%%%%%%%%%%%%%%%%%%%%%%%%%%%%%%%%%%%%%%%%%%%%%%%%%%%%%%%%%%%%%%%%%%%%%%%%%%%%%%%%%%%%%%%%%%%%%%%%%%%%%%%%%%%%%%
\subsection{The Uniform expansion method}
\label{subsec:uniform}
%%%%%%%%%%%%%%%%%%%%%%%%%%%%%%%%%%%%%%%%%%%%%%%%%%%%%%%%%%%%%%%%%%%%%%%%%%%%%%%%%%%%%%%%%%%%%%%%%%%%%%%%%%%%%%%%%%%%%%%%%%%%%%%%%%
We now attempt to get the
interpolation formula, Eq.(\ref{elastic:eq8}), in a more systematic
but nonperturbative way. 
The basic idea
is to introduce an effective Gaussian distribution with a new
elastic constant  $\psi^{-2}$ in such a way that 
\begin{equation}
  \label{uni:eq:1}
 \langle R^2\rangle= \langle R^2 \rangle_{0}^{\prime}, 
\end{equation}
where $\langle (\ldots) \rangle_{0}^{\prime}$ is the Gaussian average with respect to the new
distribution. \cite{Doi86}   

Let us write the Edwards Hamiltonian as (absorbing $\beta$ in the
Hamiltonian)
\begin{subequations}
\begin{equation}
  \label{uni:eq:1n}
  H= H_0^{\prime}+ (H-H_0^{\prime}) = H_0^{\prime}+ \delta H 
\end{equation}
where
\begin{eqnarray}
  \label{uni:eq:2}
 H_0^{\prime} &=& \frac{1}{2\psi^2} \int_{0}^{L} ds \left( \frac{\partial \mathbf{r}}{\partial s}\right)^2,\\
  \delta H &=& \frac{1}{2}\left(1-\psi^{-2}\right) \int_{0}^{L} ds
                \left( \frac{\partial \mathbf{r}}{\partial s}\right)^2  
                + \frac{1}{2!} u  \int_{0}^{L} ds_1  \int_{0}^{L} ds_2 \;
              \delta\left(\mathbf{r}\left(s_2\right)-\mathbf{r}\left(s_1\right)\right)\\
            &\equiv& \delta H_1 + \delta H_2.
\end{eqnarray}
\end{subequations}
Here $\psi^2$ is a dimensionless ``elastic constant'' such that for
$H_0^{\prime}$ which is a Gaussian Hamiltonian, $\langle R^2\rangle_0^{\prime}=d
\psi^2 L$.  Admittedly, $\psi$ is reminiscent of the swelling factor
of the previous section, $\psi^2=\langle R^2\rangle_0^{\prime}/R_0^2$,
and in fact it is the actual one if the condition expressed by Eq.
\eqref{uni:eq:1} is satisfied.

Expanding $\langle R^2\rangle$,  to first
order in $\delta H$, we have
\begin{equation}
  \label{uni:eq:3}
  \langle R^2\rangle
  = \frac{\int d^d \mathbf{R} \; R^2
       G_{L}\left(\mathbf{R}\right)}{\int d^d \mathbf{R} \;
       G_{L}\left(\mathbf{R}\right)}
  =  \langle R^2\rangle_0^{\prime} -\left( \langle R^2 \delta H\rangle_0^{\prime}  -  \langle R^2\rangle_0^{\prime} \langle \delta H\rangle_0^{\prime}\right), 
\end{equation}
where the prime denotes an averaging with $H_0^{\prime} $.
The resulting Gaussian integrations
can be performed to obtain
\begin{eqnarray}
\label{uniform:eq1}
\left \langle R^2 \right \rangle 
   &=& \left \langle R^2 \right \rangle_{0}^{\prime} -
\left \langle R^2 \right \rangle_{0}^{\prime} 
\left[1-\psi^2 + \frac{1}{\psi^3}\ \frac{4}{3} z 
+ \ldots \right].
\end{eqnarray}
The details of the calculations are similar to those reported in Appendix B, with the relevant result given by Eq.(\ref{gaussian:eq14}), but the
occurrence of the $\psi$-terms can be understood from a
transformation.  Since $H_0^{\prime}$ represents a Gaussian with the
extra $\psi$ factor, we have $G_0^{\prime}({\bf r})=\psi^{-d/2}
G_0({\bf r}/\psi)$.  Therefore, by scaling $R,{\bf r}\to R/\psi, {\bf
  r}/\psi$, one gets the factor $\psi^4$ coming from the correlation
of $R^2$ and $\delta H_1$.  As $H$ is dimensionless the correlation is
dimensionally $\sim L$ because of $R^2$.  When these are combined with
the factor $(1-\psi^{-2})$ of $\delta H_1$, we find the contribution
$\sim L \psi^2 (1-\psi^2)$.  A change in variable gives $\delta({\bf
  r})\to \psi^{-d} \delta({\bf r}/\psi)$, so that the $\delta H_2$
correlation would be $\sim (L \psi^2) (\psi^{-3}) (uL^{1/2})$,  in three
dimensions, as in  Eq.  \eqref{uniform:eq1}.

The swelling factor analog of Eq.(\ref{eq:21}) now reads
\begin{eqnarray}
\label{uniform:eq2}
\psi^2 &=& 
  \frac{\left \langle R^2 \right \rangle}{\left \langle R^2 \right \rangle_{0}}
  = 
 \frac{\left \langle R^2 \right \rangle_{0}^{\prime}}{\left \langle R^2 \right \rangle_{0}}
\end{eqnarray}
The last step stems from the condition
\begin{eqnarray}
\label{uniform:eq3}
\left \langle R^2 \right \rangle &=& \left \langle R^2 \right \rangle_{0}^{\prime}
\end{eqnarray}
which is the requirement of the uniform expansion method, equating the
correction term in the square bracket to zero. This yields
\begin{eqnarray}
\psi^{5}\left(z\right)- \psi^{3}\left(z\right) &=& \frac{4}{3} z
\label{uniform:eq4}
\end{eqnarray}
that coincides with Eq.(\ref{elastic:eq8}) with $\gamma=0$ (no $3-$body term), as it should.

Although the scheme is based on the first order perturbative result,
the method via the choice of $\psi$ makes it nonperturbative and 
applicable to a wide variety of situations.

%%%%%%%%%%%%%%%%%%%%%%%%%%%%%%%%%%%%%%%%%%%%%%%%%%%%%%%%%%%%%%%%%%%%%%%%%%%%%%%%%%%%%%%%%%%%%%%%%%%%%%%%%%%%%%%%%%%%%%%%%%%%%%%%%%%
\subsection{Extension of Flory theory to more complex systems}
\label{subsec:extensions}
%%%%%%%%%%%%%%%%%%%%%%%%%%%%%%%%%%%%%%%%%%%%%%%%%%%%%%%%%%%%%%%%%%%%%%%%%%%%%%%%%%%%%%%%%%%%%%%%%%%%%%%%%%%%%%%%%%%%%%%%%%%%%%%%%%%
As anticipated in the Introduction, Flory theory is still a widely used tool in many different soft matter systems with increasing
complexity. This is because, in spite of its simplicity and known limitations, it is able to capture the main essential competition
between entropic and energetic contributions.
There are clearly too many cases of extensions of Flory theory to these more complex systems to be reproduced here.
As representative examples, we will then confine ourselves to two important cases.

The first one is related to the possibility of having anisotropy with a preferred direction as, for instance, for
directed linear or branched polymers \cite{Redner82,Lubensky82}. The same idea will also be taken up in Sec. \ref{subsec:inclusion}
where the case of the inclusion of an external force will be discussed.

For directed systems, with a preferred direction, we introduce a transverse typical radius $R_{\perp}$ and a longitudinal radius $R_{\|}$,
along with the corresponding exponents $\nu_{\perp}$ and $\nu_{\|}$, so that $R_{\perp} \sim N^{\nu_{\perp}}$ and 
$R_{\|} \sim N^{\nu_{\|}}$.
The extension of the free energy (\ref{elementary:eq11}) to the present case then reads in general dimensionality $d$ \cite{Redner82,Lubensky82}
\begin{eqnarray}
\label{extension:eq1}
F_N\left(R_{\perp},R_{\|}\right) &=&  F_0 +e_0 \left(\frac{R_{\perp}^2}{N^{2} b^2}+ \frac{R_{\|}^2}{N b^2}\right)+e_1 
v_{exc} \frac{N^2}{R_{\|} R_{\perp}^{d-1}}, 
\end{eqnarray}
Note that the different $N$-dependence on the longitudinal and transverse Gaussian case, stems from the from the fact that
the system is directed along the longitudinal direction ($\nu_{\|}^{0}=1$) and diffusive along the transversal one ($\nu_{\perp}^{0}=1/2$).
As in the isotropic case, the upper critical dimension $d_c$ is found by assuming the repulsive part to be of the order of unit, thus 
obtaining $d_c=3$. Upon minimizing with respect to $R_{\perp}$ and  $R_{\|}$, one obtains a system of two coupled equations involving
 $\nu_{\perp}$ and $\nu_{\|}$, whose result is
\begin{eqnarray}
\label{estension:eq2}
\nu_{\perp} = \frac{5}{2\left(d+2\right)} &\qquad& \nu_{\|}=\frac{d+7}{2\left(d+2\right)}.
\end{eqnarray}

A second case of great interest and actuality concerns the case of branched polymers formed by several reacting multifunctional monomer units,
that are often referred to as starbust dendrimers \cite{Boris96}. In this case, the $N$ monomers are distributed into $g$ generations
of successive growth, so that the size of a typical strand of $g$ monomers is $R_{0} \sim b g^{1/2}$.

In this case, specialized to the three-dimensional system to be specific, the repulsive term for a system of $g$ monomers embedded into
a system of density $N/R^{3}$, will be proportional to, $g N/R^{3}$, thus leading to a further generalization of  Eq. (\ref{elementary:eq11})
\begin{eqnarray}
\label{estension:eq3}
F_N\left(R\right) &=& F_0+e_0 \frac{R^2}{N}+e_1 v_{exc} \frac{gN}{R^{3}}.
\end{eqnarray}
A minimization with respect to $R$ then leads to an equation for the linear expansion factor $R/R_{0}$ akin to Eq.(\ref{elastic:eq8}) for this case.
A recent application of this methodology to several examples of branched polymers can be found in Ref.\cite{Kroger10}.

%%%%%%%%%%%%%%%%%%%%%%%%%%%%%%%%%%%%%%%%%%%%%%%%%%%%%%%%%%%%%%%%%%%%%%%%%%%%%%%%%%%%%%%%%%%%%%
\section{Temperature induced transition and external force}
\label{sec:crossover}
%%%%%%%%%%%%%%%%%%%%%%%%%%%%%%%%%%%%%%%%%%%%%%%%%%%%%%%%%%%%%%%%%%%%%%%%%%%%%%%%%%%%%%%%%%%%%%
The scaling theory discussed in the previous section is an attempt to
go beyond the Gaussian limit in each phase of a polymer.  However, for
arbitrary $d$, the interacting polymer may never be in the Gaussian
limit.  We have seen that as the quality of the solvent is changed,
the size exponent takes only three possible values, one for the
repulsive, one for the attractive and one for the transition point.
The universality of the exponents on the repulsive and the attractive
sides (i.e. independent of the strength) suggests that the special
situation is the transition point, not the Gaussian noninteracting
one.  The scaling behaviour should then be in terms of the deviations
from the transition point.  This we do now in this section.  The law
we obtain are analogous to the scaling observed in magnetic and fluid
phase transitions near tricritical points.  This is not surprising.
To get the transition point we need to tune the two-body interaction
and $N\to\infty$.  In fact one more parameter is needed, the
concentration of the polymer in solution, which also needs to be zero
(dilute limit).  A transition point with three relevant parameters
(parameters that can destroy or change the nature of the transition),
is called a tricritical point.

As in any phase transitions, in polymer theory too, the criticality is
obtained only in the $N \to \infty$ limit.  Near a phase transition
point, a finite system then shows typical, often universal, size
dependence which are characteristic of the infinite system.  This is
called finite size scaling.  The phase transitions that occurs at
$T=T_{\theta}$, requires $N\to\infty$, but its character can be seen
in finite $N$ behaviour.  One interesting, and largely overlooked,
consequence of the polymer theory developed so far
lies in the possibility of getting this finite $N \gg 1$ behaviour as
one drives the transition upon changing the temperature.  This is
discussed in the present Section, where the Flory approach will be
cast into a more general framework of a crossover among the three
different regimes as driven by the temperature at large but fixed $N$.
As before, we will study the dimensionality dependence of the system
separately.

The polymer at the theta point is in a very special state because
finiteness of the length or any change in temperature would take it
away from the theta point.  In such situations, the transition
behaviour is expressed in terms of the theta point behaviour as
\begin{equation}
  \label{case0:eq:1}
   R= b N^{\nu_{\theta}} \Psi(\alpha N^{\phi}),
\end{equation}
where $\Psi(0)=1$ is the theta temperature behaviour $(\alpha=0)$.
For higher temperatures, $\alpha>0$, as $N\to\infty$, $\Psi(x)\sim
x^q$ in such a way that the $N$ dependence becomes the Flory value, the
characteristic of the swollen phase.  This requires
$\nu_{\theta}+q\phi=\nu_F$. This is nicely corroborated by the Flory
theory, as shown below.

%%%%%%%%%%%%%%%%%%%%%%%%%%%%%%%%%%%%%%%%%%%%%%%%%%%%%%%%%%%%%%%%%%%%%%%%%%%%%%%%%%%%%%%%%%%%%%
\subsection{Case $2\le d \le 3$}
\label{subsec:case1}
%%%%%%%%%%%%%%%%%%%%%%%%%%%%%%%%%%%%%%%%%%%%%%%%%%%%%%%%%%%%%%%%%%%%%%%%%%%%%%%%%%%%%%%%%%%%%%
We now go back to the free energy as given by Eq.(\ref{mfa:eq8}), and
consider the saddle point equation, Eq. \eqref{steepest:eq4}, rearranged as
\begin{eqnarray}
\left(\frac{R}{L^{\nu_{\theta}}}\right)^{2d+2}
  - \widetilde\alpha  L^{\frac{d-1}{d+1}} \left(\frac{R}{L^{\nu_{\theta}}}\right)^d
    -2 \widetilde\gamma
       &=&0, \qquad (\nu_{\theta}=2/(d+1)).
\label{case1:eq2}
\end{eqnarray}
We consider the $\theta$-regime $T \to T_{\theta}$ and restrict to the
case $2\le d \le 3$. To allow for non-Gaussian behaviour, let us
introduce an arbitrary scale $b$ with $N=L/b^2$ dimensionless as
before.  For the solution $R$ of the steepest descent equation
(\ref{case1:eq2}) we assume a scaling form of the type
\begin{eqnarray}
\label{case1:eq3}
R&=&b N^{\nu_{\theta}}\Psi\left(z\right),
\end{eqnarray}
where we have generalized the scaling variable 
$$z=|\alpha| N^{\phi},$$ 
involving the cross-over exponent  
$$\phi=\frac{d-1}{d+1}.$$
The scaling function $\Psi(z)$ ($>0$) is  a clear generalization of
Eq. (\ref{eq:21}). 
Eq.(\ref{case1:eq2}) can then be cast into the following form
\begin{eqnarray}
\Psi^{2d+2}\left(z\right) - \chi\left(T\right) z \Psi^{d} \left(z\right) - 2 \gamma&=&0,
\label{case1:eq4}
\end{eqnarray}
the function $\chi(T)$ which is equal to $1$ for $T>T_{\theta}$ and
$-1$ for $T<T_{\theta}$.  Because of this, we must distinguish two
cases depending on $T$, and we will denote as $\Psi_{+}$ ($\Psi_{-}$)
the solution of (\ref{case1:eq4}) when $T>T_{\theta}$
($T<T_{\theta}$).
%%%%%%%%%%%%%%%%%%%%%%%%%%%%%%%%%%%%%%%%%%%%%%%%%%%%%%
\subsubsection{Case $T> T_{\theta}$.} 
%%%%%%%%%%%%%%%%%%%%%%%%%%%%%%%%%%%%%%%%%%%%%%%%%%%%%%
Eq.(\ref{case1:eq4}) has only one solution for $\Psi_{+}>0$ which, at large $z$, behaves as
\begin{eqnarray}
\label{case1:eq4b}
\Psi_{+}\left(z\right)&=& z^{\frac{1}{d+2}} \left(1+\frac{2 \gamma}{d+2}\frac{1}{z^\frac{2d+2}{d+2}} + \ldots \right).
\end{eqnarray}
On the other hand, for $z \to 0$ ($T \to T_{\theta}^{+}$) we find $\Psi_{+}(z) \sim
(2 \gamma)^{1/(2d+2)}$. 
The scaling function has then the following behavior 
\begin{equation}
\Psi_{+}\left(z\right) =
\left\{
\begin{array}{lccc}
z^{\frac{1}{d+2}} &\qquad \qquad & z \gg \left(2 \gamma\right)^{\frac{d+2}{2d+2}}   \\
\left(2 \gamma\right)^{\frac{1}{2d+2}} &\qquad \qquad & z \ll \left(2 \gamma\right)^{\frac{d+2}{2d+2}}
\end{array}
\right. \label{case1:eq5}
\end{equation}
%%%%%%%%%%%%%%%%%%%%%%%%%%%%%%%%%%%%%%%%%%%%%%%%%%%%%%
\subsubsection{Case $T<T_{\theta}$.}
%%%%%%%%%%%%%%%%%%%%%%%%%%%%%%%%%%%%%%%%%%%%%%%%%%%%%
Now $\chi(T)<1$ and $\alpha=-\vert\alpha\vert$. Again Eq.(\ref{case1:eq4}) has only one solution
\begin{eqnarray}
\label{case1:eq5b}
\Psi_{-}\left(z\right)&=& \frac{2 \gamma}{z}^{\frac{1}{d}} \left(1-2 \gamma^{d+2}+ d z^\frac{2d+2}{d+2} + \ldots \right)
\end{eqnarray}
For $z \to 0$ we have the same behavior as before, so that
\begin{equation}
\Psi_{-}\left(z\right) =
\left\{
\begin{array}{lccc}
\frac{2 \gamma}{z}^{\frac{1}{d}} &\qquad \qquad & z \gg \left(2 \gamma\right)^{\frac{d/2}{2d+2}}   \\
\left(2 \gamma \right)^{\frac{1}{2d+2}} &\qquad \qquad & z \ll \left(2 \gamma\right)^{\frac{d/2}{2d+2}}
\end{array}
\right. \label{case1:eq6}
\end{equation}
%%%%%%%%%%%%%%%%%%%%%%%%%%%%%%%%%%%%%%%%%%%%%%%%%%%%%%%
\subsubsection{Phase diagram}
%%%%%%%%%%%%%%%%%%%%%%%%%%%%%%%%%%%%%%%%%%%%%%%%%%%%%%%
Inserting these finding for $\Psi(z)$ into the scaling \textit{Ansatz}
(\ref{case1:eq3}) we obtain 
\begin{equation}
\frac{R}{b}  \approx
\left\{
\begin{array}{lccc}
\left(T-T_{\theta}\right)^{\frac{1}{d+2}} N^{\nu_F}, 
   &\text{when}&   
       N \gg \left \vert T - T_{\theta} \right \vert^{-1/\phi},~~ T>T_{\theta} \\
\left(T_{\theta}-T\right)^{-\frac{1}{d}} N^{\nu_c}, 
   &\text{when}& 
        N \gg \left \vert T - T_{\theta} \right \vert^{-1/\phi},~~  T<T_{\theta} \\
N^{\nu_{\theta}}, 
   &\text{when}& 
        N \ll \left \vert T - T_{\theta} \right \vert^{-1/\phi}, 
                 ~{\rm  or}~ T=T_{\theta}.
\end{array}
\right. \label{case1:eq7}
\end{equation}
A schematic diagram of the three regions is shown in
Fig. \ref{fig:fig5}.  The crossover expressed by the exponent $\phi$
above is represented by the dashed (blue) lines.   The region
enclosing the x-axis is the theta region.

\figfive

This behavior is depicted in Fig.\ref{fig:fig6} in the particular case of $d=3$.

\figsix

%%%%%%%%%%%%%%%%%%%%%%%%%%%%%%%%%%%%%%%%%%%%%%%%%%%%%%%%%%%%%%%%%%%%%%%%%%%%%%%
\subsection{Case $d>3$}
\label{subsec:case2}
%%%%%%%%%%%%%%%%%%%%%%%%%%%%%%%%%%%%%%%%%%%%%%%%%%%%%%%%%%%%%%%%%%%%%%%%%%%%%%%
In this case the steepest descent solution of the modified free energy Eq.(\ref{elastic:eq6}), accounting for Flory's correction, reads
\begin{eqnarray}
\frac{R^2}{N b^2} -1 
-\left[2 \gamma b^{2d}\frac{N^3}{R^{2d}}+ \ldots \right]&=& 
\alpha N^2 \frac{b^d}{R^{d}}
\label{case2:eq1}
\end{eqnarray}
where we have not included higher correction terms $\gamma_{ij}$ that are necessary to ensure the convergence of the saddle point
in Eq.(\ref{flory:eq3}) in the case $\alpha < 0$.

In analogy with what we have attempted in the previous case, we assume a scaling of the form
\begin{eqnarray}
\label{case2:eq2}
 R&=&b N^{1/2}\Psi\left(z\right)
\end{eqnarray}
in view of the fact that $\nu_{\theta}=1/2$ when $d>3$. This yields
\begin{eqnarray}
\Psi^{2d+2}\left(z\right)- \Psi^{2d}\left(z\right)
 - \left[\frac{2 \gamma}{N^{d-3}}+ 
\ldots \right] &=&\frac{\alpha}{N^{d/2-2}}  \Psi^{d}\left(z\right)
\label{case2:eq3}
\end{eqnarray}
Note that, unlike the case $2\le d\le 3$, it is \textit{not} possible to
cast Eq.(\ref{case2:eq3}) in terms of an equation for a single scaling variable $z$
as both terms multiplying $\alpha$, $\gamma$ and higher terms depend upon $N$.
We then assume $z=\vert \alpha \vert/N^{(d-4)/2}$ and $\chi(T)=1$ when
$T>T_{\theta}$ and $-1$ for $T< T_{\theta}$, so that the left hand side of
Eq.(\ref{case2:eq3}) reads $\chi(T) z \Psi^{d}(z)$.

Again we consider two cases
%%%%%%%%%%%%%%%%%%%%%%%%%%%%%%%%%%%%%%%%%%%%%%%%%%%%%%%%%%%%%%%%%%%%%
\subsubsection{Case $T>T_{\theta}$  ($\alpha >0$)}
%%%%%%%%%%%%%%%%%%%%%%%%%%%%%%%%%%%%%%%%%%%%%%%%%%%%%%%%%%%%%%%%%%%%%
In this case, Eq.(\ref{case2:eq3}) becomes
\begin{eqnarray}
\label{case2:eq4a}
\Psi_{+}^{2d+2} \left(z\right)-\Psi_{+}^{2d} \left(z\right)-2 \gamma N^{3-d}-z \Psi_{+}^{d} \left(z\right)&=&0
\end{eqnarray}
where $z>0$ and $\Psi_{+}>0$. For small $z$ and being $d>3$, the last two terms in the above Eq.(\ref{case2:eq4a}) are negligible
and $\Psi_{+}(z) \approx 1$. On the other hand, if $z \gg 1$ the first and last terms of  Eq.(\ref{case2:eq4a}) dominate leading to
\begin{eqnarray}
\label{case2:eq4b}
\Psi_{+} \left(z\right) &\approx& \left(\alpha z \right)^{1/(d+2)}
\end{eqnarray}
Depending on dimensionality $d$, we the get, using Eq.(\ref{case2:eq2}), 

\begin{equation}
\frac{R}{b}  \approx
\left\{
\begin{array}{lccc}
N^{1/2} &1 \ll N \ll \left(T-T_{\theta}\right)^{-2/(4-d)}& 3 < d < 4 \\
\left(T-T_{\theta}\right)^{1/(d+2)} N^{\nu_{F}} & N \gg \left(T-T_{\theta}\right)^{-2/(4-d)} & 3< d < 4 \\
N^{1/2} & N \gg 1 & d>4 
\end{array}
\right. \label{case2:eq4c}
\end{equation}
Notice that for this case, the Flory's correction term (the second one in Eq.(\ref{case2:eq4a})) is important for the relatively small $N$ regime
but not for the large $N$ regime where self-avoidance dominate. The $\gamma$ term is, on the other hand, always irrelevant.

%%%%%%%%%%%%%%%%%%%%%%%%%%%%%%%%%%%%%%%%%%%%%%%%%%%%%%%%%%%%%%%%%%%%%
\subsubsection{Case $T<T_{\theta}$  ($\alpha <0$)} 
%%%%%%%%%%%%%%%%%%%%%%%%%%%%%%%%%%%%%%%%%%%%%%%%%%%%%%%%%%%%%%%%%%%%
Eq.(\ref{case2:eq3}) becomes in this case
\begin{eqnarray}
\label{case2:eq5a}
\Psi_{-}^{2d+2}\left(z\right)- \Psi_{-}^{2d}\left(z\right) -2 \gamma N^{3-d} + z  \Psi_{-}^{d}\left(z\right)&=& 0
\end{eqnarray}
Again, when $z \ll 1$ and $N \gg 1$ we have $\Psi_{-}(z) \approx 1$.

When $z \gg 1$ since $\Psi_{-}(z)>0$ we have
\begin{eqnarray}
\label{case2:eq5b}
\Psi_{-}\left(z\right) \approx \left(\frac{2 \gamma}{z N^{d-3}}\right)^{1/d}= \left(2 \gamma \right)^{1/d}\left(T_{\theta}-T\right)^{1/d} 
N^{-\frac{d-2}{2d}}
\end{eqnarray}
leading to 

\begin{equation}
\frac{R}{b}  \approx
\left\{
\begin{array}{lccc}
N^{1/2} &1 \ll N \ll \left(T_{\theta}-T\right)^{-2/(4-d)}& 3 < d < 4 \\
\left(T_{\theta}-T\right)^{-1/d} N^{\nu_{c}} & N \gg \left(T_{\theta}-T\right)^{-2/(4-d)} & 3< d < 4 \\
N^{1/2} & N \gg 1 & d>4 
\end{array}
\right. \label{case2:eq6}
\end{equation}
In this phase both terms, the Flory's correction and the $\gamma-$ term are relevant depending on the $N$ regime. Thus, the annoying
$N-$ dependence in Eq.(\ref{case2:eq4a}) for $T > T_{\theta}$ can be neglected and $\Psi_{+}$ depends on $z$ only in the large $N$ regime,
whereas when $T < T_{\theta}$, the $N-$ dependence enters through the dangerous irrelevant $\gamma-$ term that is necessary
to ensure the convergence of the integral in Eq.(\ref{flory:eq3}). 
%%%%%%%%%%%%%%%%%%%%%%%%%%%%%%%%%%%%%%%%%%%%%%%%%%%%%%%%%%%%%%%%%%%%%%%%%%%%%%%%

%%%%%%%%%%%%%%%%%%%%%%%%%%%%%%%%%%%%%%%%%%%%%%%%%%%%%%%%%%%%%%%%%%%%%%%%%%%%%%%%%%%%%%%%%%%
\subsection{Inclusion of an external force}
\label{subsec:inclusion}
%%%%%%%%%%%%%%%%%%%%%%%%%%%%%%%%%%%%%%%%%%%%%%%%%%%%%%%%%%%%%%%%%%%%%%%%%%%%%%%%%%%%%%%%%%
The standard approach to probe any system is to perturb it by a small
amount and look for the response.  One would therefore like to obtain
the response of a polymer in different phases to perturbations that
try to change its size or shape.  This would tell us about the
stability of the size and also would give us information about the
distribution function.  One such perturbation would be an external
force pulling at one end keeping the other fixed.  This is equivalent
to pulling the two ends with equal force in opposite directions.  This
is the fixed-force ensemble.

The Flory free energy given in Eq.(\ref{mfa:eq8}) can be extended to
include the effect of an external force. A detailed analysis of this
situation in the case of semiflexible polymer, will be given in
Section \ref{sec:semiflexible}, but we here discuss the Flory
result for the flexible case. Eq.(\ref{mfa:eq8}) modifies as
\begin{eqnarray}
\label{inclusion:eq1}
\beta F_{L}\left(R\right)=\frac{d}{2} \frac{R^2}{N b^2} +  \alpha
 \frac{N^2}{R^d}+ \gamma  \frac{N^3}{R^{2d}}+ \ldots   
-\mathbf{f} \cdot \mathbf{R}
\end{eqnarray}
where the last term accounts for the reduction in free energy for
chain alignment along the (reduced) external force per unit of length
$\mathbf{f}$.  From a thermodynamic point of view, we are going from a
fixed-$R$ ensemble to a fixed force ensemble.

The force introduces a cylindrical anisotropy so that
\begin{eqnarray}
\label{inclusion:eq2}
\mathbf{R} &=& \mathbf{R}_{\parallel}+\mathbf{R}_{\perp}
\end{eqnarray}
where $\mathbf{R}_{\parallel}=\left(\mathbf{R} \cdot \hat{\mathbf{f}}
\right) \hat{\mathbf{f}}$ and $\hat{\mathbf{f}}=\mathbf{f}/f$  
is the unit vector of $\mathbf{f}$.

In analogy with Eq.(\ref{flory:eq1}) we assume here a different
scaling in the directions parallel and perpendicular to the applied
force
\begin{eqnarray}
\label{inclusion:eq3}
R_{\parallel} = b N^{\nu_{\parallel}} x_{\parallel} &\qquad \qquad &
R_{\perp} = b N^{\nu_{\perp}} x_{\perp} 
\end{eqnarray}
where one expects $\nu_{\parallel} \approx 1> \nu_{\perp}$ so that $
x_{\parallel}> x_{\perp}/N^{\nu_{\parallel}-\nu_{\perp}}$.

In the $\alpha>0$ case, i.e. for $T> T_{\theta}$, both the interaction
terms in Eq.(\ref{inclusion:eq1}) (those proportional to $\alpha$ and
$\gamma$) are subdominant with respect to the Gaussian and the force
terms, so that one finds in the $N \gg 1$ limit
\begin{eqnarray}
\label{inclusion:eq4a}
\beta F_L &\approx& N \left[\frac{d}{2} x_{\parallel}^2 -f b
  x_{\parallel} \right]+\frac{d}{2}
\frac{R_{\perp}^2}{Nb^2}-\frac{\alpha d}{2 x_{\parallel}^{d+2}} 
\frac{R_{\perp}^2}{N^d b^2} + \text{less dominant terms} ,
\end{eqnarray}
whose minimization with respect to $R_{\perp}$ leads to $R_{\perp}
\sim b N^{1/2}$ implying $\nu_{\perp}=1/2$. Then
\begin{eqnarray}
\label{inclusion:eq4}
\beta F_L &\approx& N \left[\frac{d}{2} x_{\parallel}^2 -f b
  x_{\parallel} \right]  
\end{eqnarray}
This yields the saddle point
\begin{eqnarray}
x_{\parallel}^{*} &=& \frac{f b}{d}
\label{inclusion:eq5}
\end{eqnarray}
that can be inserted back into Eq.(\ref{inclusion:eq4}) to give the
minimum of the swollen phase free energy
\begin{eqnarray}
\label{inclusion:eq6}
\left(\beta F_L\right)_{S} &=& -N \frac{\left(b f\right)^2}{2d}
\end{eqnarray}

In the opposite case $\alpha=-\vert\alpha\vert$ (i.e. $T< T_{\theta}$)
the phase is compact and hence both terms in Eq.(\ref{inclusion:eq3})
coincide with
\begin{eqnarray}
\label{inclusion:eq7}
R &=& b N^{1/d} x 
\end{eqnarray}
The external force term is then subdominant and one has to match the
two interaction terms as in the absence of external force. Thus,
equation (\ref{inclusion:eq1}) yields
\begin{eqnarray}
\label{inclusion:eq8}
\beta F_L &=& N\left[-\frac{\vert \alpha \vert}{x^d} + \frac{\gamma}{x^{2d}} \right]
\end{eqnarray}
This yields the saddle point equation
\begin{eqnarray}
\label{inclusion:eq9}
x^{*}&=& \left(\frac{2\gamma}{\left \vert \alpha \right \vert}\right)^{1/d}
\end{eqnarray}
and a minimum of the compact phase free energy
\begin{eqnarray}
\label{inclusion:eq10}
\left(\beta F_L\right)_{C} &=& - N \frac{\alpha^2}{4 \gamma} 
\end{eqnarray}
A first-order transition between the swollen and compact phase occurs
when the two free energies (\ref{inclusion:eq6}) and
(\ref{inclusion:eq10}) are equal, that is at the critical force $f_c$
given by
\begin{eqnarray}
b f_c &=& \left \vert \alpha \right \vert \sqrt{\frac{d}{2\gamma}}
\label{inclusion:eq11}
\end{eqnarray}
As $\alpha$ is proportional to $(T-T_{\theta})/T_{\theta}$, Flory
theory then predicts a linear dependence of the critical force $f$ on
the reduced temperature, as schematically illustrated in
Fig.\ref{fig:fig7}.  This approach has been exploited to infer the
unzipping transition in DNA \cite{Marenduzzo02,Maren09,unzip}.

\figsept

%%%%%%%%%%%%%%%%%%%%%%%%%%%%%%%%%%%%%%%%%%%%%%%%%%%%%%%%%%%%%%%%%%%%%%%%%%%%%%%%%%%%%%%%%%%%%%

%%%%%%%%%%%%%%%%%%%%%%%%%%%%%%%%%%%%%%%%%%%%%%%%%%%%%%%%%%%%%%%%%%%%%%%%%%%%%%%%%%%%%%%%%%%%%%%%%%%%%%%%%%%%%%%%%%%%%%%%%%%%%%%%%%%
\subsection{Polymer Solution}
\label{sec:formulation}
%%%%%%%%%%%%%%%%%%%%%%%%%%%%%%%%%%%%%%%%%%%%%%%%%%%%%%%%%%%%%%%%%%%%%%%%%%%%%%%%%%%%%%%%%%%%%%%%%%%%%%%%%%%%%%%%%%%%%%%%%%%%%%%%%%%
The single chain behaviour discussed so far is for  a very dilute solution.
The monomers on different chains also interact like  monomers on the same chain.
We  discuss qualitatively the combined effect of additional chains and 
temperature.  See Fig \ref{fig:fig8}.

For polymers in good solvent, one may start from a very dilute regime
where each chain has its own size and are too far apart to have any
mutual interaction.  Taking each chain to be like a sphere of radius $R\sim
N^{\nu}$, the dilute limit corresponds to the regime where the
separation of the center of the spheres $\Lambda$ is much greater than
$R$, $\Lambda\gg R$.  Like any dilute solution, the polymers then
exert an osmotic pressure well-described by the perfect gas law,
\begin{equation}
  \label{eq:34}
  \Pi=k_BT\ c_p,\qquad (\Lambda\gg R), 
\end{equation} 
where $c_p$ is the
polymer number concentration (number of polymers per unit volume).  If
we have $n_p$ polymers in the solution of volume $V$, $c_p=n_p/V$.

\figeight
 
Polymers are not hard spheres and so they would start interacting when
$\Lambda \sim R$.  They start to interpenetrate.  Under such a
condition, monomer concentration
\begin{equation}
\label{eq:35} 
c=\frac{N n_p}{V}=N c_{p}, 
\end{equation} 
is a more appropriate variable than $c_p$ because the end points do
not matter.  In the dilute limit, the polymers are identifiable, the
end points acting as labels for them.  For the interpenetrating case,
there is no noticeable distinction between the interior of the
solution of $n_p$ polymers each of $N$ monomers and the interior of a
single chain of length $n_p N$.  The chain length ceases to be a
suitable measure to characterize the solution.  This regime is called
the {\it semi-dilute} regime or a semi-dilute solution of polymers.
The change from the dilute to the semi-dilute case is not a phase
transition but a smooth crossover involving a concentration dependent
length scale.  From the transient network created by the
interpenetration, one may identify a spatial length $\xi$ within which
a polymer segment is free and assumes the behaviour of a swollen chain
($\xi\sim n^{\nu}$).  It looks like a solution of blobs of size $\xi$.
Thanks to the interaction with other monomers, the long range
correlation of a single chain is lost.  As a result, a long polymer,
$N\gg n$, will be in a Gaussian state.  This is a screening effect ---
the repulsive interactions with other monomers screening out the long
range effect of self-repulsion.  In a $T-c$ plane for a finite $N$,
there will be a crossover line separating the dilute and the semi
dilute case. See Fig \ref{fig:fig8}a.

The scale, $c^*$ for the crossover from dilute to semi-dilute case can
be obtained from a physical picture.  This is the concentration at
which the individual spheres of size $R$ just start to touch each
other, $\Lambda\sim R$.  In a sense, the overall monomer concentration
matches the concentration inside a single polymer sphere, viz.,
\begin{equation}
  \label{eq:37}
  c_p\sim \frac{1}{\Lambda^d}\sim \frac{1}{R^d}, \quad{\mathrm{ so \
    that}}\quad c^*\sim \frac{L}{R^d}\sim N^{1-d\nu}.
\end{equation}
With this scale, the osmotic pressure would take a form
$$\Pi=k_BT\ c_p\; f(c/c^*)=k_BT\; (c/N)\; f(c/c^*)$$
where the function $f(x)$ is such that for $c\gg c^*$, $\Pi$ is
independent of $N$.  It then follows that $\Pi\sim c^{1/(d\nu-1)}$.
The nonlinear dependence, $\Pi\sim c^{5/4}$ in three dimensions (using
the Flory value) has been observed experimentally in many polymer
solutions \cite{Doi86}.

The dilute-semidilute crossover is indicated by a dash-dot line in Fig
\ref{fig:fig8}a. As $N$ is decreased the crossover line shifts to higher
values as indicated by a hashed blue line.  For infinitely long
chains any solution is in the semidilute regime ($c^*\to 0$) as in Fig
\ref{fig:fig8}b.

For attractive interaction, the theta temperature is strictly for an
infinitely long chain.  A solution of polymers of finite chains with
attractive interaction would show a phase separation between a very
dilute and a semi-dilute solutions similar to the phase separation of
any binary mixture or alloy.  Such a phase separation, in addition to
a region of coexistence, would also have a critical point in the
temperature concentration plane.  The critical point is expected at a
temperature $T_c<T_{\theta}$ with $T_c\to T_{\theta}$ at $c\to 0$ as
$N\to\infty$.  The behaviour close to the critical point (``critical
phenomenon'') is identical to other binary mixtures, controlled by
concentration fluctuations.  In a three dimensional $T$--$c$--$1/N$
phase diagram a line of these critical points ends at the theta point
at $1/N=0,c=0$. See Fig \ref{fig:fig8}b.  We see the special status of
the theta point: it is the confluence of two independent phenomena,
the criticality of phase separation in solution and the collapse of a
single long chain.  Such a point is defined as a tricritical point.  A
tricritical point requires three critical lines meeting at a point.
For $N\to\infty, c=0$, the $T>T_{\theta}$ line is a critical line
showing power law behaviour at every $T$.  We therefore see two
critical lines meeting at the theta point.  Unfortunately, $N,c$ are
strictly positive and so the complete picture of the tricritical
behaviour is not possible.  For $N\to\infty$, there is a phase
separation between the collapsed phase and a semi-dilute solution.
The phase separation line has zero osmotic pressure.

Fig. \ref{fig:fig8}a shows the various crossovers in the $T$--$c$ plane
for a fixed $N$, a slice of the three dimensional phase diagram.
There is a region close to the theta point with small $c$, marked by
the horizontal lines $T_{\theta}\pm N^{-\phi}$ in the dilute regime
where the signature of the theta point is visible.  

There are experimental attempts \cite{anisimov05} to generate such a 
phase diagram for polymers in terms of theta-point scaling but a 
Flory-like theory for this rich phenomena remains  elusive.
%%%%%%%%%%%%%%%%%%%%%%%%%%%%%%%%%%%%%%%%%%%%%%%%%%
%%%%%%%%%%%%%%%%%%%%%%%%%%%%%%%%%%%%%
\section{Semiflexible chain under tension}
\label{sec:semiflexible}
%%%%%%%%%%%%%%%%%%%%%%%%%%%%%%%%%%%%%%%%%%%%%%%%%%%%%%%%%%%%%%%%%%%%%%%%%%%%%%%%%%%%%%%%%%%%%%%%%%%%%%%%%%%%%%%%%%%%%
So far we have discussed the properties of flexible chains, where the
chain is Gaussian in the absence of any external interactions. The
discussed Edwards model is then the continuum counterpart of a
freely-jointed chain.  The two matched nicely because all the
properties were controlled by the configurations near the peak of the
distribution.  We now discuss a case where one needs the
extreme states for which the Gaussian approximation is not sufficient.

In addition to the three phase we have seen, there is the possibility
of a stretched state or a rod-like state with $\nu=1$ as, e.g. one
expects for a repulsive polymer in $d=1$.  This state can be produced
by stretching by a force or by bending rigidity.  In both cases there
is a competition with entropy.  Since an extended state would
correspond to configurations in the tail of a Gaussian distribution,
the continuum model we used would not be of much use.

In this section, for completeness and for practical usefulness, we
consider the situation of a polymer with bending rigidity, called a
semi-flexible polymer, in the presence of an external pulling force
acting on one end of the chain \cite{Ha95,Ha97}.  The small force and
the large force regions are to be determined from which an approximate
interpolation formula is derived.   

%%%%%%%%%%%%%%%%%%%%%%%%%%%%%%%%%%%%%%%%%%%%%%%%%%%%%%%%%%%%%%%%%%%%%%%%%%%%%%%%%%%%%%%%%%%%%%%%%%%%%%%%%%%%%%%%%%%%%
\subsection{Discrete approach}
\label{subsec:discrete}
%%%%%%%%%%%%%%%%%%%%%%%%%%%%%%%%%%%%%%%%%%%%%%%%%%%%%%%%%%%%%%%%%%%%%%%%%%%%%%%%%%%%%%%%%%%%%%%%%%%%%%%%%%%%%%%%%%%%%
Let us start with the FJC of Sec. II, with normalized bond
vectors by $\hat{\mathbf{T}}_j={{ \bm \tau}}_j/b$.
Instead of free joints, we admit a bending energy at every joint that
there is an energy  penalty if the two bonds are not parallel.   This
energy cost is taken as $\propto -\hat{\mathbf{T}}_{j-1}\cdot
\hat{\mathbf{T}}_j$.  If one end is fixed at origin, then the force on the
other end is equivalent to a orientational force on every bond because
of the relation ${\mathbf{R}}=\sum_j {{\bm \tau}}_j$.
The Hamiltonian can then be written as
\begin{eqnarray}
\label{discrete:eq2}
\beta H &=& -K \sum_{j=1}^N \hat{\mathbf{T}}_{j+1} \cdot
\hat{\mathbf{T}}_{j} - \sum_{j=1}^N b \mathbf{f} \cdot
\hat{\mathbf{T}}_{j}, 
\end{eqnarray}
with the partition function 
\footnote{The factor $2 \pi$ appearing in Eq.(\ref{discrete:eq1}) in place of the usual normalization 
$4 \pi$, is due to the
presence of the $\hat{\mathbf{T}}^2 -1$ in the argument of the $\delta$-function,
and to the fact that $\delta(x^2-1)=[\delta(x-1)+ \delta(x+1)]/2$
with the second term not contributing to the result of the radial integral.}
\begin{eqnarray}
\label{discrete:eq1}
e^{-\beta F} &=& Z= \int \left[\prod_{j=1}^N \frac{d^3
    \hat{\mathbf{T}}_j}{2 \pi} \delta\left( 
 \hat{\mathbf{T}}_j^2  -1 \right) \right] e^{-\beta H}.
\end{eqnarray}
The delta function in Eq.  \eqref{discrete:eq2} maintains the fixed
length constraint of each bond.  It is this constraint that prevents
the unwanted extensions (in the tail of the Gaussian distribution) of
a continuous chain.  We choose our axes such that the force is in the
$z$-direction $\mathbf{f}_{j} = f \hat{\mathbf{z}}$ and the quantity
of interest is the extension $\langle z\rangle= b \sum_i\langle
T_i^{(\parallel)}\rangle$, where $\parallel$ indicates the
z-direction.  By the way, the Hamiltonian Eq. \eqref{discrete:eq1} is
identical to a classical ferromagnetic one dimensional Heisenberg
model in a field, if ${\mathbf{T}}$ is treated as a fixed length spin
vector.

Let us first consider the small force regime, where a linear response
is expected, $\langle z\rangle=b^2 \chi_T f$, with the response function
\begin{equation}
  \label{discrete:eq:1}
\chi_T=\sum_{ij}\left\langle T_i^{(\parallel)}T_j^{(\parallel)}\right \rangle_0,
\end{equation}
where the correlations are evaluated in the zero-force condition
indicated by the subscript 0.  For the classical 1-dimensional model,
these correlations decay exponentially for all temperatures,
\begin{equation}
  \label{discrete:eq:5}
\left\langle  T_i^{(\parallel)}T_j^{(\parallel)}\right \rangle_0\sim \exp(|i-j|b/l_p).   
\end{equation}
Here $\mathbf{T}^{\parallel}=\hat{\mathbf{z}} (\hat{\mathbf{T}} \cdot \hat{\mathbf{z}})$ and
this also defines the perpendicular component $\mathbf{T}^{\perp}$ as given in Eq.(\ref{discrete:eq3}).

The decay length $l_p$ is the persistence length.  The correlation
here may be compared with the flexible case, Eq.
\eqref{elementary:eq2}.  Ignoring end-point effects (equivalent to
assuming a circular polymer), and converting the sum to an integral,
we get $\chi_T\sim N l_p$.  Therefore for small forces, we expect

\begin{eqnarray}
\label{discrete:eq2bis}
\left \langle z \right \rangle &=& N b l_p f + {\cal O}\left(f^2\right)
\end{eqnarray}

For large forces, the polymer is going to align with the force and be
completely stretched except for thermal fluctuations.  The fully
stretched condition means $z= bN$ and therefore the delta function
constraint in Eq. \eqref{discrete:eq1} is going to play an important
role.  The deviation from the fully stretched state comes because of
transverse fluctuations and it would go to zero as $f\to\infty$.  By
writing
\begin{equation}
  \label{discrete:eq3}
\hat{\mathbf{T}}_{j} = \mathbf{T}_j^{(\parallel)}+\mathbf{T}_j^{(\perp)}  
\end{equation}
with small transversal part, i.e. 
$\vert \mathbf{T}_j^{(\perp)} \vert \ll 1$ for $b\; f \gg 1$, 
we have
\begin{eqnarray}
\label{discrete:eq9}
\left \langle z \right \rangle=\sum_{l=1}^N b \left \langle
  \hat{\mathbf{T}}_{l} \cdot \hat{\mathbf{z}}  \right \rangle = 
b \sum_{l=1}^N  \left \langle  \sqrt{1-T_{l}^{{(\perp)}^{2}}}  \right \rangle
\approx N b-\frac{b}{2} \sum_{l=1}^N  \left \langle
  T_{l}^{{(\perp)}^{2}}  \right \rangle  
\end{eqnarray}
Under the same approximation for $f\gg 1$ as in Eq.
\eqref{discrete:eq9}, the Hamiltonian can be approximated, dropping
redundant terms, as
\begin{eqnarray}
  \beta H &=& - K \sum_{j=1}^N \mathbf{T}_{j+1}^{(\perp)} \cdot
  \mathbf{T}_{j}^{(\perp)} +\frac{1}{2} \sum_{j=1}^{N} \left( b f
    +K_{j}\right)   \mathbf{T}_{j}^{{(\perp)}^2},
\label{discrete:eq5}
\end{eqnarray}
where $K_j=2K$, for all $j$ except $K_1=K_N=K$.  In the following, we
neglect this boundary effect and set $K_j=2K$.  For a very large
force, the leading term of the Hamiltonian is $\beta H\approx
\frac{1}{2} \sum_{j=1}^{N} b\; f \; \mathbf{T}_{j}^{{(\perp)}^2}$.  By
the equipartition theorem, we then expect $b \;f\;\langle
\mathbf{T}_{j}^{{(\perp)}^2}\rangle=2$.  By using this result in Eq.
\eqref{discrete:eq9}, the behavior is
\begin{equation}
  \label{discrete:eq:3}
\frac{\left \langle z \right \rangle}{Nb} \approx 1 - \frac{1}{b f},
\quad f\to\infty,
\end{equation}
Both Eqs.(\ref{discrete:eq2bis}) and (\ref{discrete:eq:3}) agree with the
small and large $f$ limits obtained by the more elaborate calculation of Sec\ref{subsec:large} and Sec.\ref{subsec:marko}.
It is then possible to generate an interpolation formula that
satisfies the two asymptotes, namely $f\to 0$ and $f\to \infty$.  The
interpolation formula is derived below {\it after} taking the
continuum limit $b\to 0$ which requires a more detailed evaluation of
the large force limit.

%%%%%%%%%%%%%%%%%%%%%%%%%%%%%%%%%%%%%%%%%%%%%%%%%%%%%%%%%%%%%%%%%%%%%%%%%%%%%%%%%%%%%%%%%%%%%%%%%%%%%%%%%%%%%%%%%%%%%
\subsection{Continuum limit: A detour}
\label{subsec:semi}
%%%%%%%%%%%%%%%%%%%%%%%%%%%%%%%%%%%%%%%%%%%%%%%%%%%%%%%%%%%%%%%%%%%%%%%%%%%%%%%%%%%%%%%%%%%%%%%%%%%%%%%%%%%%%%%%%%%%%
The continuum limit of the discrete chain with bending energy does not
follow from the procedure adopted for the FJC.  The reason for this is
that in the Edwards model the length $L$ is like an area or the chain
is not a space curve.  A semiflexible polymer configuration involves
the tangent vectors ${\bf T}$ for which it has to be taken as a space
curve \cite{Coxter89,Kamien99}.  Therefore two points on the polymer
${\bf r}$ and ${\bf r}+ d{\bf r}$ separated by a contour length $ds$
has to satisfy $(\partial {\bf r}/\partial s )^2=1$.  This condition
at every point on the curve can be enforced by a $\delta$-function in
the partition function and the Gaussian term of the Edwards model does
not appear.  By writing $-2{\mathbf{T}}_i\cdot{\mathbf{T}}_j=
({\mathbf{T}}_i-{\mathbf{T}}_j)^2 -2$, a continuum limit for the
bending energy for $b\to 0$ would give a derivative of
\begin{eqnarray}
\label{semi:eq3}
\hat{\mathbf{T}}\left(s\right)&=&\frac{\partial  \mathbf{r}}{\partial s},
\end{eqnarray}
i.e., $\partial^2 {\mathbf{r}}/\partial s^2$.  

With the above introduction, let us introduce the partition function
and the free energy for a semiflexible chain under the action of an
external force\cite{Ha95,Ha97},
\begin{eqnarray}
\label{semi:eq1}
e^{-\beta F} &=& Z= \int {\cal D} {\mathbf{r}} \
\left[\prod_s  \;\delta\left(\hat{\mathbf{T}}^2\left(s\right)-1\right)\right] \ e^{-\beta H},
\end{eqnarray}
with a Hamiltonian 
\begin{eqnarray}
\label{semi:eq2}
\beta H &=& \frac{l_p}{2} \int_{0}^{L} ds \left(\frac{\partial  \hat{\mathbf{T}}}{\partial s} \right)^2 - \int_{0}^{L}
ds \; \mathbf{f}\left(s\right) \cdot \hat{\mathbf{T}}\left(s\right).
\end{eqnarray}
In Eq.(\ref{semi:eq2}) $l_p$ is the persistence length which is the
tangent tangent correlation length defined as
\begin{eqnarray}
\label{semi:eq3b}
\left \langle \hat{\mathbf{T}}\left(s\right) \cdot \hat{\mathbf{T}}\left(s^{\prime}\right) \right \rangle \sim \exp\left[\frac{\left\vert s-s^{\prime} \right \vert}{l_p} \right],
\end{eqnarray}
the continuum analog of Eq. \eqref{discrete:eq:5}.  The fact that the
$l_p$ introduced in Eq.(\ref{semi:eq2}) coincides with the actual
persistent length given in Eq.(\ref{semi:eq3b}) will be shown below.
If $\mathbf{f}$ is a constant, then the last term in
Eq.(\ref{semi:eq2}) becomes $\mathbf{f} \cdot
[\mathbf{r}(L)-\mathbf{r}(0)]$ which is the standard force term.

If one softens the rigid constraint by a Gaussian weight factor, i.e.,
the $\delta\left(\hat{\mathbf{T}}^2\left(s\right)-1\right)$ by
$\exp(-\hat{\mathbf{T}}^2/2\sigma^2)$, and absorb this extra term in
the Hamiltonian, we get
\begin{eqnarray}
\label{semi:eq2n}
\beta H &=& \frac{l_p}{2} \int_{0}^{L} ds \left(\frac{\partial^2  {\mathbf{r}}}{\partial s^2} \right)^2
  + \frac{1}{2\sigma^2} \int_{0}^{L} ds \left(\frac{\partial {\mathbf{r}}}{\partial s} \right)^2
 - \int_{0}^{L}
ds \; \mathbf{f}\left(s\right) \cdot \hat{\mathbf{T}}\left(s\right),
\end{eqnarray}
which allows discussions of a crossover from the Gaussian to the
semiflexible case\cite{Bhattacharjee87}.

%Rather than following Refs. \cite{Ha95,Ha97}, in the following, we
We follow a discrete approach of Ref.\cite{Rosa03,Marko95}, that is
simpler than a continuum formulation and yields the same results.

%%%%%%%%%%%%%%%%%%%%%%%%%%%%%%%%%%%%%%%%%%%%%%%%%%%%%%%%%%%%%%%%%%%%%%%%%%%%%%%%%%%%%%%%%%%%%%%%%%%%%%%%%%%%%%%%%%%%%
\subsection{Large $f$ limit: detailed calculations}
\label{subsec:large}
%%%%%%%%%%%%%%%%%%%%%%%%%%%%%%%%%%%%%%%%%%%%%%%%%%%%%%%%%%%%%%%%%%%%%%%%%%%%%%%%%%%%%%%%%%%%%%%%%%%%%%%%%%%%%%%%%%%%%
Let us start with Eq. (\ref{discrete:eq1}).  We further assume the
transversal part to be small, i.e. $\vert \mathbf{T}_j^{(\perp)} \vert
\ll 1$, an assumption that holds in the large $f$ limit. Then, to
leading order
\begin{eqnarray}
\label{discrete:eq4}
\delta\left(\hat{\mathbf{T}}_j^2  -1
\right)&=& \frac{1}{2 \sqrt{1-T_{j}^{{(\perp)}^{2}}}}
\delta\left(T_{j}^{(\parallel)}- 
\sqrt{1-T_{j}^{{(\perp)}^{2}}} \right),
\end{eqnarray}
where the additional term containing
$\delta\left(T_{j}^{(\parallel)}+\sqrt{1-T_{j}^{{(\perp)}^{2}}}
\right)$ has been neglected since it leads to subdominant
contributions.  To leading order, the square root term appearing in
Eq.(\ref{discrete:eq4}) can be exponentiated as
\begin{eqnarray}
\label{discrete:eq6}
( 1-T_{j}^{{(\perp)}^{2}})^{-1/2} &\approx&  e^{T_{j}^{{(\perp)}^{2}}/2},
\end{eqnarray}
and in Eq.(\ref{discrete:eq1}) we can further split $ d^3
\hat{\mathbf{T}}_j = d T_{j}^{(\parallel)} d^2
\mathbf{T}_{j}^{(\perp)}$. The integral over the longitudinal part can
be carried out immediately so that Eq.(\ref{discrete:eq1}) can be
written as
\begin{eqnarray}
\label{discrete:eq7}
e^{-\beta F} &=& (\text{const}) \left[ 
  \int \left(\prod_{j=1}^N d T_j \right) \exp\left( -\frac{1}{2} \sum_{ij=1}^N T_{i} M_{ij} T_{j} \right)
\right]^2,
\end{eqnarray}
where $M_{ij}$ is a  tri-diagonal  matrix
\begin{eqnarray}
\label{discrete:eq8}
M_{ij} &=& \left(b f +2 K+1\right) \delta_{ij} - K \left(\delta_{i j+1}+ \delta_{i j-1}\right),
\end{eqnarray}
and $T_j$ is any of the two transversal components of
$\hat{\mathbf{T}}_j$ whose range can be extended to the whole real
line, $-\infty<T_j<\infty$. The multiplicative constant appearing in
front of Eq.(\ref{discrete:eq7}) is irrelevant and can be dropped.
Being of a Gaussian form, the computation of any correlation function
of (\ref{discrete:eq8}) can be easily done as it is related to the
inverse matrix $M_{ij}^{-1}$ (see Appendix \ref{app:hubbard}).  With the use of the result
\begin{eqnarray}
\label{discrete:eq9b}
\sum_l \left \langle T_{l}^{{(\perp)}^{2}} \right \rangle &=& 2\sum_l M_{ll}^{-1},
\end{eqnarray}
one then finds from Eq. \eqref{discrete:eq9},
\begin{eqnarray}
\label{discrete:eq9c} 
\frac{\left \langle z \right \rangle}{N b} &=& 1-\frac{1}{N} {\rm Tr} \mathbf{M}^{-1}.
\end{eqnarray}

In order to compute the trace of the inverse matrix $\mathbf{M}^{-1}$ one may
switch to Fourier variables for diagonalization (since the boundary
conditions are not relevant in the large $N$ limit, we use periodic
boundary conditions, as already done for $K_j$),
\begin{eqnarray}
\label{discrete:eq10}
\sum_{l=1}^N M_{lm} e^{\mathrm{i} \omega_{n} \left(l - m \right)} &=& \lambda\left(\omega_{n}\right), \qquad \omega_n=\frac{2n \pi}{N}
\end{eqnarray}
where the eigenvalues are
\begin{eqnarray}
\label{discrete:eq11}
\lambda\left(\omega\right) &=& u -2 K \cos \omega
\end{eqnarray}
with $u=bf+2 K+1$. Then in the $N\to\infty$ limit
\begin{eqnarray}
\label{discrete:eq13}
\frac{1}{N} \sum_{l=1}^N M_{ll}^{-1} &=& \frac{1}{N} \sum_{n=0}^{N-1} \frac{1}{\lambda\left(\omega_n\right)}\stackrel{N \to \infty}{\rightarrow} 
\int_{-\pi}^{+\pi} \frac{d \omega}{2 \pi} \frac{1}{\lambda\left(\omega\right)}
=\frac{1}{\sqrt{\left(bf+1\right)^2+ 4 K \left(bf+1\right)}}.
\end{eqnarray}
The last equality in Eq.(\ref{discrete:eq13}) has been obtained in the
$N \to \infty$ limit by contour integration.  Note that condition $u
\ge 2 K$ is required to ensure positive eigenvalues of the $M_{lm}$
matrix and well defined integral in Eq. (\ref{discrete:eq7}).  This
can be inserted into Eq.(\ref{discrete:eq9}) so that (assuming $b f
\gg 1$)
\begin{eqnarray}
\label{discrete:eq15}
\frac{\left \langle z \right \rangle}{N b} &=& 1 - \frac{1}{\sqrt{b^2 f^2+ 4 K bf}},
\end{eqnarray} 
whose leading order terms agree  with Eq. (\ref{discrete:eq:3}).
%%%%%%%%%%%%%%%%%%%%%%%%%%%%%%%%%%%%%%%%%%%%%%%%%%%%%%%%%%%%%%%%%%%%%%%%
\subsection{Small force limit: detailed calculations }%Marko-Siggia interpolation formula}
\label{subsec:marko}
%%%%%%%%%%%%%%%%%%%%%%%%%%%%%%%%%%%%%%%%%%%%%%%%%%%%%%%%%%%%%%%%%%%%%%%
In the $f \to 0$ limit, we can expand the partition function $Z$ given
in Eq.(\ref{discrete:eq2}). If $Z_{0}$ is the partition function associated with the
$f=0$ Hamiltonian in eq.(\ref{discrete:eq2}), we have
\begin{eqnarray}
\label{marko:eq5a}
Z &=& Z_{0} \left[ 1+ \frac{1}{2} b^2 \sum_{ij} \sum_{\mu, \nu=x,y,z}
  f_{\mu} f_{\nu} \left \langle \hat{T}_{i}^{\mu} \hat{T}_{j}^{\nu}
  \right \rangle_{0} + \ldots \right] 
\end{eqnarray}
where the first-order term in the expansion vanishes because $\langle
\hat{\mathbf{T}}_{j} \rangle_{0}=0$ by symmetry.  The averages denoted by the subscript $0$ are
with respect to $Z_0$. Because of the rotational invariance of the zero-force hamiltionian (\ref{discrete:eq2}),
we have that $\langle \hat{T}_{i}^{\mu}  \rangle_{0}=0$ and
$\langle \hat{T}_{i}^{\mu} \hat{T}_{j}^{\nu} \rangle_{0}=
(1/3) \delta^{\mu \nu} \langle \hat{\mathbf{T}}_{i} \cdot
\hat{\mathbf{T}}_{j} \rangle_{0}$. Hence, one gets
\begin{eqnarray}
\label{marko:eq5b}
Z &=& Z_{0} \left[ 1+ \frac{1}{6} b^2 f^2 \sum_{ij}  \left \langle \hat{\mathbf{T}}_{i} \hat{\mathbf{T}}_{j} \right \rangle_{0} + \ldots \right],
\end{eqnarray}
where
\begin{eqnarray}
\label{marko:eq6}
\left \langle \hat{\mathbf{T}}_{i} \hat{\mathbf{T}}_{j} \right\rangle_{0} 
&=& \frac{1}{Z_{0}} 
\int \left[\prod_{n=1}^N \frac{d^3 \hat{\mathbf{T}}_n}{2 \pi}   
   \delta\left(\hat{\mathbf{T}}_n^2  -1 \right) \right] 
     e^{K \sum_{l=1}^N \hat{\mathbf{T}}_{l+1} \cdot \hat{\mathbf{T}}_{l}} \left( \hat{\mathbf{T}}_{i} \hat{\mathbf{T}}_{j} \right).
\end{eqnarray}
In Eq.(\ref{marko:eq6}) we have assmed periodic boundary conditions so that $\hat{\mathbf{T}}_{N}=\hat{\mathbf{T}}_{1}$.
Now, ${\mathbf{R}}=  b \sum_j\hat{\mathbf{T}}_j$ so that the quantity
\begin{eqnarray}
\label{marko:eq16n}
\sum_{i=0}^N \sum_{j=0}^{N} \left \langle \hat{\mathbf{T}}_{i} \cdot
  \hat{\mathbf{T}}_{j} \right \rangle_{0}
&=& \langle R^2\rangle/b^2,
\end{eqnarray}
the mean square end-to-end distance.  We then have from
Eq.(\ref{marko:eq5b})
\begin{eqnarray}
\label{marko:eq17}
Z &=& Z_{0} \left[ 1+\frac{1}{6}  f^2 \langle R^2\rangle  + \ldots \right],
\end{eqnarray}
and hence
\begin{eqnarray}
\label{marko:eq18}
\left \langle z \right \rangle &=& \frac{\partial}{\partial f} \ln Z=
\frac{1}{3}  f \langle R^2\rangle  + \ldots ,
\end{eqnarray}
which is consistent with the expected linear response mentioned
earlier.

The connection between the response function and the polymer size raises an interesting question on the
thermodynamic limit. This is discussed in Appendix \ref{app:issue-therm-limit}.

\subsubsection{Evaluation of $\langle R^2\rangle$}
\label{sec:evaluation-a_n}
We now evaluate $\langle R^2\rangle$.  Consider the quantity
\begin{eqnarray}
\label{marko:eq7}
I\left(K,\hat{\mathbf{T}}_{j-1}\right) &=& 
\int \frac{d^3 \hat{\mathbf{T}}_j}{2 \pi} 
\delta\left(
  \hat{\mathbf{T}}_j^2  -1 \right) e^{K \hat{\mathbf{T}}_{j-1} \cdot
  \hat{\mathbf{T}}_{j}}. 
\end{eqnarray}
This can be easily computed as an integral over the solid angle.  With
$\gamma_j$ as the angle between $\hat{\mathbf{T}}_j$ and
$\hat{\mathbf{T}}_{j-1}$, one gets
\begin{eqnarray}
\label{marko:eq9}
I\left(K,\hat{\mathbf{T}}_{j-1}\right) &=& \frac{1}{2\pi} \int d \Omega_{j}  
 e^{K \cos \gamma_{j}}
= \frac{\sinh K}{K} \equiv I_{0}\left(K\right),
\end{eqnarray}
independent of $\hat{\mathbf{T}}_{j-1}$. Then one can clearly
integrate $\mathbf{T}$'s in Eq. \eqref{discrete:eq:1} with $f=0$, term
by term, with the result
\begin{eqnarray}
\label{marko:eq11}
Z_{0} &=& \left[I_{0}\left(K\right)\right]^{N+1}.
\end{eqnarray}

Likewise one can also compute the average involved in
Eq.(\ref{marko:eq6}) as (assuming without loss of generality $j>i$)
\begin{eqnarray}
\label{marko:eq12}
\left \langle \hat{\mathbf{T}}_{i} \hat{\mathbf{T}}_{j} \right\rangle_{0} 
&=& \frac{1}{\left[I_{0}\left(K\right)\right]^{\vert j-i \vert}}
\int \frac{d^3 \hat{\mathbf{T}}_i}{2 \pi} \delta\left(
\hat{\mathbf{T}}_i^2 -1 \right) \ldots \int \frac{d^3 \hat{\mathbf{T}}_j}{2 \pi} \delta\left(
\hat{\mathbf{T}}_j^2  -1 \right) e^{K \sum_{l=i}^{j-1} \hat{\mathbf{T}}_{l+1} \cdot \hat{\mathbf{T}}_{l}} \  \hat{\mathbf{T}}_{i} \hat{\mathbf{T}}_{j} .
\end{eqnarray}
One then observes that
\begin{eqnarray}
\label{marko:eq13}
\frac{1}{I_{0}\left(K\right)} \int \frac{d^3 \hat{\mathbf{T}}_j}{2 \pi} \delta\left(
\hat{\mathbf{T}}_j^2 -1 \right) e^{K \hat{\mathbf{T}}_{j-1} \cdot
\hat{\mathbf{T}}_{j}}  \hat{\mathbf{T}}_{j}
&=&\mathcal{L}\left(K\right)  \hat{\mathbf{T}}_{j-1}, 
\end{eqnarray}
where
\begin{eqnarray}
\label{marko:eq14}
\mathcal{L}\left(K\right) &=& \frac{\partial}{\partial K} \log I_{0}\left(K\right)=\coth K-\frac{1}{K}.
\end{eqnarray}
Here $\mathcal{L}(K)$ is the Langevin function appearing in the exact
solution of the FJC subject to an external force discussed in Appendix
\ref{app:exact}.  In deriving Eqs.(\ref{marko:eq13}) and (\ref{marko:eq14}) 
we have neglected sub-dominant terms in the limit 
$N \gg \vert i -j \vert$, and exploited the rotationalinvariance of the zero-force hamiltonian (\ref{discrete:eq2}).

Then, by iteration,
\begin{eqnarray}
\label{marko:eq15}
\left \langle \hat{\mathbf{T}}_{i} \hat{\mathbf{T}}_{j} \right \rangle_{0} &=& \mathcal{L}\left(K\right) 
\left \langle \hat{\mathbf{T}}_{i} \hat{\mathbf{T}}_{j-1} \right \rangle_{0}= \left[\mathcal{L}\left(K\right)\right]^{\vert i-j \vert},
\end{eqnarray}
in the form of Eq. \eqref{discrete:eq:5}  with 
\begin{equation}
\label{marko:eq15n}
l_p=b/\vert\ln \mathcal{L}(K)\vert.
\end{equation}

For an explicit computation of $\langle R^2 \rangle$, note that
\begin{eqnarray}
\label{marko:eq19}
b^{-2} \langle R^2\rangle &=& 
\sum_{i=0}^N \sum_{j=0}^{N} \left[\mathcal{L}\left(K\right) \right]^{\vert i-j \vert}
= \sum_{l=0}^N \sum_{m=0}^l
\mathcal{L}^{l-m}\left(K\right)+\sum_{l=0}^N \sum_{m=l+1}^{N}
\mathcal{L}^{m-l}\left(K\right),
\end{eqnarray} 
With the help of the summation formula 
\begin{eqnarray}
\label{marko:eq20a}
\sum_{l=0}^N \sum_{m=0}^l x^{l-m} &=& 
\frac{1}{1-x} \left[N+1 -\frac{x}
{1-x} \left(1-x^{N+1}\right) \right],
\end{eqnarray}
and
\begin{eqnarray}
\label{marko:eq20b}
\sum_{l=0}^N \sum_{m=l+1}^N x^{l-m} &=& 
\frac{x}{1-x}      
\left[N+1 -\frac{\left(1-x^{N+1}\right)}{1-x}\right],
\end{eqnarray}
the final form is 
\begin{eqnarray}
\label{marko:eq21}
\langle R^2\rangle &=& b^2 \left(N+1\right)
\frac{1+\mathcal{L}\left(K\right)}{1-\mathcal{L}\left(K\right)}
-2 b^2 \mathcal{L}\left(K\right)
\frac{1-\mathcal{L}^{N+1}\left(K\right)}{\left(1-\mathcal{L}\left(K\right)\right)^2}.
\end{eqnarray}
Because $|\mathcal{L}|<1$, we do see $\langle R^2\rangle \sim N$ for
$N\to\infty$, as claimed in Sec. \ref{sec:elementary} (see below Eq.
\eqref{eq:9}).  For $N\ll l_p/b$, a polymer would look like a rod in an
extended state but for $N\gg l_p/b$ it would be Gaussian. The
semiflexible regime corresponds to the intermediate case $N>l_p/b$.
Since $\vert {\mathcal{L}}(K)\vert <1$, $l_p$ is always finite, except
for $K\to \infty$.  For large $K$, ${\mathcal{L}}(K)\approx
1-\frac{1}{K}$ so that, from Eq. \eqref{marko:eq15n}, $l_p\approx Kb$.
In this limit, $(1+\mathcal{L}(K))/(1-\mathcal{L}(K)) \approx 2 K$, so
that the size can be written as
\begin{eqnarray}
\label{structure:eq14n}
\left \langle R^2 \right \rangle
 = 2l_p L_c- 2l_p^2( 1  -   e^{-L_c/l_p}),
\end{eqnarray}
in terms of the length $L_c=Nb$.  Note that in the limit $N \ll l_p/b$, Eq.(\ref{structure:eq14n}) predicts
a ballistic dependence $\langle R^2 \rangle \sim N^2$, as expected.
%%%%%%%%%%%%%%%%%%%%%%%%%%%%%%%%%%%%%%%%%%%%%%%%%%%%%%%%%%%%%%%%%%%%%%%%%%%%%%%%%
\subsection{An interpolation formula}
\label{subsec:interpolation}
%%%%%%%%%%%%%%%%%%%%%%%%%%%%%%%%%%%%%%%%%%%%%%%%%%%%%%%%%%%%%%%%%%%%%%%%%%%%%%%%%
The result given in Eq.(\ref{discrete:eq15}) can be reduced to its
continuum counterpart \cite{Marko95}, by considering the $b \to 0$
limit with $N,K\to\infty$, keeping the persistence length $l_p=K b$
fixed and also the chain length $L_c=Nb$ \cite{Rosa03}. From Eq.(\ref{discrete:eq2}) we see that $\mathbf{f}$ has dimensions of the inverse of a length.  Upon
introducing the ``physical'' force $f_{\text{phys}}=f/\beta$, and the
dimensionless ratio $\zeta=\langle z \rangle/(Nb)$, one obtains from 
Eq.(\ref{discrete:eq15}) and Eq.(\ref{marko:eq18})
\begin{eqnarray}
\label{interpolation:eq1}
\zeta &=& 1- \frac{1}{2 \sqrt{l_p \beta f_{\text{phys}} }}
\end{eqnarray}
in the large force limit. Notice that we are working in the limit $l_p \ll L_c$  so that the first term in
the right hand side of Eq.(\ref{structure:eq14n}) is the dominant one.

The opposite limit $f \to 0$ , can be obtained directly in the $\beta
f_{\text{phys}} b \ll 1$ limit as given in previous section. Indeed,
from Eq.(\ref{marko:eq21}) in the $b \to 0$ and $N>>1$ limit we get
from Eq. \eqref{structure:eq14n}
\begin{eqnarray}
\label{interpolation:eq4}
\left \langle z \right \rangle &=& \frac{2}{3} l_p L_{c}\beta f_{\text{phys}}+ \ldots,
\end{eqnarray}
and hence 
\begin{eqnarray}
\label{interpolation:eq5}
\zeta &=& \frac{2}{3} \beta f_{\text{phys}}  l_p + \ldots.
\end{eqnarray}
Similar results can be obtained by considering a Gaussian chain
subject to an external force.  Within the discrete limit, we have the
FJC model that can be solved exactly, as discussed in Appendix
\ref{app:exact}.

Notice that it can be shown that $b=2 l_p$ within the WLC model \cite{Grosberg94,Rubinstein03},
so that this relation agrees with Eq.(\ref{discrete:eq2bis}).

The two above relations (\ref{interpolation:eq1}) and
(\ref{interpolation:eq5}) can be inverted to yield
\begin{equation}
\label{interpolation:eq6}
l_p \beta f_{\text{phys}}
 =
\left\{
\begin{array}{ccc}
\frac{3}{2} \zeta & \text{if} & \zeta \ll 1 \\
\frac{1}{4\left(1-\zeta\right)^2} & \text{if} & \zeta \lesssim 1 
\end{array}
\right .
\end{equation}
Both regimes can be embodied into an interpolation formula
\cite{Marko95}
\begin{eqnarray}
\label{interpolation:eq7}
l_p \beta f_{\text{phys}}&=& \zeta + \frac{1}{4\left(1-\zeta\right)^2} - \frac{1}{4}
\end{eqnarray}
that reduces to the two limits given in Eq.(\ref{interpolation:eq6})
in the respective regimes.  Additional discussions can be found in
Refs.\cite{Odijk95,Bosch85,Livadaru03,Stepanow04,Doniach96}, while a
recent discussion on the numerical supporting results can be found in
Ref.\cite{Becker10}.
%%%%%%%%%%%%%%%%%%%%%%%%%%%%%%%%%%%%%%%%%%%%%%%%%%%%%%%%%%%%%%%%%%%%%%%%%%%%%%%%%%%%%%%%%%%%%%%%%%%%%%
\subsection{Structure factor and end-to-end distance}
\label{subsec:structure}
%%%%%%%%%%%%%%%%%%%%%%%%%%%%%%%%%%%%%%%%%%%%%%%%%%%%%%%%%%%%%%%%%%%%%%%%%%%%%%%%%%%%%%%%%%%%%%%%%%%%%
A very useful quantity to connect with experiment is given by the
structure factor. In the absence of external force, this was obtained
by Shimada et al \cite{Shimada88}.

In the discrete representation, the structure factor is defined as
\begin{eqnarray}
\label{structure:eq1}
S\left(\mathbf{k}\right) &=& \frac{1}{N} \left \langle \sum_{ij=1}^{N} \exp\left[\mathrm{i} \mathbf{k} \cdot \left(\mathbf{r}_i-
\mathbf{r}_j \right) \right] \right \rangle
\end{eqnarray}
For a Gaussian FJC chain this can be easily evaluated by Gaussian
integrals, as reported in Appendix \ref{app:structure_gaussian} with
the result \cite{Doi86}
\begin{eqnarray}
\label{structure:eq2}
S_0\left(k \right) &=& N F_D\left(3 R_g^2 k^2\right)
\end{eqnarray}
where $R_g=(N b^2/6)^{1/2}$ is the radius of gyration and $F_D(x)$ is the Debye function
\begin{eqnarray}
\label{structure:eq3}
F_D\left(x\right) &=& \frac{2}{x^2} \left[e^{-x}-1+x\right]
\end{eqnarray}
This structure factor was used in the study of a semi-dilute solution.
Now consider the WLC model in the continuum formulation. Unlike the
case of Edwards model, we can consider the limit $b \to 0$ and $N\gg1$
with $L_c=Nb$ fixed. Then $S(\mathbf{k})$ reads
\begin{eqnarray}
\label{structure:eq4}
S\left(\mathbf{k}\right) &=& \frac{1}{L_c b} \int_{0}^{L_c} ds \int_{0}^{L_c} ds^{\prime} \left \langle 
  \exp\left[\mathrm{i} \mathbf{k} \cdot \left(\mathbf{R}\left(s\right)-
      \mathbf{R}\left(s^{\prime}\right) \right)\right] \right \rangle
\end{eqnarray}

Consider now the case $f=0$. As discussed, the continuum limit of
Eq.(\ref{marko:eq15}) is
\begin{eqnarray}
\label{structure:eq6b}
\left \langle \hat{\mathbf{T}} \left(s\right) \cdot\hat{\mathbf{T}} \left(s^{\prime} \right) \right \rangle_{0} &=&
\exp\left[-\frac{1}{l_{p}}\left \vert s - s^{\prime} \right \vert \right]
\end{eqnarray} 

This can be alternatively viewed by using a different scheme as
detailed in Appendix \ref{app:integral} (see Eq.(\ref{integral:eq3}).

Next we consider the exact evaluation of the structure factor for the
WLC model as an expansion in powers of $k$, that can be computed terms
by terms.

The discretized version of Eqs.(\ref{semi:eq1}) and (\ref{semi:eq2})
when $f=0$ is Eq.(\ref{discrete:eq1}), that is
\begin{eqnarray}
\label{structure:eq7}
Z&=& \int \left[ \prod_{j=0}^{N} \frac{d^3 \mathbf{T}_{j}}{2 \pi} \delta\left(\mathbf{T}_j^2-1\right) \right] \exp\left[-\frac{1}{2} \frac{l_p}{b}
  \sum_{j=1}^N \left(\mathbf{T}_j-\mathbf{T}_{j-1} \right)^2 \right]
\end{eqnarray} 
As before, $\mathbf{T}_j$ is a vector tangent to the polymer axis at
position $\mathbf{r}_j$ and $d^3\mathbf{T}_j=d \hat{\mathbf{T}}_j d
T_j T_j^2$ so that the integrals over all $dT_j$ can be carried out
immediately because of the delta function appearing in
Eq.(\ref{structure:eq7}).  Upon introducing the Green function
\begin{eqnarray}
\label{structure:eq8}
G_{0L_c}\left(\hat{\mathbf{T}}_0,\hat{\mathbf{T}}_N \right) &=& \int \left[ \prod_{j=0}^{N} \frac{d \hat{\mathbf{T}}_{j}}{4 \pi} \right]
\exp\left[-\frac{1}{2} \frac{l_p}{b}
  \sum_{j=1}^N \left(\mathbf{T}_j-\mathbf{T}_{j-1} \right)^2 \right]
\end{eqnarray}  
we then have
\begin{eqnarray}
\label{structure:eq9}
Z&=& \int  \frac{d \hat{\mathbf{T}}_{0}}{4 \pi} \int  \frac{d \hat{\mathbf{T}}_{N}}{4 \pi} G_{0L}\left(\hat{\mathbf{T}}_0,\hat{\mathbf{T}}_N \right)
\end{eqnarray}
The Green function (\ref{structure:eq8}) has the following form in the
$b \to 0$ limit with $L_c=N b$ fixed as remarked. This is done in
Appendix \ref{app:green}) with the result
\begin{eqnarray}
\label{structure:eq10}
G_{0L_c}\left(\hat{\mathbf{T}}_0,\hat{\mathbf{T}}_N \right) &=& 4 \pi \sum_{l=0}^{+\infty} \sum_{m=-l}^{+l} e^{-L_c l\left(l+1\right)/(2 l_p)}
Y_{lm}\left(\hat{\mathbf{T}}_0\right) Y_{lm}^{*} \left(\hat{\mathbf{T}}_N \right)
\end{eqnarray}

Next, we expand the exponential in the structure factor
(\ref{structure:eq4}) in powers of $k$ up to second order
\begin{eqnarray}
\label{structure:eq11}
S\left(\mathbf{k}\right) &=& \frac{1}{L_c b} \int_{0}^{L_{c}} ds \int_{0}^{L_{c}} d s^{\prime} \left[1-\frac{1}{2} 
\left \langle \left[r_z\left(s\right)-r_z\left(s^{\prime}\right) \right] \right \rangle k^2 + \ldots \right]
\end{eqnarray}
where we have assumed $\mathbf{k}=k \hat{\mathbf{z}}$, and where (for
$s> s^{\prime} $) we have defined
\begin{eqnarray}
\label{structure:eq12}
\left \langle \left[r_z\left(s\right)-r_z\left(s^{\prime}\right) \right] \right \rangle &=& \int_{{s}^{\prime}}^{s} d s_1 
\int_{{s}^{\prime}}^{s} d s_2 \left \langle \hat{T}_z\left(s_1\right)  \hat{T}_z\left(s_2\right)
\right \rangle
\end{eqnarray}
Note that all odd powers vanish by symmetry, so the power expansion is
formed by even powers only. The integral (\ref{structure:eq12}) is
computed in Appendix \ref{app:integral} (see Eq.(\ref{integral:eq4})),
and can be inserted back into the expansion Eq.(\ref{structure:eq11}).
Elementary integrations, along with the relation $b=2 l_p$
\cite{Rubinstein03}, then lead to
\begin{eqnarray}
\label{structure:eq13}
S\left(k\right) &=& \frac{L_{c}}{2 l_p} \left\{
1-\frac{4}{3}\frac{l_p^5}{L_{c}} \left[\frac{1}{6} \left(\frac{L_{c}}{l_p}\right)^3
-\frac{1}{2}\left(\frac{L_{c}}{l_p}\right)^2+ \frac{L_{c}}{l_p}-1+ e^{-L_{c}/l_p} \right] k^2 + \ldots 
\right\}
\end{eqnarray}

As a by-product of this calculation, we can obtain the end-to-end
distance that can be compared with the mean-field calculation reported
in Ref \cite{Ha97}.  Indeed, using Eq.(\ref{structure:eq12}), the
end-to-end distance is given by
\begin{eqnarray}
\label{structure:eq14}
\left \langle R^2 \right \rangle \equiv \left \langle \left[\mathbf{r}\left(L_{c}\right) - \mathbf{r}\left(0\right) \right]^2 \right \rangle &=&
3  \left \langle \left[r_z\left(L_{c}\right) - r_z\left(0\right) \right]^2 \right \rangle = 2l_p^2 \left[\frac{L_{c}}{l_p}-1 + e^{-L_{c}/l_p}\right]
\end{eqnarray}
where the last equality again stems from Eq.(\ref{integral:eq4}) (see
Appendix \ref{app:integral}).  This agrees with the direct result
obtained in Eq. \eqref{structure:eq14n}.  Note that this result
coincides also with that obtained in Ref.\cite{Ha97} with $f=0$, provided that a mean-field
translation $2l_p/3 \to l_p$ is carried out.
%%%%%%%%%%%%%%%%%%%%%%%%%%%%%%%%%%%%%%%%%%%%%%%%%%%%%%%%%%%%%%%%%%%%%
%%%%%%%%%%%%%%%%%%%%%%%%%%%%%%%%%%%%%%%%%%%%%%%%%%%%%%%%%%%%%%%%%%%%%%%%%%%%%%%%%%%%
\section{Outlook}
\label{sec:outlook}
%%%%%%%%%%%%%%%%%%%%%%%%%%%%%%%%%%%%%%%%%%%%%%%%%%%%%%%%%%%%%%%%%%%%%%%%%%%%%%%%%%%%%
The aim of this review was to introduce some well-known and
less-well-known features of polymer physics, within a unified
framework hinging upon the Flory theory as a pillar.  In doing this,
we have reviewed some formalisms, approximations, and results briefly,
but in a self-contained way, so that it could be used as a first
approach to these methods at the graduate student level.

Starting with the simplest and well-known version of the Flory
approach given in Sec. \ref{sec:elementary}, we have proceeded by
introducing the Edwards continuum approach in Sec.\ref{sec:edwards}
that is used as a toolbox for field-theoretical approaches to polymer
physics.

One of the reasons that stimulated us to review this topic derives
from the fact that the Flory theory is frequently exploited, in
different forms, as a theoretical tool to tackle remarkably complex
systems.  Mean field theories are the generic tools to handle
interacting systems in a nonperturbative way, especially in problems
without any small parameter.  It ignores fluctuations and so provides
results too coarse to distinguish the subtle effects of dimensionality
and correlations.  As a result the predictions of the nature of phase
transitions, or of the emergent phases become questionable.  Although
technically the Flory theory uses the saddle point, steepest descent
method associated with mean field theories, it remarkably provides us
with signatures of dimensionality dependence.  This is a point that
often gets glossed over.  Except for rare exact solutions and full
fledged renormalization group calculations, no approaches other than
the Flory theory give $d$-dependent results.  Often the Flory results
are very close to the correct ones.  

The Flory theory can be used for systems with long-range correlations
or with no relevant length scale other than the large one determined
by the size.  In this respect the approach is expected to be
applicable to problems faced by different communities that hardly
communicate one another.  Hence, our aim here was to focus on some
specific aspects of the Flory theory that we regarded as the most
useful for graduate students, rather than performing an exhaustive
review.  As a result, many important aspects and contributions on this
topic have not been covered, nor cited, by the present work. One
example of that is polymer solutions that have been synthesized in a
short summary in Section \ref{sec:crossover}.  The trade-off lies in
the fact that we could stress some nuances and details.  For instance
the case of  Sections \ref{sec:flory}, \ref{sec:additional}, and
\ref{sec:crossover}, where we have discussed in some detail the
steepest-descent approach to the Flory theory (Sec. \ref{sec:flory}),
the interpolation formula (Sec. \ref{sec:additional}), and an
interesting crossover effect related to finite size effects and
tricritical point.  We have also tried to cast the Flory theory within
some modern perspective (see Sec.\ref{sec:flory-theory-modern}) that
included the scaling theory, and critical exponents.

The Flory mean-field approach can be simply modified by the addition
of an external force, as described in Section \ref{sec:semiflexible}),
and this technique has become particularly useful in the last two
decades due to the remarkable improvements in the experimental control
of the single-molecule stretching, with far reaching consequences in
various biological systems, most notably DNA.
 
All in all, the Flory theory, and its variants, continue to be a very
powerful tool in the study of polymer systems.  We hope that this
review will help to convey this message and to understand many
different scale-invariant problems.

%%%%%%%%%%%%%%%%%%%%%%%%%%%%%%%%%%%%%%

\appendix

%%%%%%%%%%%%%%%%%%%%%%%%%%%%%%%%%%%%%%%%%%%%%%%%%%%%%%%%%%%%%%%%%%%%%%%%%%%%%%%%%%%%%%%%%%%%%%%%%%%%%%%%%%%%%%%%%%%%%%%%%
\section{Gaussian integrals and the Hubbard-Stratonovich transformation}
\label{app:hubbard}
%%%%%%%%%%%%%%%%%%%%%%%%%%%%%%%%%%%%%%%%%%%%%%%%%%%%%%%%%%%%%%%%%%%%%%%%%%%%%%%%%%
Consider the following Gaussian identity
\begin{eqnarray}
\label{hubb:eq1}
\int_{-\infty}^{+\infty} d \phi e^{-\frac{1}{2} a \phi^2 - \mathrm{i} \phi x} &=& \sqrt{\frac{2\pi}{a}} e^{-\frac{1}{2} \frac{x^2}{a}}
\end{eqnarray}
that can be easily proved by completing the square. 

A generalization of this to $n$ variables reads
\begin{eqnarray}
\label{hubb:eq2}
e^{\frac{1}{2} \sum_{ij} \xi_i K_{ij} \xi_j} &=& \left[\left(2\pi\right)^{n} \det \mathbf{K} \right]^{-1/2}
\int \left[\prod_{l} d \phi_{l} \right] e^{-\frac{1}{2} \sum_{ij} \phi_i K_{ij}^{-1} \phi_j + \sum_{j} \xi_j \phi_j}
\end{eqnarray}
where $\mathbf{K}$ is any symmetric matrix with positive eigenvalues, and it
is the basis of the so-called Hubbard-Stratonovich transformation.

For Gaussian variables, correlation functions of the the type $\langle \phi_i \phi_j \rangle$ are
related to the inverse matrix appearing in the interactions. This can be seen as follows.  From Eq.(\ref{hubb:eq2})
we have
\begin{eqnarray}
\label{hubb:eq3}
\left \langle \phi_i \phi_j \right \rangle &=& \frac{\int \left[\prod_{l} d \phi_{l} \right] \phi_i \phi_je^{-\frac{1}{2} \sum_{lm} \phi_l K_{lm}^{-1} \phi_m}}{\int \left[\prod_{l} d \phi_{l} \right] e^{-\frac{1}{2} \sum_{lm} \phi_l K_{lm}^{-1} \phi_m}} \\ \nonumber
&=& \left[\left(2\pi\right)^{n} \det \mathbf{K} \right]^{1/2}
\int \left[\prod_{l} d \phi_{l} \right] \phi_i \phi_j e^{-\frac{1}{2} \sum_{lm} \phi_l K_{lm}^{-1} \phi_m} \\ \nonumber
&=& \frac{\partial^2}{\partial \xi_i \partial \xi_j} \left[e^{\frac{1}{2} \sum_{lm} \xi_l K_{lm} \xi_m} \right]_{\left\{ \xi=0 \right\}} = K_{ij}
\end{eqnarray}

%%%%%%%%%%%%%%%%%%%%%%%%%%%%%%%%%%%%%%%%%%%%%%%%%%%%%%%%%%%%%%%%%%%%%%%%%%%%%%%%%%%%%

%%%%%%%%%%%%%%%%%%%%%%%%%%%%%%%%%%%%%%%%%%%%%%%%%%%%%%%%%%%%%%%%%%%%%%%%%%%%%%%%%%
\section{Distribution of the end-to-end distance in $d$ dimensions}
\label{app:distribution}
%%%%%%%%%%%%%%%%%%%%%%%%%%%%%%%%%%%%%%%%%%%%%%%%%%%%%%%%%%%%%%%%%%%%%%%%%%%%%%%%%%
Introduce the bond ${\bm \tau}_j=\mathbf{r}_j-\mathbf{r}_{j-1}$,
$j=1,\ldots,N$, as depicted in Fig.\ref{fig:fig1}, and let $p({\bm
  \tau}_j)$ be the probability distribution of the $j$-th bond.

Then the probability distribution function for the end-to-end
distance, $P(\mathbf{R},N)$, as given in Eq.(\ref{elementary:eq1})
reads
\begin{eqnarray}
\label{distribution:eq2}
P\left(\mathbf{R},N\right) &=& \int \prod_{i=1}^N d^d {\bm \tau}_i\ \  \delta^d\left(\mathbf{R} - \sum_{j=1}^N {\bm \tau}_j \right)\, \prod_{l=1}^N p \left({\bm \tau}_l \right),
\end{eqnarray}
where one can use the integral representation
\begin{eqnarray}
\label{distribution:eq3}
\delta^d\left({\bm \tau}_j \right) &=& \int \frac{d^d \mathbf{k}}{\left(2 \pi\right)^d} e^{\mathrm{i} \mathbf{k} \cdot {\bm \tau}_j},
\end{eqnarray}
to obtain
\begin{eqnarray}
\label{distribution:eq4}
P\left(\mathbf{R},N\right) &=& \int \frac{d^d \mathbf{k}}{\left(2 \pi\right)^d}
e^{\mathrm{i} \mathbf{k} \cdot \mathbf{R}}
   \left[\hat{p}\left(\mathbf{k}\right) \right]^N
=\int \frac{d^d \mathbf{k}}{\left(2 \pi\right)^d}  
     e^{\mathrm{i} \mathbf{k} \cdot \mathbf{R}}\,\, e^{N \ln \hat{p}\left(\mathbf{k}\right)},
\end{eqnarray}
where 
\begin{eqnarray}
\label{distribution:eq5}
\hat{p}(\mathbf{k})&=&\int d^d {\bm \tau}_j e^{-\mathrm{i} \mathbf{k} \cdot \mathbf{r}_j} p\left(\mathbf{\tau}_j\right),
\end{eqnarray}
is the Fourier transform of $ p({\bm \tau}_j)$. 

The normalization condition guarantees that $\hat{p}({\bf k}=0)=1$, while a
spherically symmetric distribution implies
$\hat{p}(\mathbf{k})=\hat{p}(k)$.  With these, a Taylor series
expansion yields
\begin{eqnarray}
\label{distribution:eq10}
\hat{p}(\mathbf{k})&=& 1-\frac{1}{2} \left(k\sigma\right)^2 + O(k^4),
\end{eqnarray}
where $\sigma^2$ is the variance of the distribution $p({\bm \tau})$.

For $N \gg 1$, our interest is in the overall description of the
polymer set by the scale $1/k$ which is  much larger than the
microscopic scale set by $\sigma$,  i.e.,  we can assume $ k \sigma
\ll 1$.
Therefore,
\begin{equation}
\label{distribution:eq11}
\ln \hat{p}({\bf k})\approx -\frac{1}{2} \left(k\sigma\right)^2, \ ( k \sigma \ll 1).
\end{equation}
Substituting in Eq.(\ref{distribution:eq4}) we then get
\begin{eqnarray}
\label{distribution:eq12}
P\left(\mathbf{R},N\right) &\approx&   \int \frac{d^d \mathbf{k}}{\left(2 \pi\right)^d}  
e^{\mathrm{i} \mathbf{k} \cdot \mathbf{R}}\  \exp\left[-\frac{N}{2} \left(k\sigma\right)^2 \right] \approx \left(\frac{1}{2 \pi N \sigma^2} \right)^{d/2}
\exp\left[-\frac{1}{2} \frac{R^2}{N \sigma^2} \right]
\end{eqnarray}
In $d=3$ this reduces to Eq.(\ref{elementary:eq4}).

\subsection{Examples}
\label{sec:examples}
We consider two examples.
One is the example of the distribution for the FJC
\begin{eqnarray}
\label{distribution:eq1}
p\left({\bm \tau}_j\right) &=& \frac{1}{S_d \tau_j^{d-1}} \delta\left(\tau_j-b\right)
\end{eqnarray}
where $S_d=2 \pi^{d/2}/\Gamma(d/2)$ is the surface of a unit sphere in
$d$-dimensions, and $\Gamma(z)$ is the Gamma function
\cite{Abramowitz72}.  Note that this choice ensures $\int d^d {\bm
  \tau}_j p({\bm \tau}_j)=1$.    Another possibility is a Gaussian distribution
\begin{eqnarray}
\label{distribution:eq1b}
p\left({\bm \tau}_j\right) &=& \left(\frac{1}{2 \pi b^2} \right)^{d/2} \exp \left[- \frac{1}{2} \frac{\tau_j^2}{b^2} \right].
\end{eqnarray}
There is the obvious difference between the two, the first one has a
fixed length but the second one has no fixed length.  Many other
choices are possible.

By using Eq.(\ref{distribution:eq1}), one can easily compute that
\begin{eqnarray}
\label{distribution:eq6}
\hat{p}(\mathbf{k})&=& \int_{0}^{+\infty} dr \; r^{d-1} \int d \Omega_{d} e^{-i k r \hat{\mathbf{k}}\cdot \hat{\mathbf{r}}} \frac{1}{S_d r^{d-1}} 
\delta\left(r-b \right) \\ \nonumber
&=& \frac{1}{S_d} \int_{0}^{2 \pi} d \theta_{1} \int_{0}^{\pi} d \theta_{2} \sin \theta_{2} \ldots \int_{0}^{\pi} d \theta_{d-1} \sin^{d-2} \theta_{d-1}
  e^{-i k b \cos \theta_{d-1}} \\ \nonumber
&=& \frac{S_{d-1}}{S_d} \int_{0}^{\pi} d \theta_{d-1} \sin ^{d-2} \theta_{d-1}  e^{-i k b \cos \theta_{d-1}}
\end{eqnarray}
In the limit of small $kb$, we expand the exponential $\exp(x)=1 + x +
x^2/2...$.  The first order term will vanish by symmetry.  The second
order term involves an integral $\int_0^{\pi} \sin^{d-2}\theta
\cos^2\theta d\theta= \sqrt{\pi} \Gamma[(d-1)/2]/(2\Gamma[1+d/2])$.
This matches Eq. \eqref{distribution:eq10} with $\sigma^2=b^2/d$.

The integral in Eq. \eqref{distribution:eq6} can be handled exactly.
We use the following result \cite{Gradshteyn00}
\begin{eqnarray}
\label{distribution:eq7}
\int_{0}^{\pi} d \theta \sin^{2 \nu} \theta e^{\mathrm{i} \beta \cos \theta} &=& \sqrt{\pi} \left(\frac{2}{\beta} \right)^{\nu} \Gamma\left(\nu+\frac{1}{2} \right)
J_{\nu}\left(\beta\right)
\end{eqnarray}
where $J_{\nu}(z)$ is a Bessel function with the property
$J_{\nu}(-z)=(-1)^{\nu} J_{\nu}(z)$ \cite{Abramowitz72}, to obtain
\begin{eqnarray}
\label{distribution:eq8}
\hat{p}(\mathbf{k})&=& \left(\frac{2}{k b} \right)^{d/2-1} \Gamma\left(\frac{d}{2} \right) J_{d/2-1} \left(k b\right)
\end{eqnarray}
By making use of the expansion \cite{Abramowitz72}
\begin{eqnarray}
\label{distribution:eq9}
J_{\nu}\left(z\right) &=& \left(\frac{1}{2} z \right)^{\nu} \sum_{n=0}^{+\infty} \frac{1}{n! \Gamma\left(\nu+n+1\right)} \left(-\frac{1}{4} z^2 \right)^n
\end{eqnarray}
and the property of the Gamma function $\Gamma(z+1)=z \Gamma(z)$, one
obtains Eq. \ref{distribution:eq10} with $\sigma^2= b^2/d$.

For the Gaussian distribution,
$\hat{p}({\bf k})=\exp(-k^2 b^2/2)$ and its Taylor series expansion
around $k=0$ matches with Eq. \ref{distribution:eq10}. Here $\sigma=b$.

\subsection{Non-gaussian case}
The above derivation, a version of the central limit theorem, is valid
only if $\sigma<\infty$, otherwise the expansion in Eq. \ref{distribution:eq10} is useless.
There are important distributions which may
not have finite variances.  In those cases, a Gaussian distribution is
not expected. An example is the Cauchy distribution 
\begin{equation}
  \label{distribution:eq13}
p(x)= \frac{1}{\pi}\ \frac{b}{x^2+b^2},\quad{\rm(in\  1\text{--}dimension)},
\end{equation}
with infinite mean and variance.  
$P(R,N)$, in Eq. \eqref{distribution:eq13}, for large $N$, does not
converge to  a Gaussian but to  another  Cauchy distribution. 
The difference with the Gaussian distribution lies mainly in the tail
(large $|x|$ behaviour) of this distribution - the large $|x|$
behaviour of Eq. \ref{distribution:eq13} is responsible for the
divergent mean and variance.   It is precisely for
this reason, we do not consider such distributions in this review.
Our interest is in the behaviour of a polymer whose properties do
not require special or exceptional contributions from very large
sizes.

%%%%%%%%%%%%%%%%%%%%%%%%%%%%%%%%%%%%%%%%%%%%%%%%%%%%%%%%%%%%%%%%%%%%%%%%%%%%%%%%%%%%%

%%%%%%%%%%%%%%%%%%%%%%%%%%%%%%%%%%%%%%%%%%%%%%%%%%%%%%%%%%%%%%%%%%%%%%%%%%%%%%%%%%%%%%%%%%%%%%%%%
\section{Perturbation Theory}
\label{sec:perturbation}
%%%%%%%%%%%%%%%%%%%%%%%%%%%%%%%%%%%%%%%%%%%%%%%%%%%%%%%%%%%%%%%%%%%%%%%%%%%%%%%%%%%%%%%%%%%%%%%%%
Instead of the Flory approach that explores the large $z$ region directly, 
we here consider the small $z$ case which in principle can be handled in a 
perturbative way.   The ultimate difficulty is in tackling the series which 
in most cases turns out to be asymptotic in nature.

We go back to Eq.(\ref{edwards:eq12}) and expand the right-hand side
in powers of $u$ (for $v=0$)  
\begin{eqnarray}
\label{perturbation:eq1}
G_{L}\left(\mathbf{R}\right) &=& G_{L}^{(0)}\left(\mathbf{R}\right) - u \mathcal{G}_{L}^{(1)}\left(\mathbf{R}\right)
%-\beta v \mathcal{G}_{L}^{(13)}\left(\mathbf{R}\right) 
+ \ldots
\end{eqnarray}
where the first order terms in the expansion of the two-body term shown in Eq.(\ref{perturbation:eq1}) is
\begin{eqnarray}
\label{perturbation:eq2}
\mathcal{G}_{L}^{(1)}\left(\mathbf{R}\right)&=& G_{L}^{(0)}\left(\mathbf{R}\right) \frac{1}{2!} b^{d-2} \int_{0}^{L}
ds_{1} \int_{0}^{L} ds_{2} \, \left \langle \delta^{d} \left(\mathbf{R}\left(s_2\right)-\mathbf{R}\left(s_1\right)
\right)\right \rangle_{\mathbf{R}}.
\end{eqnarray}
The delta function ensures that there is one contact along the chain.
The series has the interpretation that the first term is the partition
function without any concern about the interactions while the second
term ${\mathcal{G}}_L^{(1)}$ is the sum over all configurations that have
one interaction along the chain.

The calculation of the end-to-end distance 
\begin{eqnarray}
\label{perturbation:eq4}
\left \langle R^2 \right \rangle &=& \frac{\int d^d \mathbf{R} \; R^2
  G_{L}\left(\mathbf{R}\right)}{\int d^d \mathbf{R} \;
  G_{L}\left(\mathbf{R}\right)}. 
\end{eqnarray}
also involves an expansion in $u$, coming from both the numerator and
the denominator. 
It is more or less straight-forward to calculate for the free case
\begin{eqnarray}
\label{perturbation:eq4b}
\left \langle R^2 \right \rangle_{0} &=& \frac{\int d^d \mathbf{R} \;
  R^2 G^{(0)}_{L}\left(\mathbf{R}\right)}{\int d^d \mathbf{R} \;
  G^{(0)}_{L}\left(\mathbf{R}\right)}=Lb. 
\end{eqnarray}

For generality, especially for higher order corrections,
two possible procedures to compute the first order
correction are discussed below.

%%%%%%%%%%%%%%%%%%%%%%%%%%%%%%%%%
\subsection{Direct evaluation}
\label{sec:direct-evaluation}
%%%%%%%%%%%%%%%%%%%%%%%%%%%%%%%%%
The convolution property \cite{Doi86} of the Gaussian distribution 
\begin{eqnarray}
\label{laplace:eq4}
G_{L}^{(0)} \left(\mathbf{R}\right) &=& \int d^d \mathbf{R'} G_{s}^{(0)} \left(\mathbf{R'}\right)
G_{L-s}^{(0)} \left(\mathbf{R}-\mathbf{R'}\right),
\end{eqnarray}
states that the probability of a Gaussian polymer reaching
$\mathbf{R}$ at length $L$ can be written as a product of its being at
any point $\mathbf{R}^{\prime}$ at an intermediate length $s$ and then
from $\mathbf{R}^{\prime}$ to $\mathbf{R}$ in the remaining $L-s$
length, with an integration over $\mathbf{R}^{\prime}$.
 
With repeated use of the convolution property, Eq. (\ref{laplace:eq4}).
the relevant average required for the two-body correction term is
\begin{subequations}
\begin{eqnarray}
\left \langle  \delta^{d} \left( \mathbf{R}\left(s_1\right)-
    \mathbf{R} \left(s_2 \right) \right)   \right \rangle_{\mathbf{R}}^{(0)} 
&=& 
  \frac{1}{G_{L}^{(0)} \left(\mathbf{R}\right)} 
     \int_{\mathbf{R}\left(0\right)=\mathbf{0}}^{\mathbf{R}\left(L\right)=\mathbf{R}}
            {\cal D} \mathbf{R}\left(s\right) \, 
                   \delta^{d} \left(\mathbf{R}\left(s_1\right)- \mathbf{R} \left(s_2 \right)
                                 \right)
                        e^{-\frac{d}{2b} \int_{0}^{L} ds \, 
                                 \left(\frac{\partial \mathbf{R}}{\partial s} \right)^2} \\ 
&=& \frac{1}{G_{L}^{(0)} \left(\mathbf{R}\right)} 
         \int d^d \mathbf{R}_1 \int d^d \mathbf{R}_2 \, 
                 \delta^d\left(\mathbf{R}_1  -\mathbf{R}_2 \right)
                    G_{L-s_{2}}^{(0)}\left(\mathbf{R}-\mathbf{R}_2\right)\times\nonumber\\
&&\qquad\qquad \qquad\qquad \qquad\qquad 
             G_{s_{2}-s_{1}}^{(0)} \left(\mathbf{R}_2-\mathbf{R}_1\right)
                              G_{s_{1}}^{(0)} \left(\mathbf{R}_1\right) \label{eq:3pert} \\ 
&=&\frac{1}{G_{L}^{(0)} \left(\mathbf{R}\right)} 
         \int d^d \mathbf{R'} \, G_{L-s_{2}}^{(0)}\left(\mathbf{R}-\mathbf{R'}\right) 
                              G_{s_{2}-s_{1}}^{(0)}
                              \left(\mathbf{0}\right) 
                                 G_{s_{1}}^{(0)} \left(\mathbf{R'}\right).\label{laplace:eq5}\\
&=&\frac{G_{L-s_2+s_1}^{(0)}\left(\mathbf{R}\right)}{G_{L}^{(0)}\left(\mathbf{R}\right)}
          G_{s_{2}-s_{1}}^{(0)}\left(\mathbf{0}\right),\label{eq:2pert}  
\end{eqnarray}
\end{subequations}
Eq. (\ref{eq:3pert}) has the interpretation of a polymer reaching
$\mathbf{R}^{\prime}$ at length $s_1$ from the origin and then
returning to $\mathbf{R}^{\prime}$ at length $s_2$ from where it goes
to the desired endpoint $\mathbf{R}$. Since $s_1,s_2$ could be any two
points, there are integrals over each of them.  The occurrence of
$G_{s_2-s_1}^{(0)}(\mathbf{0})$ is the signature of a loop formation that contains the main
aspect of the polymer correlations because it involves contact of two
monomers which may be nearby ($s_2-s_2$ small) or far-apart ($s_2-s_1$
large) along the chain.  The
eventual Gaussian integrals can be done. However the $s_1,s_2$
integrals are divergent.   The integrals over $s_1,s_2$
involve a term of the type $\int_0^L ds s^{1-d/2}$ which is  divergent
for $d\le 4$.  Such
divergent integrals can be handled by analytic continuation in $d$ 
by performing the integration where it is convergent and then
analytically continued to other dimensions. 
If $d$ is such that the integral converges, then
\begin{eqnarray}
\label{gaussian:eq13}
\frac{1}{2!} \int_{0}^{L} d s_1 \int_{0}^{L} d s_2 \left \vert s_2
  -s_1 \right \vert^{1-d/2} &=& \frac{4}{\left(4-d\right) 
\left(6-d\right)} L^{3-d/2},
\end{eqnarray}
which can then be extended to all $d$.  The poles at $d=4$ and $d=6$
are responsible for the divergence at other values of $d$.

A similar expansion in $z$ can be performed for the end-to-end
distance given by Eq.(\ref{perturbation:eq4}), by collecting terms of
similar order from both the numerator and the denominator.
To first order the correction would look like
\begin{equation}
  \label{eq:13R2}
  \langle R^2\rangle= \langle R^2\rangle_0 - u \int dR\  R^2 \ {\mathcal G}_L^{(1)}({\mathbf{R}}) + u
  \int dR \  R^2\  G_L^{(0)}\left(\mathbf{R}\right)  \int dR\  {\mathcal G}_L^{(1)}({\mathbf R})+ ... .
\end{equation}
With the use of Eqs. (\ref{perturbation:eq2}) and (\ref{eq:2pert}),
and the standard results of Gaussian integrals,  the two $u$-dependent
terms can be written as 
\begin{eqnarray}
  \label{eq:1R2}
  \int dR R^2 {\mathcal G}_L^{(1)}({\mathbf{R}}) 
 &=& \int_0^Lds_1\int_0^L ds_2  (L-s_2+s_1) G_{|s_2-s_1|}^{(0)}(0)\\ 
  \int dR \ R^2 G_L^{(0)} \left(\mathbf{R}\right) \int dR {\mathcal G}_L^{1}({\mathbf R})
 &=&   L \int_0^Lds_1\int_0^L ds_2  G_{|s_2-s_1|}^{(0)}(0),
\end{eqnarray}
so that we are left with the integral of Eq. (\ref{gaussian:eq13}).
With the analytic
continuation,  the end-to-end distance is  given by
\begin{eqnarray}
  \label{gaussian:eq14}
\left \langle R^2 \right \rangle &=& \left \langle R^2 \right \rangle_{0} \left[
1 + \frac{4}{\left(4-d\right) \left(6-d\right)} z+ \ldots \right],
\end{eqnarray}
with $z$ as in Eq. \eqref{eq:16} with  $c_1=(1/2\pi)^{d/2}$.
The divergence as $d\to 4$ is an important outcome of this
perturbative analysis and its
handling is part of the renormalization group machinery.

%%%%%%%%%%%%%%%%%%%%%%%%%%%%%%%%%%%%%%%%%%%%%%%%%%%%%%

%%%%%%%%%%%%%%%%%%%%%%%%%%%%%%%%%%%%%%%%%%%%%%%%%%%%%%%%%%%%%%
\subsection{Laplace-Fourier approach}
\label{subsec:laplace}
%%%%%%%%%%%%%%%%%%%%%%%%%%%%%%%%%%%%%%%%%%%%%%%%%%%%%%%%%%%%%%%
The same result can be obtained by using the Laplace-Fourier approach
\cite{Muthukumar84}. % following a remarkable piece of work by Muthukumar and Nickel \cite{Muthukumar84}.
This method requires an integral over the length from zero to infinity
and therefore may be called ``grand canonical'' compared to the
approach of the previous section, which may be termed as ``canonical''.

The Laplace-Fourier transform is defined by
\begin{eqnarray}
\label{laplace:eq1}
\widetilde{F}_{E} \left(\mathbf{k} \right) = \int_{0}^{+\infty} dL \, e^{-E L} \widehat{F}_{L} 
\left(\mathbf{R}\right) &=&   \int_{0}^{+\infty} dL \, e^{-E L} \int d^d \mathbf{R} \, e^{-\mathrm{i} \mathbf{k} \cdot \mathbf{R}}
F_{L} \left(\mathbf{R} \right)
\end{eqnarray}
along with its inverse
\begin{eqnarray}
\label{laplace:eq2}
F_L\left(\mathbf{R}\right) = \int  \frac{d^d \mathbf{k}}{\left(2\pi\right)^{d}} \, e^{+\mathrm{i} \mathbf{k} \cdot \mathbf{R}}
\widehat{F}\left(\mathbf{k}\right)
 &=&  \int \frac{d^d \mathbf{k}}{\left(2\pi\right)^d} \, e^{+\mathrm{i} \mathbf{k} \cdot \mathbf{R}} 
\int_{\gamma - \mathrm{i} \infty}^{\gamma + \mathrm{i} \infty} \frac{dE}{2\pi \mathrm{i}} \, e^{E L} 
\widetilde{F}_{E} \left(\mathbf{k} \right)
\end{eqnarray}
As usual, in Eq.(\ref{laplace:eq2}) $\gamma$ is a real constant that exceeds the real part of all the singularities of 
$\widetilde{F}_{E} \left(\mathbf{k} \right)$. 

We now go back to the expansion (\ref{perturbation:eq1}) that can be Laplace-Fourier transformed to obtain 
\begin{eqnarray}
\label{laplace:eq2a}
\widetilde{G}_{E}(\mathbf{k}) &=&  \widetilde{G}_{E}^{(0)}\left(\mathbf{k}\right) - 
u \widetilde{\mathcal{G}}_{E}^{(12)}\left(\mathbf{k}\right)+ \ldots
\end{eqnarray}
For simplicity, we limit here the discussion to the two-body interactions, but additional terms can be also considered.

Given that, the end-to-end distance can be computed from 
\begin{eqnarray}
\label{laplace:eq3}
\left \langle R^2 \right \rangle &=& \left\{
\frac{\int_{\gamma-\mathrm{i} \infty}^{\gamma+\mathrm{i} \infty} dE \, e^{EL} 
\left[- \nabla_{\mathbf{k}^2} \widetilde{G_{E}} \left(\mathbf{k} \right)\right]}
{\int_{\gamma-\mathrm{i} \infty}^{\gamma+\mathrm{i} \infty} dE \, e^{EL}
\widetilde{G_{E}} \left(\mathbf{k} \right)}
\right\}_{\mathbf{k}=\mathbf{0}}
\end{eqnarray}

% We then can use the convolution property \cite{Muthukumar84}
% \begin{eqnarray}
% \label{laplace:eq4}
% G_{L}^{(0)} \left(\mathbf{R}\right) &=& \int d^d \mathbf{R'} G_{s}^{(0)} \left(\mathbf{R'}\right)
% G_{L-s}^{(0)} \left(\mathbf{R}-\mathbf{R'}\right)
% \end{eqnarray}
% to obtain for the two-body term 
% \begin{eqnarray}
% \label{laplace:eq5}
% \left \langle  \delta^{d} \left( \mathbf{R}\left(s_1\right)- \mathbf{R} \left(s_2 \right) \right)   \right \rangle_{\mathbf{R}} &=&
% \frac{1}{G_{L}^{(0)} \left(\mathbf{R}\right)} \int_{\mathbf{R}\left(0\right)=\mathbf{0}}^{\mathbf{R}\left(L\right)=\mathbf{R}}
% {\cal D} \mathbf{R}\left(s\right) \, \delta^{d} \left( \mathbf{R}\left(s_1\right)- \mathbf{R} \left(s_2 \right) \right)
% e^{-\frac{d}{2b} \int_{0}^{L} ds \, \left(\frac{\partial \mathbf{R}}{\partial s} \right)^2} \\ \nonumber
% &=& \frac{1}{G_{L}^{(0)} \left(\mathbf{R}\right)} \int d^d \mathbf{R}_1 \int d^d \mathbf{R}_2 \, \delta^d\left(\mathbf{R}_1
% -\mathbf{R}_2 \right)  G_{L-s_{2}}^{(0)} \left(\mathbf{R}-\mathbf{R}_2\right) 
% G_{s_{2}-s_{1}}^{(0)} \left(\mathbf{R}_2-\mathbf{R}_1\right)
% G_{s_{1}}^{(0)} \left(\mathbf{R}_1\right) \\ \nonumber
% &=&\frac{1}{G_{L}^{(0)} \left(\mathbf{R}\right)} \int d^d \mathbf{R'} \, G_{L-s_{2}}^{(0)} \left(\mathbf{R}-\mathbf{R'}\right) 
% G_{s_{2}-s_{1}}^{(0)} \left(\mathbf{0}\right) G_{s_{1}}^{(0)} \left(\mathbf{R'}\right)
% \end{eqnarray}
The great advantage of the Laplace-Fourier transform is clearly that both the $\mathbf{R}$ and $s$ convolutions
appearing in Eq.(\ref{laplace:eq5}) can be decoupled so that
\begin{eqnarray}
\widetilde{\mathcal{G}}_{E}^{(12)} \left(\mathbf{k} \right) &=& \frac{1}{2!}\int_{0}^{+\infty} dL \;e^{-E L}b^{d-2}  
\int \frac{d^d \mathbf{q}}{\left(2\pi\right)^d} \, e^{+\mathrm{i} \mathbf{k} \cdot \mathbf{R}}
\int_{0}^{L} d s_2 \, \int_{0}^{L} d s_1 \, \widehat{G}_{L-s_{2}}^{(0)} \left(\mathbf{k} \right) 
\widehat{G}_{s_{2}-s_{1}}^{(0)} \left(\mathbf{q} \right) \widehat{G}_{s_{1}}^{(0)} \left(\mathbf{k} \right) \nonumber \\
&=& b^{d-2} \int \frac{d^d \mathbf{q}}{\left(2\pi\right)^d} \, \widetilde{G}_{E}^{(0)} \left(\mathbf{k} \right)
\widetilde{G}_{E}^{(0)} \left(\mathbf{q} \right) \widetilde{G}_{E}^{(0)} \left(\mathbf{k} \right) 
\label{laplace:eq8}
\end{eqnarray}
that is
\begin{eqnarray}
\label{laplace:eq9}
\widetilde{\mathcal{G}}_{E}^{(12)} \left(\mathbf{k} \right) &=&  b^{d-2} \left[\widetilde{G}_{E}^{(0)} 
\left(\mathbf{k} \right)\right]^2
\int \frac{d^d\mathbf{q}}{\left(2\pi\right)^d} \widetilde{G}_{E}^{(0)} \left(\mathbf{q} \right) 
\end{eqnarray}
where \cite{Muthukumar84}
\begin{eqnarray}
\label{laplace:eq10}
 \widetilde{G}_{E}^{(0)} \left(\mathbf{k} \right)&=& \lim_{\epsilon \to 0} \frac{1}{E+ b k^2/(2d)+\epsilon k^4} 
\end{eqnarray}
In Eq. (\ref{laplace:eq10}) we have included an $\epsilon k^4$ term to keep all integrals convergent, with the
understanding that the limit $\epsilon \to 0$ will be taken at the end of the calculation \cite{Muthukumar84}.
The integral appearing in Eq.(\ref{laplace:eq9}) is given by
\begin{eqnarray}
\label{laplace:eq11}
I_d\left(E,\epsilon\right)&\equiv&\int \frac{d^d\mathbf{q}}{\left(2\pi\right)^d} \widetilde{G}_{E}^{(0)} \left(\mathbf{q} \right) =
\frac{S_d}{\left(2 \pi\right)^d} \int_{0}^{+\infty} dq \, \frac{q^{d-1}}{E+b q^2/(2d)+ \epsilon q^4}
\end{eqnarray}

Let us now compute the first correction $\widetilde{\mathcal{G}}_{E}^{(12)} \left(\mathbf{k} \right)$ explicitly. Eq.(\ref{laplace:eq2a}) yields
\begin{eqnarray}
\label{laplace:eq14}
\widetilde{G}_{E}(\mathbf{k}) &=&  \widetilde{G}_{E}^{(0)}\left(\mathbf{k}\right) - 
 b^{d-2} \widetilde{\Sigma}_{E}^{(12)} \left[\widetilde{G}_{E}^{(0)}\left(\mathbf{k}\right)\right]^2 + \ldots
\end{eqnarray}
where we have introduced the ``self-energy''
\begin{eqnarray}
\label{laplace:eq15}
\widetilde{\Sigma}_{E}^{(12)}&=& u I_d\left(E,\epsilon\right)
\end{eqnarray}
Then we next note that because the Green function is always a function of  $k^2$ rather than
the wavevector $\mathbf{k}$ itself, we can write
\begin{eqnarray}
\label{laplace:eq16}
\mathbf{\nabla}_{\mathbf{k}} \widetilde{G}_{E}\left(\mathbf{k}\right) &=& 2 \mathbf{k} \frac{\partial}{\partial k^2}
\widetilde{G}_{E}\left(\mathbf{k}\right)
\end{eqnarray}
and
\begin{eqnarray}
\label{laplace:eq17}
\mathbf{\nabla}_{\mathbf{k}}^2 \widetilde{G}_{E}\left(\mathbf{k}\right) \vert_{\mathbf{k}=0}&=& 2 d \frac{\partial}{\partial k^2}
\widetilde{G}_{E}\left(\mathbf{k}\right)\vert_{\mathbf{k}=0}
\end{eqnarray}
Using Eqs.(\ref{laplace:eq14}), (\ref{laplace:eq15}), and (\ref{laplace:eq17}) into Eq.(\ref{laplace:eq3}) one gets
\begin{eqnarray}
\label{laplace:eq18}
\left \langle R^2 \right \rangle &=& b
\frac{
\int_{\gamma-\mathrm{i} \infty}^{\gamma+\mathrm{i} \infty} dE \, \frac{e^{EL}}{\left[E+ u I_d\left(E,\epsilon\right)\right]^2}
} 
{\int_{\gamma-\mathrm{i} \infty}^{\gamma+\mathrm{i} \infty} dE \, \frac{e^{EL}}{\left[E+ u I_d\left(E,\epsilon\right)\right]}} 
\end{eqnarray}
This completes the scheme for the solution.% In the following, we will perform an explicit computation in the case $d=3$.
%%%%%%%%%%%%%%%%%%%%%%%%%%%%%%%%%%%%%%%%%%%%%%%%%%%%%%%%%%%%%%%%%%%%%%%%%%%%%%%%%%%%%%%%%%%%%%%%
\subsubsection{First order correction from perturbation theory}
\label{subsection:first}
%%%%%%%%%%%%%%%%%%%%%%%%%%%%%%%%%%%%%%%%%%%%%%%%%%%%%%%%%%%%%%%%%%%%%%%%%%%%%%%%%%%%%%%%%%%%%%%%%%%%%%%%%%%%%%%%%%%%%%%%%%%%%%%%%%%%%%%%%%%%%%%%%%%%%%%%%%%%
%%%%%%%%%%%%%%%%%%%%%%%%%%%%%%%%%%%%
%The alternative procedure hinges on the formalism obtained in Section \ref{subsec:laplace}.

In  $d=3$, the relevant  integral (\ref{laplace:eq11}) reads
\begin{eqnarray}
\label{first:eq1}
I_3\left(E,\epsilon\right) &=& \frac{1}{4 \pi^2} \int_{-\infty}^{+\infty} dq \, \frac{q^2}{E+b q^2/6 + \epsilon q^4}
\end{eqnarray}
where the integral has been extended to negative values by taking advantage of the parity of the integrand. The integral
can be easily computed by contour method by extending the contour in the upper plane and noting that only two of the four poles
are then included. These are to lowest order in $\epsilon$, $q_1=+\mathrm{i} \sqrt{6 E}{b}$ and $q_2=+\mathrm{i} \sqrt{b}{6 \epsilon}$.
This produces  the result 
\begin{eqnarray}
\label{first:eq2}
I_3\left(E,\epsilon\right) &=& -\frac{3}{2 \pi b} \sqrt{\frac{6 E}{b}}+ \frac{1}{4\pi} \sqrt{\frac{6}{b \epsilon}}
\end{eqnarray}
Once again, only the lowest correction in $\epsilon$ has been included. Clearly the integral is divergent for $\epsilon \to 0$
but this divergence can be accounted for using a renormalizing procedure, as explained in Ref.\cite{Muthukumar86} and they
turn out to be irrelevant for the computation of the $\langle R^2 \rangle$ as it should.

On dropping the $\epsilon$ dependent term in Eq. \ref{first:eq2},%(\ref{perturbation:eq2}), 
this can be inserted into Eq.(\ref{laplace:eq18}),
that can then be expanded in powers of $\beta u$ to first order. The result is
\begin{eqnarray}
\label{first:eq3}
\left \langle R^2 \right \rangle &=& b 
\frac{ 
\left[
\int_{\gamma-\mathrm{i} \infty}^{\gamma+\mathrm{i} \infty} dE \, \frac{e^{EL}}{E^2}
+ 2 u \frac{3}{2 \pi} \sqrt{\frac{6}{b}}
\int_{\gamma-\mathrm{i} \infty}^{\gamma+\mathrm{i} \infty} dE \, \frac{e^{EL}}{E^{5/2}} + \ldots \right]
}
{ 
\left[
\int_{\gamma-\mathrm{i} \infty}^{\gamma+\mathrm{i} \infty} dE \, \frac{e^{EL}}{E}
+ u \frac{3}{2 \pi} \sqrt{\frac{6}{b}}
\int_{\gamma-\mathrm{i} \infty}^{\gamma+\mathrm{i} \infty} dE \, \frac{e^{EL}}{E^{3/2}} + \ldots \right]
}.
\end{eqnarray}
All integrals can then be performed  by using the result
\begin{eqnarray}
\label{first:eq4}
\int_{\gamma-\mathrm{i} \infty}^{\gamma+\mathrm{i} \infty} dE \, \frac{e^{EL}}{E^{\nu}}&=& \frac{L^{\nu-1}}{\Gamma\left(\nu\right)}.
\end{eqnarray}
Higher orders and additional details  can be found in Ref.\cite{Muthukumar84}.
The final result has  been quoted in Eq. (\ref{first:eq5}).
%%%%%%%%%%%%%%%%%%%%%%%%%%%%%%%%%%%%%%%%%%%%%%%%%%%%%%%%%%%%%%%%%%%%%%%%%%%%%%%%%%%%%%%%%%%%%%%%%%%%%%%%%%%%%%%%%%%%%%%%%%%%%%%%%%%

%%%%%%%%%%%%%%%%%%%%%%%%%%%%%%%%%%%%%%%%%%%%%%%%
\section{Issue of thermodynamic limit}
\label{app:issue-therm-limit}
The size of a polymer $R$ is a geometric quantity which is generally
not a conventional thermodynamic variable.  However the discrete
polymer model introduced here allows one to translate the polymer
problem to a more familiar language for which one may associate
standard thermodynamic quantities.

The bond variables introduced in Eqs. \eqref{elementary:eq1}, and
\eqref{discrete:eq2} can be taken as spin like variables whose allowed
orientations depend on the dimensionality and the topology of the
space (e.g., continuum or lattice).  The interactions of the monomers
can also be expressed as interactions among the spins, not necessarily
restricted to simple two spin interactions as in Eq.
\eqref{discrete:eq2}.  The polymer problem is then exactly equivalent
to a statistical mechanical problem of a collection of spins at a
given temperature $T$.  The response function of such a collection of
spins is the susceptibility which measures the response of the total
spin (i.e. total magnetization) to a uniform magnetic field.  The end
to end distance of the polymer ${\mathbf R}$ turns out to be the total
spin ${\mathbf M}=\sum_i {\mathbf r}_i$, as noted is Sec
\ref{subsec:discrete}.

The fluctuation-response theorem connects the susceptibility $\chi_N$
to the fluctuation of the total spin, (see Sec {subsec:marko}) as
\begin{equation}
  \label{eq:29}
  \chi_N\sim <{\mathbf M}^2> - <{\mathbf{M}}>^2\sim <R^2> - <{\mathbf{R}}>^2,
\end{equation}
and by symmetry, $<{\mathbf{R}}>=0$.  Therefore the susceptibility of
the spin system, as a magnetic model, corresponds to the mean square
end-to-end distance of the polymer.  As a magnetic system, the primary
requirement is to have an extensive susceptibility which means
$\chi_N\propto N$ for $N$ spins, at least for large $N$.  The
stringent requirement of a thermodynamic limit as a magnetic model
would enforce only the Gaussian behaviour of the polymer.  In
contrast, the susceptibility per spin would behave as
\begin{equation}
  \label{eq:30}
 \chi\equiv \lim_{N\to\infty}\frac{\chi_N}{N}\sim N^{2\nu
   -1}\to \left\{ \begin{array}{lcl}
                        \infty&\quad&{\rm (good\ solvent)},\\
                        0      &\quad&{\rm (poor \ solvent)},
                         \end{array}\right. ,
\end{equation}
for the spin models that correspond to an interacting discrete
polymer.  Interestingly, the polymer size exponent is linked to the
finite size behaviour of the spin-problem as $N\to\infty$.

This points towards the care needed in using thermodynamics and
extensivity in polymer problems.

%%%%%%%%%%%%%%%%%%%%%%%%%%%%%%%%%%%%%%%%%%%%%%%%%%%%%%%%%%%%%%%%%%%%%%%%%%%%%%%%%%%%%%%%%%%%%%%%%%%%%%%%%%%%%%%%%%%%%%%%%%%%%%
\section{The structure factor of a Gaussian chain}
\label{app:structure_gaussian}
%%%%%%%%%%%%%%%%%%%%%%%%%%%%%%%%%%%%%%%%%%%%%%%%%%%%%%%%%%%%%%%%%%%%%%%%%%%%%%%%%%
Consider the structure factor 
\begin{eqnarray}
\label{gaus-struct:eq1}
S_0\left(\mathbf{k}\right)&=& \frac{1}{N} \sum_{ij=1}^N \left \langle e^{\rm{i} \mathbf{k}\cdot \left(\mathbf{r}_{i}-\mathbf{r}_{j}\right)}
\right \rangle_{0}.
\end{eqnarray}
For a Gaussian chain, we know that
\begin{eqnarray}
\label{gaus-struct:eq2}
\left \langle \left(\mathbf{r}_{i}-\mathbf{r}_{j} \right)_{\mu} \left(\mathbf{r}_{i}-\mathbf{r}_{j} \right)_{\nu} \right \rangle_{0} 
&=& \frac{\delta_{\mu \nu}}{d} 
\left \langle \left(\mathbf{r}_{i}-\mathbf{r}_{j} \right)^2 \right \rangle_{0} = \frac{\delta_{\mu \nu}}{d} \left \vert i-j \right \vert b^2,
\end{eqnarray}
and hence
\begin{eqnarray}
\label{gaus-struct:eq3}
\left \langle e^{\rm{i} \mathbf{k}\cdot \left(\mathbf{r}_{i}-\mathbf{r}_{j}\right)} \right \rangle_{0} &=& 
\exp\left[-\frac{1}{2} \sum_{\mu \nu} k_{\mu} k_{\nu} \left \langle \left(\mathbf{r}_{i}-\mathbf{r}_{j}\right)_{\mu} 
\left(\mathbf{r}_{i}-\mathbf{r}_{j}\right)_{\nu} \right \rangle_{0} \right] = e^{-\frac{1}{2} k^2 \frac{\left\vert i-j \right \vert}{d}}. 
\end{eqnarray}
Therefore we find
\begin{eqnarray}
\label{gaus-struct:eq4}
S_0\left(\mathbf{k}\right)&\sim& \frac{1}{N} \int_{0}^{N} d n_{1}  \int_{0}^{N} d n_{2} 
e^{-\frac{1}{2} k^2 \frac{\left\vert n_1-n_2 \right \vert}{d}} = N F_{D}\left(N \frac{k^2 b^2}{2d} \right),
\end{eqnarray}
where we have introduced the Debye function
\begin{eqnarray}
\label{gaus-struct:eq5}
F_D\left(x\right)&=& \frac{2}{x^2}\left(x-1+e^{-x}\right).
\end{eqnarray}
Dimensionally, $k$ is like an inverse of length and we see that the structure factor involves 
the dimensionless variable $kR_0$.  The scale for $k$ is set by the overall size of the polymer, 
not its microscopic scales.
%%%%%%%%%%%%%%%%%%%%%%%%%%%%%%%%%%%%%%%%%%%%%%%%%%%%%%%%%%%%%%%%%%%%%%%%%%%%%%%%%%%%%

%%%%%%%%%%%%%%%%%%%%%%%%%%%%%%%%%%%%%%%%%%%%%%%%%%%%%%%%%%%%%%%%%%%%%%%%%%%
\section{Exact solution of the freely jointed chain model with external force}
\label{app:exact}
%%%%%%%%%%%%%%%%%%%%%%%%%%%%%%%%%%%%%%%%%%%%%%%%%%%%%%%%%%%%%%%%%%%%%%%%%%%%%%%% 
The partition function Eq.(\ref{discrete:eq1}) can be solved exactly in the
absence of the interaction term ($K=0$), when the model reduces to the
freely jointed chain (FJC) \cite{Doi86}. In this case each term of Eq.(\ref{discrete:eq1})
decouples and we can use the result 
\begin{eqnarray}
\label{exact:eq1}
\int d^3 \hat{\mathbf{T}} \frac{1}{2 \pi } \delta\left(\hat{\mathbf{T}}^2 -1\right) e^{b \mathbf{f}\cdot\hat{\mathbf{T}}} &=& 
\frac{\sinh\left(f b\right)}{f b}
\end{eqnarray}
so that the configurational partition function becomes
\begin{eqnarray}
\label{exact:eq2}
Z&=& \left[\frac{\sinh\left(f b\right)}{f b}\right]^N
\end{eqnarray}
Introducing the physical force $f_{\text{phys}}=f/\beta$, we then have that
\begin{eqnarray}
\label{exact:eq3}
\left \langle z \right \rangle &=& \frac{\partial}{\partial\left(\beta f_{\text{phys}} \right)} \ln Z\left(f_{\text{phys}}\right)
\end{eqnarray}
This gives the well known result
\begin{eqnarray}
\label{exact:eq4}
\frac{\left \langle z \right \rangle}{N b} &=& \mathcal{L}\left(\beta f_{\text{phys}} b\right)
\end{eqnarray}
where the Langevin function $\mathcal{L}$ is defined as
\begin{eqnarray}
\label{exact:eq5}
\mathcal{L}\left(x\right) &=& \coth \left(x\right)-\frac{1}{x}
\end{eqnarray}

In the $\beta f_{\text{phys}} b \ll 1$ limit, Eq.(\ref{exact:eq3}) can be expanded and gives to leading order \cite{Fixman78}
\begin{eqnarray}
\label{exact:eq6}
\frac{\left \langle z \right \rangle}{ N b} &=& \frac{1}{3} \beta f_{\text{phys}} b + \ldots
\end{eqnarray} 
%%%%%%%%%%%%%%%%%%%%%%%%%%%%%%%%%%%%%%%%%%%%%%%%%%%%%%%%%%%%%%%%%%%%%%%%%%%%%%%%%%%%%%%%%%%%%%
\section{Derivation of the Green function for semi-flexible polymer}
\label{app:green}
%%%%%%%%%%%%%%%%%%%%%%%%%%%%%%%%%%%%%%%%%%%%%%%%%%%%%%%%%%%%%%%%%%%%%%%%%%%%%%%%%%%%%%%%%%%%%
We start from the following addition theorem \cite{Kleinert90}
\begin{eqnarray}
\label{green:eq1}
e^{\mu \hat{\mathbf{T}} \cdot \hat{\mathbf{T}}^{\prime}} &=&  4 \pi 
\sqrt{\frac{\pi}{2 \mu}} \sum_{l=0}^{+\infty} \sum_{m=-l}^{+l} I_{l+1/2} \left(\mu\right)
Y_{lm}\left(\hat{\mathbf{T}}\right) Y_{lm}^{*} \left(\hat{\mathbf{T}}^{\prime} \right) 
\end{eqnarray}
where $I_{\nu}(z)$ is the modified Bessel function \cite{Abramowitz72}
so that Eq.(\ref{structure:eq8}) becomes
\begin{eqnarray}
\label{green:eq2}
G_{0L}\left(\hat{\mathbf{T}}_0,\hat{\mathbf{T}}_N \right) &=& e^{-N l_p/b} \frac{1}{\left(4 \pi\right)^{N-1}} \left( \frac{\pi b}{2 l_p} \right)^{N/2}
\left(4 \pi\right)^N \sum_{l_1,\ldots,l_N} \sum_{m_1,\ldots,m_N} I_{l_1+1/2}\left(\frac{l_p}{b}\right) \ldots I_{l_N+1/2}\left(\frac{l_p}{b}\right) 
Y_{l_1 m_1}\left(\hat{\mathbf{T}}_0\right) Y_{l_N m_N}^{*} \left(\hat{\mathbf{T}}_N \right)  \nonumber \\
&& \int d \hat{\mathbf{T}}_1  Y_{l_1 m_1}^{*}\left(\hat{\mathbf{T}}_1\right) Y_{l_1 m_1} \left(\hat{\mathbf{T}}_1 \right) \ldots
\int d \hat{\mathbf{T}}_{N-1}  Y_{l_{N_1} m_{N-1}}^{*}\left(\hat{\mathbf{T}}_{N-1}\right) Y_{l_N m_N} \left(\hat{\mathbf{T}}_{N-1} \right)
\end{eqnarray}
Using the orthogonality relation \cite{Abramowitz72}
\begin{eqnarray}
\label{green:eq3}
\int d \hat{\mathbf{T}} Y_{l_1 m_1}\left(\hat{\mathbf{T}}\right) Y_{l_2 m_2}^{*} \left(\hat{\mathbf{T}} \right) &=& \delta_{l_1 l_2} \delta_{m_1 m_2} 
\end{eqnarray}
Eq.(\ref{green:eq2}) reduces to 
\begin{eqnarray}
\label{green:eq4}
G_{0L_c}\left(\hat{\mathbf{T}}_0,\hat{\mathbf{T}}_N \right) &=& 4 \pi \left(\frac{\pi b}{2 l_p}\right)^{N/2} \sum_{l=0}^{+\infty} \sum_{m=-l}^{+l}
I_{l+1/2}^{N} \left(\mu\right) Y_{l m}\left(\hat{\mathbf{T}}_0\right) Y_{l m}^{*} \left(\hat{\mathbf{T}}_N \right) 
\end{eqnarray}
In the limit $b \to 0$ we can use the asymptotic expansion for the Bessel function for $\vert z \vert \gg 1$ \cite{Abramowitz72}
\begin{eqnarray}
\label{green:eq5}
I_{\nu}\left(z\right) &=& \frac{e^z}{\sqrt{2 \pi z}} \left[ 1-\frac{4 \nu^2-1}{8z}+ \ldots \right] 
\end{eqnarray}
to obtain
\begin{eqnarray}
\label{green:eq6}
G_{0L_c}\left(\hat{\mathbf{T}}_0,\hat{\mathbf{T}}_N \right) &=& 4 \pi \sum_{l=0}^{+\infty} \sum_{m=-l}^{+l} e^{-L_c l(l+1)/(2 l_p)}
Y_{l m}\left(\hat{\mathbf{T}}_0\right) Y_{l m}^{*} \left(\hat{\mathbf{T}}_N \right)
\end{eqnarray}
that is the result given in Eq.(\ref{structure:eq10}). Note that in obtaining (\ref{green:eq4}) and (\ref{green:eq6}), 
we have set $L_c=Nb$ and used the relation $b=2 l_p$ between the Kuhn and the persistence length for the WLC model \cite{Rubinstein03}.%%%%%%%%%%%%%%%%%%%%%%%%%%%%%%%%%%%%%%%%%%%%%%%%%%%%%%%%%%%%%%%%%%%%%%%%%%%%%%%%%%%%%%%%%%%%%%%%%%%%%%%%%%%%%%%%%%%%%%%%%%%%%%%%%%%%%%%%
\section{Calculation of $\langle [r_z(s)-r_z(s^{\prime})]^2 \rangle$}
\label{app:integral}
%%%%%%%%%%%%%%%%%%%%%%%%%%%%%%%%%%%%%%%%%%%%%%%%%%%%%%%%%%%%%%%%%%%%%%%%%%%%%%%%%%%%%%%%%%%%%%%%%%%%%%%%%%%%%%%%%%%%%%%%%%%%%%%%%%%%%%%%
To compute $\langle [r_z(s)-r_z(s^{\prime})] \rangle$ given in Eq.(\ref{structure:eq12}), we need to compute the average quantity
\begin{eqnarray}
\label{integral:eq1}
\left \langle \hat{T}_z\left(s_1\right)  \hat{T}_z\left(s_2\right) \right \rangle &=& \frac{
\int d \hat{\mathbf{T}}_1 \int d \hat{\mathbf{T}}_2
G_{s_1 s_2} \left(\hat{\mathbf{T}}_1,\hat{\mathbf{T}}_2\right) \hat{T}_{1z} \hat{T}_{2z}}
{\int d \hat{\mathbf{T}}_1 \int d \hat{\mathbf{T}}_2
G_{s_1 s_2} \left(\hat{\mathbf{T}}_1,\hat{\mathbf{T}}_2\right)}
\end{eqnarray}
Using the first two spherical harmonics\cite{Abramowitz72}
\begin{eqnarray}
\label{integral:eq2}
Y_{00}\left(\hat{\mathbf{T}}\right) = \frac{1}{\sqrt{4 \pi}} &\qquad& Y_{01}\left(\hat{\mathbf{T}}\right) = \sqrt{\frac{3}{4 \pi}} \hat{T}_z
\end{eqnarray}
and the orthogonality relations (\ref{green:eq3}), Eq.(\ref{integral:eq1}) reduces after few steps to
\begin{eqnarray}
\label{integral:eq3}
\left \langle \hat{T}_z\left(s_1\right)  \hat{T}_z\left(s_2\right) \right \rangle &=& \frac{1}{3} 
\exp\left[\frac{\left \vert s_2 -s_1 \right \vert}{l_p} \right]
\end{eqnarray}
that coincides with the expected result Eq.(\ref{structure:eq6b}), taking into account the other two components $x$ and $y$.

Upon inserting this result into Eq.(\ref{structure:eq12}), one can use a $s-$ordering procedure so that ($s>s^{\prime}$)
\begin{eqnarray}
\label{integral:eq4}
 \left \langle \left[r_z\left(s\right)-r_z\left(s^{\prime}\right) \right]^2 \right \rangle &=&  2 \int_{{s}^{\prime}}^{s} d s_1 
\int_{{s}^{\prime}}^{s_1} d s_2 \frac{1}{3} e^{\left(s_1-s_2\right)/l_p}=
\frac{2}{3} l_p^2 \left[ \frac{s-s^{\prime}}{l_p}-1+e^{-\left(s-s^{\prime}\right)/l_p}\right]
\end{eqnarray}

%%%%%%%%%%%%%%%%%%bibliography %%%%%%%%%%%%%%%%%%%%%%%%%%%%%%%%%%%%
%\references
%\bibliography

%%%%%%


\begin{thebibliography}{99}


%1%
\bibitem{Flory53} P.J. Flory, \textit{Principles of Polymer Chemistry}, (Cornell Univ. Press 1953)

%2%
\bibitem{Flory69} P.J. Flory, \textit{Statistical Mechanics of Chain Molecules}, (Wiley 1969)

%3%
\bibitem{yamakawa} H. Yamakawa, \textit{Modern Theory of Polymer Solutions} (Harper $\&$ Row, New York 1971).

%4%
\bibitem{deGennes79} P.G de Gennes, \textit{Scaling Concepts in Polymer Physics}, (Cornell Univ. Press 1979)

%5%
\bibitem{Doi86} M. Doi and S.F. Edwards, \textit{Theory of Polymer Dynamics} (Oxford Univ. Press 1986)
(see expecially Section 2.5.3)

%6%
\bibitem{Freed87} K. F. Freed,\textit{Renormalization Group theory of Macromolecules} (Wiley 87)

%7%
\bibitem{desCloizeaux90} J. des Cloizeaux and G. Jannik, \textit{Polymer in Solutions}, (Clarendon Press 1987) 

%8%
\bibitem{lifshitz78} I.M.Lifshitz, A.Yu.Grosberg, and  A.R. Khokhlov,  Rev. Mod. Phys. {\bf 50}, 683 (1978).
 
%9%
\bibitem{Grosberg94} A. Yu. Grosberg and A. R. Khokhlov, \textit{Statistical
Physics of macromolecules} (AIP press 1994)

%10%
\bibitem{Rubinstein03} M.Rubinstein and R.H. Colby, \textit{Polymer Physics} (Oxford University Press 2003)

%11%
\bibitem{Hughes}    B.D. Hughes, \textit{Random Walks and Random
    Environments. Vols 1: Random Walks, Volume 2: Random Environments.}
    (Clarendon Press, Oxford, 1996)

%12%
\bibitem{Vanderzande}    C. Vanderzande, \textit{ Lattice Models of Polymers} 
           (Cambridge University Press, 1998)
%13%
\bibitem{Giacomin} G. Giacomin, \textit{Random Polymer Models}  (Imperial College Press, London,  2007). 

%14%
\bibitem{Raphael92} E. Rapha\"el, G.H. Fredrickson and P. Pincus,
J. Phys. II France \textbf{2}, 1811 (1992)

%15%
\bibitem{Kamien93} R.D. Kamien, J. Phys. I France \textbf{3}, 1663 (1993)

%16%
\bibitem{Orland94} H. Orland, J. Phys. I France \textbf{4}, 101 (1994)

%17%
\bibitem{bhattacharjee05} S. M. Bhattacharjee   in {\it Statistics of Linear Polymers in Disordered Media}, ed. by B.K. Chakrabarti(Elsevier 2005)

%18%
\bibitem{Edwards65} S.F. Edwards, Proc. Phys. Soc. \textbf{85}, 613 (1965)

%19%
\bibitem{SMB91} S. F. Edwards in \textit{Polymer Physics: 25
    years of the Edwards Hamiltonian} Ed. by S. M. Bhattacharjee
  (World Scientific, Singapore, 1992)

%20%
\bibitem{Ptitsyn68} O. B. Ptitsyn, A. K. Kron, and Y. Y. Eizner, J. Polym. Sci. Part
C 16, 3509 (1968).

%21%
\bibitem{deGennes75} P.G. de Gennes, J.Phys. France \textbf{36}, L-55 (1975)

%22%
\bibitem{Edwards79a} S. F. Edwards and P. Singh, J. Chem. Soc. Faraday Trans. II \textbf{75}, 1001 (1979) (see Appendix A)

%23%
\bibitem{Muthukumar84} M. Muthukumar and B. G. Nickel, J. Chem. Phys. \textbf{80}, 5839 (1984)

%24%
\bibitem{Muthukumar86} M. Muthukumar, J. Chem. Phys. \textbf{85}, 4722 (1986)

%25%
\bibitem{Ha95} B.Y. Ha and D. Thirumalai J. Chem. Phys \textbf{103}, 9408 (1995)

%26%
\bibitem{Ha97} B.Y. Ha and D. Thirumalai J. Chem. Phys \textbf{106}, 4243 (1997)

%27%
\bibitem{Rosa03} A. Rosa, T. X. Hoang, D. Marenduzzo and A. Maritan, Macromolecules \textbf{36}, 10095 (2003)

%28%
\bibitem{Marko95} J.F. Marko and E.D. Siggia, Macromolecules \textbf{28}, 8759 (1995)

%29%
\bibitem{Fisher67} M.E. Fisher, Rep. Prog. Phys. \textbf{30}, 615 (1967).

%30%
\bibitem{Nienhuis} B. Nienhuis Phys. Rev. Letts. {\bf 49} 1062 (1982); J. Stat. Phys. {\bf 34}, 731 (1984).

%31%
\bibitem{Zinn-Justin90} J. Zinn-Justin \textit{Quantum Field theory and Critical Phenomena} (Clarendon Press, Oxford 1990)

%32%
\bibitem{Kleinert90} H. Kleinert, \textit{Path Integrals in Quantum Mechanics, Statistics and Polymer Physics} (World Scientific 1990) (see Chapt. 15.7)

%33%
\bibitem{note1} Note that, within the de Gennes' mapping between spin models
and polymers with the number of components going to zero, this quantity is, in fact a
compressibility (or a susceptibility).  See V. J. Emery Phys. Rev.B \textbf{11} 239 (1975)

%34%
\bibitem{Tintah99} J.T. Tintah, C. Pierleoni, J.-P. Ryckaert, Phys. Rev. E \textbf{60}, 7010 (1999)

%35%
\bibitem{kapri} R. Kapri, S. M. Bhattacharjee and F. Seno   Phys. Rev. Lett. 93, 248102 (2004).

%36%
\bibitem{Mukherji}  S. Mukherji and S. M. Bhattacharjee,  J. Phys. A {\bf 26}, L1139 (1993); Phys. Rev. E {\bf 48}, 3427 (1993); Phys. Rev. E {\bf 63}, 051103 (2001).

%37%
\bibitem{poland} D. Poland and H. A. Scheraga, J.Chem.Phys.  {\bf 45}, 1464 (1966).

%38%
\bibitem{flavio91} C. Vanderzande, A. L. Stella and F. Seno, Phys, Rev. Lett. {\bf 67}, 2757 (1991).

%39%
\bibitem{flavio88} F. Seno and A. L. Stella, J. Phys, France. {\bf 49}, 739 (1988).

%40%
\bibitem{flavio91b} S. L. A. Dequeiroz, F. Seno,  A. L.   Stella, 
 J.  de Phys. I   {\bf 1} 339 (1991).

%41%
\bibitem{bkc95} K. Barat, B. K. Chakrabarti,  Phys. Rept.  {\bf 258}, 377 (1995).

%42%
\bibitem{Swislow80} G. Swislow, S. T. Sun, I. Nishio, and T. Tanaka, Phys. Rev.
Lett. {\bf 44}, 796 (1980).

%43%
\bibitem{nakata97}  M. Nakata  and T. Nakagawa, Phys. Rev.  E {\bf 56}, 3338 (1997)

%44%
\bibitem{Maritan87} R. Dekeyser, A. Maritan and A. Stella, Phys. Rev. A {\bf 36}, 2338 (1987).

%45%
\bibitem{Schulman} L. S. Schulman, \textit{Techniques and Applications of Path Integrals} (Dover 2005)
see Eq.(7.19).

%46%
\bibitem{Redner82} S. Redner, and A. Coniglio, J.Phys. A: Math. Gen \textbf{15}, L273 (1982)

%47%
\bibitem{Lubensky82} T.C. Lubensky, and J. Vannimenus, J. Phys.(Paris) \textbf{43}, L377 (1982)

%48%
\bibitem{Boris96} D. Boris, M. Rubinstein, Macromolecules \textbf{29}, 7251 (1996)

%49%
\bibitem{Kroger10} M. Kr\"oger, O. Peleg, and A. Halperin, Macromolecules \textbf{43},6213 (2010)

%50%
\bibitem{Marenduzzo02} D. Marenduzzo, S.M. Bhattacharjee, A. Maritan, E. Orlandini and F. Seno, Phys. Rev. Lett. \textbf{88}, 028102 (2001)

%51%
\bibitem{Maren09} D. Marenduzzo, A. Maritan, E. Orlandini, F. Seno, A. Trovato, J. Stat. Mech. L04001 (2009).

%52%
\bibitem{unzip} For DNA unzipping transition, see, e.g., S. M. Bhattacharjee, J. Phys. A: Math. Gen. {\bf 33}, L423 (2000); D. Marenduzzo, A. Maritan and A. Trovato, Phys. Rev E {\bf 64}, 031901 (2001),  

%53%
\bibitem{anisimov05} M. A. Anisimov, A. F. Kostko, J. V. Sengers, and I. K. Yudin, J. Chem. Phys. 123, 164901 (2005)

%54%
\bibitem{Coxter89} H.S.M. Coxter \textit{Introduction to Geometry} (Wiley 1989) (see Sect. 17.6)

%55%
\bibitem{Kamien99} R. Kamien, Rev. Mod. Phys. \textbf{74}, 953 (1999)

%56%
\bibitem{Bhattacharjee87} S.M. Bhattacharjee and M. Muthukumar J. Chem. Phys \textbf{86}, 411 (1987)

%57%
\bibitem{Odijk95} T. Odijk, Macromolecules \textbf{28}, 7016 (1995)

%58%
\bibitem{Bosch85} A. Ten Bosh and P. Sixou, J. Chem. Phys. \textbf{83}, 899 (1985)

%59%
\bibitem{Livadaru03} L. Livadaru, R.R. Netz, and H.J. Kreuzer, Macromolecules \textbf{36}, 3732 (2003)

%60%
\bibitem{Stepanow04} S. Stepanow, Eur. Phys. J. B \textbf{39}, 499 (2004)

%61%
\bibitem{Doniach96} S. Doniach, T. Garel, and H. Orland, J. Chem. Phys. \textbf{105}, 1601 (1996)

%62%
\bibitem{Becker10} N.B. Becker, A. Rosa, and R. Everaers, Eur. Phys. J. E \textbf{32}, 53 (2010)

%63%
\bibitem{Shimada88} T. Shimada, M. Doi, and K. Okano, J. Chem. Phys. \textbf{88}, 2815 (1988)

%64%
\bibitem{Gradshteyn00} I. S. Gradshteyn and I.M. Ryzhik, \textit{Table of Inbtegrals, Series, and Product}, (Academic Press 2000) (see 1.145.2)

%65%
\bibitem{Abramowitz72} M. Abramowitz and I. Stegun, \textit{Handbook of Mathematical Functions} (Dover 1972)



%\bibitem{Mazenko03} G.F. Mazenko \textit{Fluctuations, Order, and Defects} (Wiley 2003)
%(see Appendix E)

%\bibitem{Banavar05} J.R.Banavar, T.X. Hoang, and A. Maritan, J. Chem. Phys. \textbf{122} 234910 (2005)

%66%

\bibitem{Fixman78} M. Fixman, J. Kovac, J. Chem. Phys. \textbf{58}, 1564 (1978)


%\bibitem{Fixman55} M. Fixman, J. Chem. Phys. \textbf{23}, 1656 (1955)

%\bibitem{Saito67} N. Saito, K. Takahashi and Y. Yunoki, J. Phys. Soc. Jpn. \textbf{22}, 219 (1967)



%\bibitem{Edwards79b} S. F. Edwards and E. F. Jeffers, J. Chem. Soc. Faraday Trans. II \textbf{75}, 1020 (1979) (see Appendix )

%\bibitem{Muthukumar82} M. Muthukumar and S.F. Edwards J. Chem. Phys. \textbf{76}, 2720 (1982)

%\bibitem{Bhattacharjee91}  S. M. Bhattacharjee and J. J. Rajasekaran, Phys. Rev. A \textbf{44},  6202 (1991) 



\end{thebibliography}
\end{document}